\documentclass[12pt]{article}
\usepackage{geometry}
\usepackage[hang]{footmisc}
\usepackage{authblk}
\usepackage{amsmath,mathtools,amsthm,amssymb}
\usepackage{tabularx}
\usepackage{graphicx,graphics}
\usepackage{color,xcolor}
\usepackage{enumitem}
\usepackage{hyperref}
\usepackage{physics}
\usepackage{braket}
\usepackage{cite}

\usepackage{caption}
\definecolor{darkgreen}{RGB}{0,128,0}
\usepackage{float}

\newcommand{\id}{\mathbb{I}}

\newtheorem{proposition}{Proposition}
\newtheorem{definition}{Definition}
\newtheorem{theorem}{Theorem}

\newcommand{\cA}{{\mathcal A}}
\newcommand{\cB}{{\mathcal B}}
\newcommand{\cC}{{\mathcal C}}
\newcommand{\cD}{{\mathcal D}}

\newcommand{\cF}{{\mathcal F}}

\newcommand{\cH}{{\mathcal H}}

\newcommand{\cJ}{{\mathcal J}}

\newcommand{\cL}{{\mathcal L}}
\newcommand{\cM}{{\mathcal M}}
\newcommand{\cN}{{\mathcal N}}
\newcommand{\cO}{{\mathcal O}}
\newcommand{\cP}{{\mathcal P}}

\newcommand{\cT}{{\mathcal T}}

\newcommand{\cV}{{\mathcal V}}

\newcommand{\cY}{{\mathcal Y}}
\newcommand{\cZ}{{\mathcal Z}}

\def\a{\alpha}
\def\b{\beta}
\def\c{\chi}

\def\d{\delta}
\def\D{\Delta}
\def\e{\epsilon}

\def\g{\gamma}
\def\G{\Gamma}

\def\k{\kappa}
\def\l{\lambda}

\def\p{\phi}

\def\Q{\Theta}
\def\r{\rho}
\def\s{\sigma}
\def\S{\Sigma}
\def\t{\tau}

\def\w{\omega}

\def\y{\psi}
\def\Y{\Psi}

\begin{document}

\title{\LARGE\textbf{Searching for emergent spacetime in spin glasses}\\ \large An algebraic perspective}
\author[1,$*$]{Dimitris Saraidaris}
\author[1,$*$]{Leo Shaposhnik}

\affil[1]{\normalsize Department of Physics, Freie Universität Berlin, 14195 Berlin, Germany}

\maketitle

\renewcommand*{\thefootnote}{\fnsymbol{footnote}}
\footnotetext[1]{Contributed equally. 
d.saraidaris@fu-berlin.de,
l.shaposhnik@fu-berlin.de}
\abstract{Recent work on algebraic formulations of holographic dualities in terms of large $N$ algebras has revealed a deep connection between the properties of the associated spectral functions and the emergence of a semiclassical spacetime and causal horizons therein. One of the main lessons is that, for a radial direction to emerge, the spectral function has to exhibit non-compact support. Furthermore, there exist conjectures upon a possible duality between complex gravitational configurations and glassy systems. The goal of this paper is to combine these ideas by studying many-body quantum-mechanical systems and assess in which parameter regimes they could potentially be holographic. Thus, we compute the spectral functions of three many-body systems with quenched disorder, the SYK model, the $p$-spin model and the SU$(M)$ Heisenberg chain in the large $N$ limit and present results in different parameter regimes. Our main finding is that in the quantum spin glass phase of the SU$(M)$ Heisenberg model, the spectral function develops an exponential tail, similar to the large $q$ limit of SYK. Furthermore, we demonstrate the presence of an infinite family of quasiparticle excitations deep in the spin liquid phase of the $p$-spin model, which could point towards an emergent type I von Neumann algebra. In addition, we demonstrate the presence of an exponential tail in the spectral function for all cases without compact support and conformal symmetry. Motivated by this observation, we prove that no low-energy operator can detect a nontrivial bulk causal structure, if the spectral function has exponentially decaying tails.}

\renewcommand*{\thefootnote}{\arabic{footnote}}
\thispagestyle{empty}
\newpage
\thispagestyle{empty}
\tableofcontents 

\pagenumbering{arabic}   % start arabic page numbers
\setcounter{page}{1}  
\section{Introduction}
Since the advent of holographic dualities \cite{tHooft:1993dmi,Susskind:1994vu,Maldacena:1997re,Witten:1998qj}, there has been a surge of interest into the connection between emergent properties in many-body quantum-mechanical systems and gravitational phenomena \cite{Hartnoll:2009sz,Herzog:2009xv,Sachdev:2010uj,Faulkner:2010tq,Faulkner:2009wj,Hartnoll:2011fn,Sachdev:2011wg,Maldacena:2016hyu}. With the discovery of the SYK model \cite{KitaevTalks,Maldacena:2016hyu} and its relation to the low energy description of two-dimensional Dilaton gravity \cite{Kitaev:2017awl, Trunin:2020vwy}, it has become clear that the apparent similarities can be put into very concrete terms. Since then, a large body of work has been produced, relating gravity and the dynamics of quantum-mechanical systems and explaining gravitational phenomena from the perspective of many-body physics and quantum computation, see e.g.~\cite{Ryu:2006bv,Sekino:2008he,Shenker:2013pqa,Maldacena:2013xja,Susskind:2014rva,Pastawski:2015qua, Maldacena:2015waa, Parker:2018yvk, Brown:2017jil, Cotler:2017jue,Song:2017pfw,Saad:2018bqo}.

One of the outcomes of this development is the proposal that complex configurations of gravitating bodies, for example multihorizon solutions in supergravity, are related to states in holographic systems that exhibit glassy dynamics \cite{Anninos2011,Kachru:2009xf,Anous2021,Denef:2011ee}. A similar proposal was made in \cite{Anninos2011, Anous2021} regarding a possible connection of glassy physics to the fragmentation instability of AdS$_2$ \cite{Maldacena1999}. While these suggestions have not manifested themselves in a concrete duality, the quantum-mechanical systems describing spin glasses \cite{Sachdev:1992fk,Georges_2001,Cugliandolo1993,Cugliandolo2001,Parcollet:1999itf} are structurally very similar to SYK, so that they form a collection of systems that can function as a testing ground for ideas that attempt to deepen our understanding of holographic dualities. Motivated by this similarity, we will use these models as a framework to investigate novel ideas that have emerged in recent years.

The techniques we apply to these models arose from the use of operator-algebraic methods to study holographic systems \cite{Witten:2021jzq,Leutheusser:2021frk,Witten:2021unn,Furuya:2023fei,Gesteau:2024dhj,Gesteau:2023rrx,Lashkari:2024lkt,Jensen:2024dnl}. First attempts towards such a formulation were already present in the initial stages of AdS/CFT \cite{Rehren:1999jn,Duetsch:2002hc}. This approach attempts to understand the emergence of an extra spatial dimension by understanding the properties of algebras that give a full description of the system in the large $N$ limit. Initial studies of the connection between the behavior of two-point functions in frequency space and emergent bulk quantities were already noted nearly two decades ago in \cite{Festuccia:2005pi, Festuccia:2006sa}. A breakthrough in this direction was made in \cite{Gesteau:2024dhj}, which proposed a framework that makes concrete statements about the existence of a nontrivial bulk causal structure, based on the two-point functions that describe the system in question at $N=\infty$. The main guiding principle for us is that, if the spectral function is compactly supported, i.e. it contains nonzero values only up to a maximum frequency, then no nontrivial bulk causality can emerge. The goal of this project is to apply this framework on large $N$ glassy systems and see whether these show signs of holographic behavior according to the framework of \cite{Gesteau:2024dhj}. Concretely, we study three large $N$ systems with quenched disorder: The SYK model \cite{KitaevTalks}, the spherical $p$-spin model \cite{Cugliandolo2001} and the SU($M$) Heisenberg model \cite{Sachdev:1992fk}, which served as the initial motivation for SYK. Concretely, we numerically solve the large $N$ Schwinger-Dyson equations of those systems and extract their spectral functions. 
Our analysis and findings can be summarized as follows:
\begin{itemize}
    \item[1)] \textbf{Spin liquid states:} The SYK model and the spin liquid phases of the $p$-spin and SU$(M)$ Heisenberg model show completely supported spectral functions that exhibit exponential decay. This behavior is robust in all parameter regimes of each model but with one unique exception. Deeper into the spin liquid phase, the spherical $p$-spin model exhibits an infinite set of quasiparticle excitations, that could signal an emergent type $I$ factor.
    \item[2)] \textbf{Spin glass states:} The spherical $p$-spin model shows signs of a compactly supported spectral function in the spin glass state, which suggests the exclusion of the emergence of nontrivial bulk causality in any regime of finite parameters. Similarly, a large part of the spin glass phase of the SU($M$) Heisenberg model gives spectral functions with compact support. However, there is a parameter regime -- the quantum spin glass phase -- that appears to exhibit non-compact support, opening the possibility for a nontrivial bulk. 
    \item[3)]  \textbf{Exponential decay:} Away from the conformal limit, in all of the studied instances, states that have non-compactly supported spectral functions exhibited asymptotically exponential decay of the spectral function. This point is reminiscent of the universal operator growth hypothesis \cite{Parker:2018yvk}, which asserts exponentially decaying spectral functions for chaotic systems in the infinite temperature state. The framework of \cite{Gesteau:2024dhj} is not yet applicable to this case, as their theorems only make concrete statements about spectral functions with \textit{at most} polynomial decay. As a small step towards extending this framework to include spectral functions with exponential decay, we prove that in that case, no operator that was smeared with a function that grows at most polynomially with frequency can detect a nontrivial bulk causal structure, if it exists.
\end{itemize}

The structure of our paper is as follows: In Section \ref{sec:glasses_and_black_holes}, we review previous observations in high-energy physics that suggested a connection between complex gravitational configurations with glassy systems. This is followed by a basic review of spin glasses and the concept of \textit{replica-symmetry breaking} which we use as a diagnostic of the spin glass state in our numerics. In Section \ref{sec:gff}, we review the algebraic approach to holographic dualities and the framework proposed in \cite{Gesteau:2024dhj}. In Section \ref{sec:numerics},
we present our findings for the three models in question. In Section \ref{sec:no_em_low_energy}, we prove that conventional smearing functions that decay with large frequencies are not capable of detecting a nontrivial bulk causality, if the spectral function shows exponential decay with large frequencies. We end in Section \ref{sec:discussion} with a discussion and possible future directions.

\section{Complex systems and glasses}\label{sec:glasses_and_black_holes}
\subsection{Motivation from high energy physics}

The study of states of high complexity in high energy physics has led to a proposal \cite{Denef:2011ee} that tries to connect systems with a large landscape of allowed configurations to the physics of glassy systems. One paradigmatic example studied in \cite{Anninos2011}
is concerned with the study of black hole molecules.
These are stationary multi-horizon bound states, which are solutions to Einsteins equations and a subset of a vast variety of multi-horizon solutions studied over the years, see \cite{Anninos2011} and references therein. The term ``molecules" refers to the arbitrarily large number of probes that form highly complex but stable configurations under the influence of gravitational and electromagnetic forces.

In \cite{Anninos2011, Anninos2015} configurations were studied, in which a large number of probe BPS black holes are bound to a supermassive black hole. In \cite{Anninos2011} an explicit counting of allowed configurations was performed that demonstrated that the total number is exponential in the charge of the large black hole. This leads to a free-energy landscape with exponentially many minima, which results to phenomena like long relaxation times and memory/``aging" effects \cite{Anninos2015}. This property is also found in glassy systems, as we will discuss in the following section. This leads the authors to suspect that the multi-horizon configurations are dual to states of CFT's that exhibit glass-like properties. A second argument towards this similarity is that, as one heats up the system, the molecules collapse into a single black hole configuration \cite{Anninos2011} which mimics the phenomenon of glass melting into a liquid. 

A second line of thought concerns the phenomenon of AdS$_2$ to exhibit a \textit{fragmentation instability} \cite{Maldacena1999}, where a geometry with a single horizon breaks into a multihorizon geometry. It was argued in \cite{Anous2021}, that the spherical $p$-spin model might represent a dual description of this phenomenon, due to its possibility to preserve the SL$(2,\mathbb{R})$ symmetry across the spin-glass transition which is true for the gravitational system. A similar conjecture was made in \cite{Christos:2021wno} that argues that the vanishing entropy at zero temperature of the SU($M$) Heisenberg model indicates the possibility of the bulk to transit to a horizonless geometry.

\subsection{A review of spin glasses}
In the following sections, we review basic properties of spin glasses, in particular why these systems exhibit behavior that justifies the name of a glass and how this is connected to the notion of \textit{replica-symmetry breaking}. A naive definition of a spin glass \cite{mezard1986spin} is a system of localized spins in the presence of disorder and frustration. In order to get a first idea of what frustration means, we can imagine a triangular lattice whose vertices are occupied by spins. Obviously, it is impossible for all pairs of spins to anti-align at the same time, leading to a quite complex \textit{frustrated} system and to an energy landscape with exponentially many local minima. A real-life example of such geometrical configurations can be found in the CuMn spin glass \cite{Mulder1981}, where the manganese atoms live in random positions and have randomly either ferromagnetic or antiferromagnetic interactions, leading to \textit{frustrated} triangles. 

Now let us discuss a simpler spin model, the Edwards Anderson \cite{Edwards1975} model,
which is a paradigmatic example of a spin glass and has the Hamiltonian
\begin{equation}\label{eq:EA_HAMILTONIAN}
    H = \sum^N_{ij}J_{ij}s_is_j.
\end{equation}
In this model, the $N$ spins $s_i=\pm 1$ interact via the nearest-neighbor couplings $J_{ij}$. In order to model a spin glass, Edwards and Anderson imposed disorder to the system, by drawing the couplings $J_{ij}$ from a Gaussian distribution with zero mean and unit standard deviation. 
Such models with time-independent random couplings
    are usually described as models with \textit{quenched disorder}, which are
    supposed to model the situation, where the timescale on which the couplings
    between individual constituents change is much larger than the dynamics of the
    system, so that they can be considered time-independent. The randomness is
    supposed to mimic systems that are cooled down very rapidly, so that the
    atoms could not place themselves in an ordered location and their interactions are expected to be essentially random.

In the thermodynamic limit $N\rightarrow\infty$, the magnetization $M\equiv \frac{1}{N}\sum_im_i$, where $m_i=\langle s_i \rangle$ is self-averaged to 0 at all temperatures. However, for temperatures lower than a critical temperature $T_c$ each spin freezes into a preferred direction, something that cannot be captured by the magnetization. Thus, the authors introduced a new order parameter $q_{EA} \equiv \frac{1}{N}\sum_im_i^2$, closely related to the magnetic susceptibility, whose nonzero value functions as an order parameter for a spin glass phase as we will see in the next section.

\subsubsection{Replica symmetry breaking and spin glasses}

\begin{figure}[ht]
\centering
\includegraphics[width=0.5\linewidth]{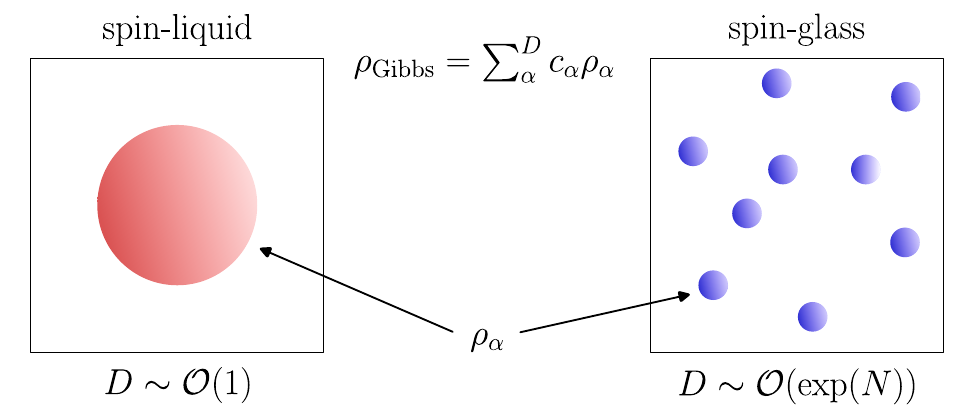}
\caption{Illustration of the Gibbs states in a spin liquid and a spin glass phase. In the spin liquid phase, the Gibbs state is given by a single ergodic state, which when transitioning to the spin glass state ``shatters" into exponentially many ergodic components.}
\label{fig:SLSG_states}
\end{figure}

We now depart from the nearest-neighbor Hamiltonian \eqref{eq:EA_HAMILTONIAN} and consider the Sherrington-Kirkpatrick (SK) model \cite{Sherrington:75}, where the nearest-neighbor
couplings are replaced by all-to-all interactions, i.e., the indices $(i,j)$ can label any two pairs of spins and the couplings $J_{ij}$ are still Gaussian with unit variance. In the following, we want to motivate the definition of a spin glass as a state with replica-symmetry breaking and how this relates to the free-energy landscape of such a system. Our
    presentation closely follows \cite{mezard1986spin}. The
    intuitive picture of a glass is that of a system which, even when coupled to a
    thermal bath, it has very long relaxation times towards thermal equilibrium.
    In systems such as the SK model, one usually finds that, below a transition
    temperature $T_C$, the thermal state becomes a convex combination of an
    exponential number of \textit{ergodic states}  \cite{mezard1986spin}. One intuitive notion
    of a (translation-invariant) ergodic state $\r$ is that at large distances connected correlation functions vanish, i.e., for every pair of local operators $A_x,B_y$ at the locations $x,y$ one has
    \begin{equation}
    \lim_{x \rightarrow \infty} \Tr \r A_x B_y = \Tr(\r A_x)  \Tr(\r B_y).
    \end{equation}
    This property is often referred to as \textit{clustering}.
    One can show
    \cite{Friedli_Velenik_2017} that every translation invariant state uniquely
    decomposes into a convex combination of ergodic components. Thus, the Gibbs state $\r^J =
    \frac{1}{Z_J} e^{-\b H_J}$, considered as a time-translation invariant state, can be written as
    \begin{equation}\label{eq:spin_glass_state}
        \r^J = \sum_\a c^J_\a \r^J_\a,
    \end{equation}
    with $\r_\a$ being ergodic. In the previous language, we would therefore consider a glass to be a system in a state for which the index $\a$ takes $\cO(\exp(N))$ many values, where $N$ is the number of degrees of freedom. 
    Therefore, the Gibbs state ``shatters" below the transition temperature into an
    exponential number of ergodic subspaces. For an illustration, see Fig.~\ref{fig:SLSG_states}.
    A heuristic argument for long relaxation times towards thermal equilibrium
    is then as follows: Common lore suggests \cite{mezard1986spin} that, if one would start close to
    an ergodic state, the system would decay down to the state very fast. However, the transition
    \textit{between} ergodic states takes a time exponentially long in the number of
    degrees of freedom. Therefore, if one starts out in a generic state and the
    goal is to transition to the thermal state through coupling to a bath, this
    takes a very long time, since one has to equilibrate towards an exponential
    number of states, each of which is far from the other. 

    With this shattering into an exponential number of ergodic components comes
    another feature, commonly used as a defining property of spin glasses,
    namely \textit{replica-symmetry breaking}
    \cite{Parisi:79, Parisi:83, mezard1986spin}.
    Since each state in the decomposition \eqref{eq:spin_glass_state} is
    different, the local spins will have different average values, i.e.,
    defining
    \begin{equation}
        m_i^\a := \Tr \r_\a \s_i,    
    \end{equation}
    one has 
    \begin{equation}
        m_i^\a \neq m_{i}^\b,\ \text{if}\ \a \neq \b.
    \end{equation}
    We can then define the \textit{overlap} between states $\r_\a,\r_\b$ as 
    \begin{equation}
        q_{\a \b} := \frac{1}{N} \sum_i m_i^\a m_i^\b.
    \end{equation}
    The self overlap $q_{\a \a} := q_{EA}$ is the Edwards-Anderson parameter mentioned in the previous section.
    Given these overlaps, we can define the function
    \begin{equation}
        P_J(q) = \sum_{\a \b} c^J_\a c^J_\b \d(q_{\a \b}-q),
    \end{equation}
    which provides a measure for the occurrence of different states. If there is
    a single ergodic component $\r_\a = \r$, the overlap matrix $q_{\a \b}$
    will be trivial and a single number, so that $P_J(q)$ is just a single
    delta function. If however there are many states, $P_J(q)$ will be a highly
    nontrivial function. Similarly we can compute the average
     \begin{equation}
         P(q) = \braket{P_J(q)},
     \end{equation}
     where the bracket denotes the ensemble average over $J$ 
     \begin{equation}
         \braket{A_J} = \int dJ \cP(J) A_J,
     \end{equation}
    where $\cP(J)$ is the probability measure on the $J$'s and $dJ = \prod_{i<j}
    dJ_{ij}$. Now we want to argue that this function is related to the notion
    of replica-symmetry breaking. One natural quantity to compute in disordered
    systems is the average free energy
    \begin{equation}
        \cF = - \braket{ \log \Tr e^{-\b H_J} }.
    \end{equation}
    The computation of this average is a very hard task, and a simpler
    computation is facilitated by the so-called \textit{replica trick}, where
    one considers $n$ non-interacting copies of the system whose partition
    function for a single instance of couplings is 
    \begin{equation}
        Z_J^n = \Tr e^{-\b\sum_{a=1}^n H^a_j},
    \end{equation}
    where the trace is taken over the Hilbert space $\cH_n =  \bigotimes_{a = 1}^n
    \cH$.
    Now it is easy to see that \cite{Anous2021}
    \begin{equation}\label{eq:replica_trick}
        \cF = \lim_{n \rightarrow 0} \partial_n \braket{Z_J^n},
    \end{equation}
    so the computation of the average free energy has been reduced to the
    computation of the average of powers of the partition function. Here we
    assume that by computing this average at integer values of $n$, one obtains a function that is differentiable in $n$. In practice,
    the disorder average will lead to the situation that the r.h.s of
    \eqref{eq:replica_trick} depends on the following matrix in the limit $N
    \rightarrow \infty$
    \begin{equation}
        G_{a b} := \braket{\frac{1}{Z^n_J}\Tr \Big( e^{-\b\sum_{a = 1}^n H^a_J}
            \s_i^a \s_i^b \Big)},  
    \end{equation}
    which computes correlations of spins between replicas inside the disorder
    average. The usual approach to determine it, is to take a saddle point
    approximation of \eqref{eq:replica_trick}. Naively one would think that
    $G_{a b}$ is the same number for $a \neq b$, but as it turns out for some
    systems, the matrix that minimizes the action is not symmetric under
    exchanges of replica indices. If this happens, one says that the replica
    symmetry is broken (RSB) and the dominant configuration is not replica-symmetric. We present in Appendix \ref{app:analytical_calculations}, how replica symmetry breaking parameters appear in the effective action of the SU$(M)$ Heisenberg model in detail.
    The matrix $G_{ab}$ is defined for each integer n. It then turns out
    \cite{mezard1986spin} that
    \begin{equation}
        P(q) = \lim_{n \rightarrow 0} \frac{2}{n(n-1)} \sum_{\{ab \}}
        \d(G_{a,b}-q).
    \end{equation}
    We see, if $P(q)$ is a nontrivial function, replica symmetry must
    be broken, since only in that case the $\d$-function in the sum can lead to
    nontrivial values of $P(q)$. This gives a direct connection between the existence of a large
    number of ergodic components in the Gibbs state to replica
    symmetry breaking in the SK model.
    In the following analysis we will therefore use the existence of a replica-symmetry breaking saddle point as the indication that a point on the phase diagram lives in the spin glass phase \footnote{In \cite{Zlokapa:2025bbm} a computational perspective on quantum spin glasses is given that provides a more detailed understanding of the mechanism leading to replica symmetry breaking. In particular, it is argued that before RSB, the Gibbs state can shatter into exponentially many components without exhibiting RSB and that this shattering should already indicate the spin glass state. In this work, replica symmetry breaking is argued to arise from a reduction of the number of ergodic components from exponentially many to $\cO(1)$. These developments are beyond the scope of this work.}.
    
\subsection{Towards a duality}
To summarize, we gather the similarities between complex systems in high energy physics and glasses again:
\begin{itemize}
\item \textit{Number of local minima}: As we mentioned in the section above, the models of black hole molecules consist of an exponentially large number of states in configurations of high complexity. This leads to free energy landscapes with exponentially many local minima, a behavior that is characteristic of spin glasses too.
\item \textit{Relaxation times}: The existence of an exponentially large number of local minima in the energy landscape leads to exponentially large relaxation times. Assuming that the black hole molecule lives in a configuration away from the equilibrium state, the transition towards the minimum free energy configuration will be interrupted by exponentially many local energy minima. This will require exponentially many transitions and large time in order to surpass them. We see similar behavior in spin glass models. 
\item \textit{Memory effects}: Black hole molecules are expected to have exponentially suppressed transition rates, which results in logarithmic time evolution with a universal timescale. This behavior resembles the memory effects found in spin glasses.
\end{itemize}
The natural question at this point is: \textit{Can we make the connection of glasses and complex systems in holography more precise?} 
The above similarities are qualitative in nature and a precise correspondence has not been established yet. A weaker version of the above question is: \textit{Can large $N$ models of spin glasses have an emergent spatial dimension?}
The goal of this work is to make a first step towards this question with techniques using operator algebras that arose in recent years.

\section{Bulk emergence and generalized free fields}\label{sec:gff}
    Now that we have motivated a possible connection between spin glasses and complex systems in quantum gravity, it remains unclear how to probe, whether the common models of spin glasses can give rise to a holographic extra dimension. While in systems such as SYK \cite{Kitaev:2017awl} it is possible to reduce the effective action in specific limits to the effective action of a gravitational model in a higher dimension, this is a hard question for more general models to find a match. It is therefore favorable to find necessary conditions that a given system has to satisfy to be holographic. While arguments for such conditions in conformal field theories have been put forward in \cite{Heemskerk:2009pn, El-Showk:2011yvt, Perlmutter:2016pkf}, which most prominently feature a large gap in the conformal dimension of primary operators, for theories that lack conformal symmetry not much is known. In recent years, studying holographic systems from an algebraic perspective \cite{Leutheusser:2021frk, Chandrasekaran:2022eqq, Gesteau:2024rpt} has led to a formulation of the problem that can probe features of an emergent bulk causality that is independent from the presence of conformal symmetry. This is the perspective we take in this article and based on which we will study the spin glasses below.

\begin{figure}[H]\label{fig:wedge_reconstruction}
\centering
\includegraphics[width=0.35\linewidth]{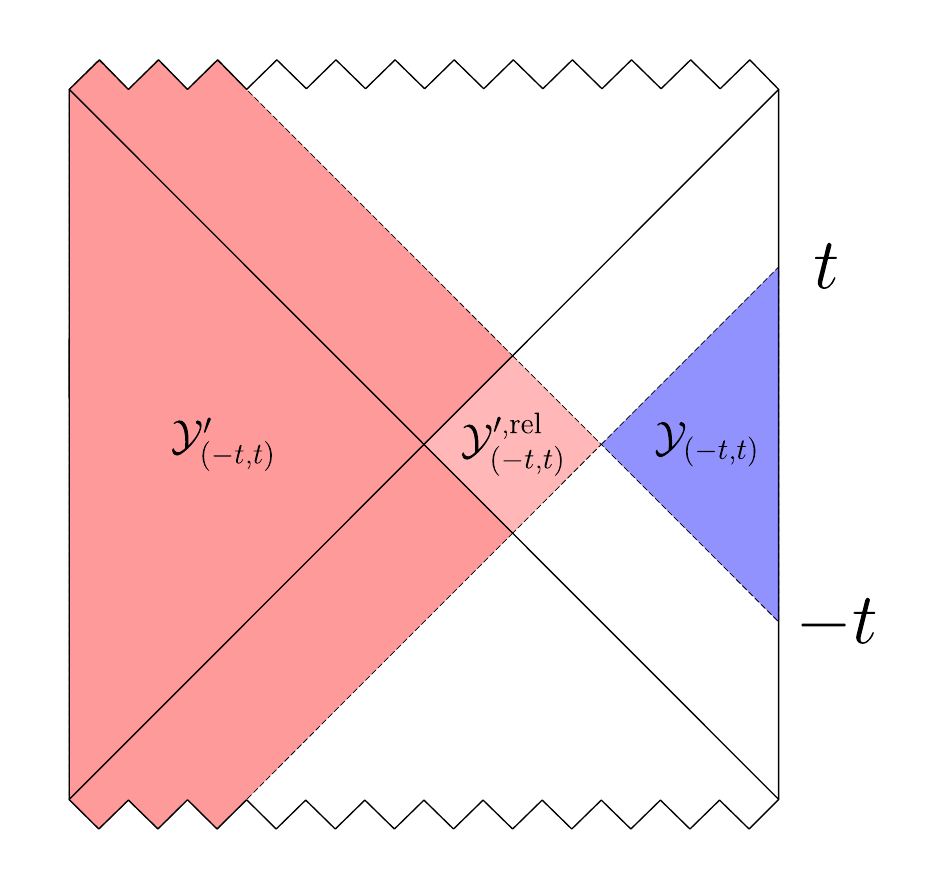}
\caption{Illustration of causal wedge reconstruction for a timeband algebra in the intervale $(-t,t)$. The algebra $\cY_{(-t,t)}$ of the causal wedge of the timeband $(-t,t)$ is depicted by the blue shaded region. The light red diamond $\cY^{',\text{rel}}_{(-t,t)}$ is the relative commutant of the bulk algebra $\cY_{(-t,t)}$. The dark red shaded region, which includes the light shaded region, describes the commutant $\cY'_{(-t,t)}$.}
\label{fig:causal_wedge_reconstruction}
\end{figure}

    The large $N$ systems under study in holographic dualities have the common
    feature that they exhibit large $N$ factorization \cite{tHooft:1973alw}. This
    entails the existence of a parameter $N$, a sequence of states $\ket{\Y_N}$ and a set of operators $O$ (usually referred to as single-trace operators due
    to their identification in gauge theory) that are defined for each value of $N$, and for which the correlation functions factorize at large $N$, i.e., each connected $n$-point function
    satisfies
    \begin{equation}
        \braket{\Y_N|O_1 \hdots O_n|\Y_N}_{C} \sim \cO(\frac{1}{N^{c_n}}),
    \end{equation}
    for some $c_n > 0$ and $n > 2$. In the case of single-trace operators in
    gauge theory, one finds in the vacuum $\ket{\Y_N}  = \ket{\Omega}$ that 
    \begin{equation}
        c_n = n-2,
    \end{equation}
    so correlation functions computed by summing up only leading contributions satisfy Wick's theorem and the theory becomes free at $N \to \infty$. In the dual bulk theory,
    one typically identifies $G_N \sim \frac{1}{N^2}$ so that the large $N$ expansion corresponds to
    a bulk perturbative expansion in Newton's constant. After a decomposition of these
    operators into spherical harmonics in $D>1$ \cite{Festuccia:2006sa}, one obtains a set of operators
    $O(t)$ that only depend on time and that individually satisfy large $N$
    factorization between them, given rotational invariance of the underlying
    theory. A good proxy for such systems where one does not have the
    complication of decomposing operators into spherical harmonics are large $N$
    quantum-mechanical systems, where one has only a time coordinate from the
    beginning. The large $N$ system is then fully specified by the set of single-trace operators and the limiting two-point functions in the sequence $\ket{\Y_N}$, i.e. the tuple
    $\{O(t),G_O(t_1,t_2)\}$, where 
    \begin{equation}\label{eq:large_N_greens}
        G_O(t_1,t_2) = \lim_{N \rightarrow \infty}\braket{\Y_N|O(t_1)O(t_2)|\Y_N},
    \end{equation}
    which contains all the dynamical information of the theory at infinite $N$ for the sequence $\ket{\Y_N}$. Note that we take the sequence of states $\ket{\Y_N}$ as part of the definition of the limiting system. This is motivated by the fact that, in AdS/CFT, the state $\ket{\Y_N}$ gives rise to the dual geometry and the limiting systems describe a quantum field theory in curved spacetime. For different spacetimes, one would consider the resulting theory also as a different system.
    An elegant framework to study AdS/CFT from an algebraic perspective was
    proposed by Leutheusser and Liu \cite{Leutheusser:2021frk}.
    If one smears the operators $O(t)$ with some function 
    \begin{equation}\label{eq:def_smeared_op}
        O(f) = \int dt f(t) O(t),
    \end{equation}
    where $f(t)$ is supported in an interval $\D T$, one obtains smeared operators which can be added and multiplied and
    therefore form an algebra $\cA_{\D T}$. We will discuss the allowed set of smearing functions below in Eq.~\eqref{eq:L_2_rho}. Considering the full timeband $\D T
    = (-\infty,\infty)$ we can consider the algebra of smeared single-trace
    operators with smearing functions supported in $\D T$ which we denote by $\cA^*$. More generally, before specifying the state, the above algebra
    can be viewed as an abstract C*-algebra as described in
    \cite{Faulkner:2022ada}. For a review of basic notions of von Neumann and C$^*$-algebras, see e.g.~\cite{Chemissany:2025vye}. Note that $\cA^*$ is supposed to encode the
    theory at $N = \infty$ because at finite $N$, products of traces are not
    independent and related by trace relations. Since we allow for arbitrarily
    high products, this only make sense if no trace relations relate higher
    powers of the respective operator to lower powers. Given a sequence of states
    $\ket{\Y_N}$, one can build a Hilbert space for the large $N$ system using the
    GNS construction \cite{Leutheusser:2021frk}. One associates a vector
    $\ket{\Y}$ with the ``vacuum'', i.e., the sequence $\ket{\Y_N}$, and for each $O \in \cA^*$ one
    defines a vector $\ket{O}$. For the set of such vectors we define an addition by linearity,
    which defines a vector space $\cV_\Y$.
    Given the structure of a vector space $\cV_\Y$, we can define a
    representation $\pi_\Y:\cA^* \rightarrow \cL(\cV_\Y)$ onto linear maps
    in $\cV_\Y$ via
    \begin{equation}
        \begin{aligned}
            \pi_\Y(O)\ket{\Y} &= \ket{O},\\
            \pi_\Y(O) \ket{O'}&=  \pi_\Y(O)\pi_\Y(O')\ket{\Y} = \pi_\Y(O O') \ket{\Y} = \ket{O
          O'}.
        \end{aligned}
    \end{equation}
    The inner product is defined by expanding $O$ in smeared operators and then
    using Wick's theorem to reduce it to the elementary inner product by defining
    \begin{equation}
        \braket{O(t)|O'(t')} = \braket{\Y|O(t)^\dagger O'(t')|\Y} :=\lim_{N \rightarrow \infty} \braket{\Y_N|O(t)^\dagger O'(t')|\Y_N},
    \end{equation}
    which reduces to
    \begin{equation}
        \braket{O(t)|O'(t')} = \d_{O,O'}G(t,t'),
    \end{equation}
    due to \eqref{eq:large_N_greens}. The factor $\delta_{O,O'}$ appears because we assume that a convenient choice of operators $O$ was made, such that they independently satisfy large $N$ factorization. This way one obtains an inner product space \footnote{One has to additionally quotient out the space of operators
        annihilated by the state. For more details on the GNS construction see
    \cite{sunder_invitation1987}.}
    which can be completed to form a Hilbert space $\cH_{\Y}$. The proposal of
    AdS/CFT at infinite $N$ is then that 
    \begin{equation}
        \cH_{\Y} = \cH_{\cM},
    \end{equation}
    where $\cH_{\cM}$ is the Hilbert space of the dual quantum fields
    propagating on the manifold $\cM$ that is dual to the state $\Y$.
    Bulk causality and locality are then encoded in the proposal of causal wedge
    reconstruction
    \cite{Hamilton:2006az,Hamilton:2005ju,Gubser:1998bc,Witten:1998qj}, see
    \cite{Kajuri:2020vxf} for a review, which determines which bulk region can
    be reconstructed from a given boundary region. In this context, a bulk operator
    $W$ is reconstructable from a region $A$, if there is an operator $O$ in the
    boundary theory localized in $A$, i.e., it is a sum of products of local operators in $A$, such that every correlation function of $W$ can be
    represented from the correlation functions of $O$. This is equivalent to the statement that $O$ is an operator in the local
    algebra of $A$. In AdS/CFT, one region
    that is reconstructable from a boundary region $A$ is the \textit{causal
    wedge}, see Fig.~\ref{fig:causal_wedge_reconstruction} for an illustration in a black hole spacetime.
    Based on the equality of boundary and bulk relative entropy \cite{Jafferis:2015del}, it was recognized that there exist a larger bulk region called the \textit{entanglement
        wedge} \cite{Dong:2016eik} that
        is reconstructable from $A$. For the states and regions we consider
    in this article the distinction is not relevant. The causal wedge of a boundary region $A$ is determined as follows:
    First one determines the boundary domain of dependence $D[A]$ of $A$, which
    is the largest boundary region such that every causal curve in it
    intersects $A$. Then one computes the bulk causal past (future) $J^{-(+)}(D[A])$, i.e., the
    set of points that can be reached from $D[A]$ by past (future) directed causal
    curves. The causal wedge $C[A]$ is then given by 
    \begin{equation}
        C[A] = J^+(D[A]) \cap J^-(D[A]).
    \end{equation}
    The notion of reconstructability of a bulk region from the
    algebra of operators in a boundary region was then lifted to the proposal
    of \textit{subalgebra-subregion
    duality} \cite{Leutheusser:2021frk,Leutheusser:2021qhd,Leutheusser:2022bgi}, which informally states that a
    subalgebra of the boundary should be dual to a subregion in the bulk.
    At $N= \infty$, one incarnation of this idea is that for a timeband algebra
    $\cA_{(t_0,t_1)} \subset \cA^*$, the
    von Neumann algebra $\cY_{(t_0,t_1)} := \pi(\cA_{(t_0,t_1)})''$ defined via a double commutant\footnote{For a set of operators $A$, the commutant $A'$ are all bounded operators that commute with all operators in $A$.} is identified with the von Neumann
    algebra of the respective dual quantum fields on $\cM$ in the causal wedge
    of the timeband $(t_0,t_1)$, i.e.,
    \begin{equation}
        \cY_{(t_0,t_1)} \sim \cA_{C[(t_0,t_1)]},
    \end{equation}
    with appropriate generalizations for non-timeband boundary regions such as
    causal diamonds. Above, given a bulk region $D$, $\cA_D$ is the von Neumann algebra generated by quantum
    fields in $D$ \cite{Haag:1996hvx}.
    In the following we denote the
    von Neumann algebra associated to the timeband by 
    \begin{equation}
        \cY_{(t_0,t_1)} := \pi(\cA_{(t_0,t_1)})''
    \end{equation}
    and the full algebra by $\cY := \cY_{(-\infty,\infty)}$.
    Therefore, if the interval $(t_1-t_0)$ is small enough, the operators
    that can be reconstruction from $(t_0,t_1)$ only belong to a wedge region as in Fig.~\ref{fig:causal_wedge_reconstruction}. Most
    importantly, if $(t_0,t_1)$ is small, there exist bulk operators that are
    spatially separated from the operators reconstructable from $\cY_{(t_0,t_1)}$ but still belong to the full boundary algebra $\cY$. The corresponding algebra is called the \textit{relative commutant} and is indicated as $\cY^{',\text{rel}}_{(t_0,t_1)}$ in Fig.~\ref{fig:causal_wedge_reconstruction}.
    This gives a first hint towards how to think about the properties of the
    tentative bulk dual in terms of algebraic properties of the respective
    boundary algebras.
    The question then asked by Gesteau and Liu (GL) \cite{Gesteau:2024rpt} was: Given just a set of two-point functions, what
    geometric structures in a higher-dimensional spacetime are compatible with
    them? In particular, what properties of the two-point functions allow one
    to infer compatibility with an emergent bulk spacetime? In the next section we will summarize the theoretical
    framework proposed by GL to approach these questions.

    \subsection{Operator algebras, bulk reconstruction and stringy connectivity}
    In the previous section, we have motivated the idea that in order to study the
    emergence of a bulk spacetime, one should try to reformulate notions of a
    bulk spacetime in terms of properties of the corresponding algebras of
    observables. Here we will recapitulate the framework set up in \cite{Gesteau:2024rpt}. For
    simplicity we will restrict ourselves to time-independent setups, such as
    the thermal state with a time-independent Hamiltonian.
    We will only consider systems with large $N$ factorization and the large $N$ von Neumann algebras $\cY_\D$ associated to boundary regions $\D$. We will refer
    to them as \textit{large $N$ systems} and \textit{large $N$ algebras}.
    The main idea is that the full timeband algebra $\cY$ should encode the
    maximal region causally accessible to the boundary theory. If there are no
    horizons in the bulk, there will be a minimal time interval $\cT$ for which
    the causally accessible region incorporates a whole Cauchy surface and
    therefore the algebra $\cY_{-\cT/2,\cT/2}$ already should be the whole
    algebra $\cB(\cH_\Y)$. A formal notion of this is 
    \begin{definition}[cf.\ Definition 2.2 in \cite{Gesteau:2024rpt}]\label{def:depth_parameter}
        The causal depth parameter $\cT$ is the largest value $t$ such
        that $\cY_{(-t,t)}$ is a proper subalgebra of $\cY$.
    \end{definition}
    If the bulk region accessible to the boundary in question does have a
    horizon, then the causal region accessible from the boundary is bounded by
    the horizon. In particular, no time interval will be able to reconstruct a
    full Cauchy surface. This motivates
    \begin{definition}[cf.\ Definition 2.6 and 2.7 in \cite{Gesteau:2024rpt}]\label{def:horizon_1}
        A large $N$ system has a future (past) horizon after (before) time $t$ if
        $\cY_{(t,\infty)}(\cY_{(-\infty,t)})$ is a proper subalgebra of $\cY$.
    \end{definition}
    This was also dubbed a \textit{stringy} horizon in \cite{Gesteau:2024dhj} due to the applicability of the definition to large $N$ gauge theory at weak coupling, which should correspond to a bulk dual with large strings, since the string length is inversely proportional to the coupling \cite{Maldacena:1997re}.
    We now want to describe a different characterization of horizons that is
    connected to the existence of an algebraic structure called \textit{half-sided
    modular inclusion} \cite{Wiesbrock:1992mg}.
    For a thermal state $\r = \frac{e^{-\b H}}{Z}$ one has
    \begin{equation}
        \b H \sim -\log \r,
    \end{equation}
    up to an additive constant, so that the standard time evolution can be identified with the evolution generated by the state in question. For a general quantum state
    $\r$, its negative logarithm is called the \textit{local modular
    Hamiltonian} \cite{Witten:2018zxz}. If $\r_B = \Tr_A \ket{\Y}\bra{\Y}$ in some bipartite system
    $\cH = \cH_A \otimes \cH_B$, then given the respective local modular
    Hamiltonians
    $H_{A(B)} = -\log \r_{A(B)}$ one defines the \textit{full modular
    Hamiltonian} for
    region $B$ as
    \begin{equation}
        H = H_B -H_A.
    \end{equation}
    While in a continuum quantum field theory, the local modular Hamiltonian
    $H_A$ is not defined, the full modular Hamiltonain H is a well defined
    operator\footnote{Here a well-defined operator $O$ means
        that it is an unbounded operator with a dense domain, i.e., there is a
        dense set of states such that $||O\ket{\Y}|| < \infty$. This is
        contrasted by objects such as the local operators $H_L$, which usually
        have $\braket{\Y|O^2|\Y} = \infty$. This is usually described as $H_L$
    having infinite fluctuations, see e.g.~\cite{Kudler-Flam:2023hkl} for a
recent discussion.} \cite{Haag:1996hvx}. The full operator
    \begin{equation}
        \D = e^{-H}
    \end{equation}
    is called the modular operator. The standard example of a causal horizon of
    4-dimensional Minkowski space is the horizon seen by a constantly accelerated observer following the
    trajectory 
    \begin{equation}
        t(\t) = \r \sinh(\t),\ x(\t) = \r \cosh(\t),
    \end{equation}
    for which the lines $ \{|t| = 0, x > 0 \}$ demarcate the horizon. The
    algebra of operators accessible to this observer is the algebra associated to a Rindler wedge $W = \{(t,x,y,z)|t^2-r^2 < 0, x>0\}$
    \cite{Haag:1996hvx}. For this algebra, the modular Hamiltonian is known
    \cite{Bisognano:1975ih} and given by $2 \pi K$, where
    \begin{equation}
        K = \int \text{d}^3x\, x\, T_{00}
    \end{equation}
    is the generator of boosts and $T_{00}$ is the local energy density. Defining null coordinates $u = t-x,
    v = t+x$, the right Rindler wedge has the
    property that if one performs a translation along $v$,
    \begin{equation}
        T_a((u,v)) = (u,v + a),\ a \geq 0,
    \end{equation}
    one ends up with a region $W_a := T_a(W)$, that is a subregion of the original wedge $W$.
    One can check that boosts act as 
    \begin{equation}
        B_s((u,v)) = (ue^{-s}, ve^{s}),
    \end{equation}
    and therefore the associated algebras satisfy 
    \begin{equation}
        \cA_{T_a(W)} \subset \cA_W,\ \forall a \geq 0
    \end{equation}
    and 
    \begin{equation}\label{eq:hsmi}
       e^{itK} \cA_{T_a(W)} e^{-itK} \subset \cA_W,\ \forall t \geq 0 \ .
    \end{equation}
    Therefore the algebras $\cA_{T_a(W)}$ are included in the algebra $\cA_W$
    under modular flow for all $t\geq 0$. Proper subalgebras $\cA_{T_a(W)}$ that satisfy
    Eq.~\eqref{eq:hsmi} constitute what is called a
    \textit{half-sided modular inclusion} of $W$ \cite{Wiesbrock:1992mg}. It can be shown that, given
    any bifurcate killing horizon in a curved spacetime \cite{Sewell:1982zz}, the algebra in a wedge
    has an associated half-sided modular inclusion. It is therefore natural in
    the context of the algebraic bulk reconstruction we are developing, to
    associate horizons with the existence of half-sided modular inclusions. We
    see that Definition \ref{def:horizon_1} incorporates the definition of
    a half-sided modular inclusion. In particular, because in a thermal state,
    modular flow coincides with time translation, one has 
    \begin{equation}
        \D^{-it} \cY_{(0,\infty)} \D^{it} = \cY_{(t,\infty)} \subset \cY,
    \end{equation}
    so that if $\cY_{(0,\infty)}$ is a proper subalgebra, one automatically has
    a half-sided modular inclusion. Now that we have associated geometric
    properties in the bulk with properties of the respective algebras, the
    question remains what structure gives rise to them.
    \subsection{A bulk from the spectral function}\label{sec:theorems}
    We saw above that the relation between different boundary timeband algebras provides information about causal
    structure in a potential bulk. Now we want to make this statement more
    precise. As we learned above, if the interval $(t_1,t_2)$
    is small enough, the algebra $\cY_{(t_1,t_2)}$ will not reconstruct a whole bulk Cauchy
     slice and therefore there have to be ``bulk operators'' spatially separated
     from the reconstructed algebra. As it is known from standard quantum field
     theory \cite{Peskin:1995ev}, if two operators are spacelike separated, they
     commute. Thus, if there are operators $O'$ in $\cY$ that are spacelike separated from the
     operators in $\cY_{(t_1,t_2)}$, then we need 
\begin{equation}
    [O',\cY_{(t_1,t_2)}]=0.
\end{equation}
    The set of such operators is called the \textit{relative commutant} and can
    be defined as the intersection $ \cY^{',\text{rel}}_{(t_1,t_2)} := \cY_{(t_1,t_2)}' \cap \cY$. This is illustrated in Fig.~\ref{fig:causal_wedge_reconstruction}.
    The presence of a relative commutant in a non-abelian algebra
    determines, whether a given von Neumann algebra is a subalgebra of a larger algebra if
    both algebras are factors. In this case, only operators outside of the smaller algebra
    can commute with all of its members. In the following, we will consider only
    factors.
    Because large $N$ algebras define free $(0+1)$-dimensional quantum field theories, standard tools of algebraic QFT
 apply. The existence of a relative commutant can then be relatively easily determined. In particular GL argued
 that in order to prove the existence of a relative commutant, it is enough to
 determine whether there exists a function $g(t)$ such that 
 \begin{equation}
     \braket{[O(f),O(g)]} = 0,\ \forall f \subset (t_0,t_1),
 \end{equation}
  where $f \subset (t_0,t_1)$ denotes that the support of $f$ is contained in
  $(t_0,t_1)$, the bracket is the large $N$ expectation value and $O(f)$ denotes the smeared operators from Eq.~\eqref{eq:def_smeared_op}. The relative commutant is then contains $O(g)$ and is nontrivial. We have here abused
  notation and written $O$ for its representative $\pi_\Y(O)$ to reduce
  clutter. From now on, we will always talk about the bulk representative
  of $O$, never its abstract counterpart.
  One can expand the equation above and find \begin{equation}\label{eq:commutator}
  \braket{[O(f),O^\dagger(g)]_{\pm}} = \int dt \int dt' f(t)\r^{\pm}(t,t')g^*(t') =  \int \frac{dw}{2\pi} \r^\pm(\w) f(\w) g^*(\w),
  \end{equation}
  where $\r^\pm(\w)$ is the Fourier transform of 
  \begin{equation}
      \r^\pm(t,t') = \braket{[O(t),O^\dagger(t')]_\pm}.
  \end{equation}
    the $\pm$ indicates that the bracket is the (anti-)commutator for (fermionic) bosonic operators. The first equation makes clear that for $O(g)$ to commute with all operators $O(f)$ that arise from smearing inside $(t_0,t_1)$, the function
    \begin{equation}
        \cF(t) := \int dt' g^*(t') \r(t,t')
    \end{equation} has to be supported outside of the support of $(t_0,t_1)$. The function $\r$ is called the spectral
    function. After an expansion in energy eigenstates and a Fourier transform,
    one finds for a thermal state, up to an overall sign,\footnote{The
        above discussion took place in the Fourier transform convention of GL.
        The following definition of the spectral function is the one we adopt
    in the main text.}
    \begin{equation}
        \r^{\pm}(\w) = \frac{2\pi}{Z} \sum_m |O_{m+\w,m}|^2 e^{-\b e_m}(1 \pm e^{-\b \w}
        ),
    \end{equation}
    where $Z = \Tr e^{-\b H}$ is the partition function.
    From that explicit form one sees that 
    \begin{equation}
        \begin{aligned}
            \r(w) &\geq 0 \ \text{for fermions},\\
            \w \r(\w) &\geq 0 \ \text{for bosons}
        \end{aligned}
    \end{equation}
    and if $O$ is Hermitian, $\r(\w)$ is antisymmetric for bosons and symmetric
    for fermions\footnote{There is an implicit assumption in deriving these last symmetry statements that one has a continuous spectrum, so that one can shift energies by $\w$ and the
    respected states exist.}. As reviewed in \cite{Gesteau:2024dhj}, the spectral function determines also the set of allowed smearing functions. This is given by functions that have finite norm in the inner product
    \begin{equation}\label{eq:L_2_rho}
    \langle f, f' \rangle_\beta = \int \frac{d\omega}{2\pi} \,
f^*(\omega)\, \theta(\omega) \, \frac{\rho(\omega)}{1 - e^{-\beta \omega}} \, f'(\omega) < \infty.
\end{equation}
    One can show \cite{Gesteau:2024rpt}, that for generalized free fields, the depth parameter $\cT$ of Def.~\ref{def:depth_parameter} is given by the length of
    the smallest time interval, so that the algebra of the timeband has no
    relative commutant anymore. The relevance of the spectral functions for emergent bulk properties was already conjectured in \cite{Festuccia:2005pi,Festuccia:2006sa}. In the bulk language, this corresponds to
    a bulk reconstruction of a full Cauchy surface. This is due to the fact that if one can
    reconstruct a full Cauchy surface, no nontrivial operator can commute with
    all operators in the algebra anymore, since the algebra associated to the Cauchy surface contains all operators\footnote{This hinges technically on the statement that the timeband
    algebras are \textit{factors}, which means that every operator that is not
the identity and commutes with every operator in the algebra is not itself part
of the algebra.}. GL then proved the following statements (which we repeat in an
informal fashion).

\begin{figure}[h]
\centering
\includegraphics[width=0.25\linewidth]{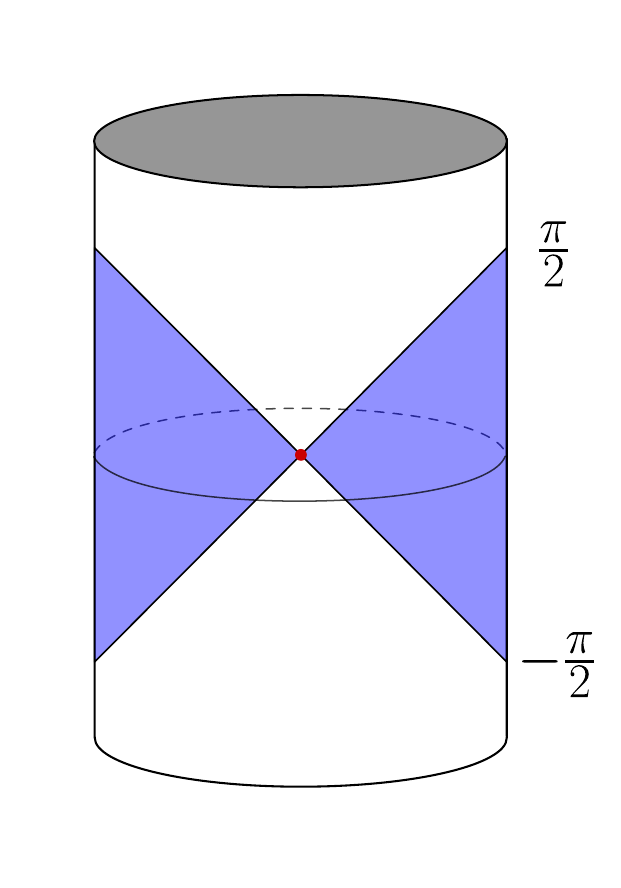}
\caption{Illustration of causal wedge reconstruction for a timeband algebra in the interval $(-\pi/2,\pi/2)$ in empty AdS, which reconstructs a whole Cauchy surface and thus contains the entire algebra $\cB(\cH)$.}
\label{fig:TDF}
\end{figure}

\begin{theorem}[cf.~Proposition 2.12 in \cite{Gesteau:2024rpt}]\label{thm:compact_support}
    If the support of $\r(\w)$ is compact, then $\cT= 0$.
\end{theorem}
\begin{theorem}[cf.~Proposition 2.13 in \cite{Gesteau:2024rpt}]\label{thm:poly_for_horizon}
    If $\r(\w)$ does not vanish for any $\w$, except $\w = 0$, and decays at
    most polynomially in $\w$, then $\cT = \infty$.
\end{theorem}
The intuitive reason behind the first theorem is that if $\r(\w)$ has compact
support, $\r(t)$ is analytic and therefore one can infer the whole function
from an arbitrarily small timeband, i.e., one only needs to observe the
dynamics for an arbitrarily small amount of time to know all the development. The
proof of the second theorem proceeds by constructing an explicit function
$g(t)$ that can be made to have support outside of the timeband, so that the
integrals are on orthogonal subspaces when computing the commutator.
Furthermore GL proved that 
\begin{theorem}[cf.~Proposition 2.9 in \cite{Gesteau:2024rpt}]
    If there exist a half-sided modular inclusion, $\r(\w)$ has to be nonzero
    on every open intervall.
\end{theorem}
The above theorem gives the necessary conditions that are needed for a nonzero
semiclassical emergent radial direction. Here by semiclassical we mean that
there is a nontrivial notion of bulk causality and that a finite boundary
interval can only probe a finite distance into the bulk. The remaining questions is whether there is an example of a system with a finite $\cT$. This was also answered affirmatively by GL for spectral functions of the form
\begin{equation}\label{eq:rho_delta}
    \r(\w) = \sum_n a_n [\d(\w-2n)-\d(\w+2n)]
\end{equation}
for which GL showed the following
\begin{theorem}[cf.~Proposition C.9 in \cite{Gesteau:2024rpt}]\label{thm:finite_depth_delta}
If $\r(\w)$ is of the form of Eq.~\eqref{eq:rho_delta} and the coefficients $a_n$ are nonzero and decay at most polynomially, then $\cT = \pi$.
\end{theorem}
The above spectral function is exhibited by conformal field theories on a sphere in the vacuum state and its bulk dual is illustrated in Fig.~\ref{fig:TDF}.

\subsection{Type classification in large $N$ systems}
The previous discussion focused on fine-grained information of the system, such as the
exact behavior of the spectral function at $N = \infty$, which gives
information about more detailed structure of the bulk, such as the existence of
causal horizons through the appearance of half-sided modular inclusions.
Here we want to comment on more coarse-grained information that can be obtained from
the spectral function, namely the type of the large $N$ algebras. This perspective was the
initial insight of \cite{Leutheusser:2021frk}, to quantify the emergent bulk
through the properties of the large $N$ two-point function. It is well known
\cite{sunder_invitation1987} that von Neumann algebras come in three types,
Type I, Type II and Type III, which can be split further into the subtypes I$_n$,
I$_\infty$, II$_1$, II$_\infty$ and a continuum of types III$_\l,\ 0 \leq \l \leq
1$. In the case of hyperfinite factors, which is the class relevant for quantum field theories \cite{Haag:1996hvx}, there
is up to isomorphisms only one algebra of each subtype. The cases relevant for
quantum field theories in curved spacetime are the types I$_\infty$ and III$_1$. The former is
the algebra of bounded operators acting on an infinite-dimensional separable
Hilbert space. An physical example is the algebra generated by operators acting on the full
Cauchy slice of a globally hyperbolic spacetime. The latter type III$_1$ algebras are for example given by
the algebra of operators associated with causally complete subregions, such as
causal diamonds and wedges \cite{Haag:1996hvx, Fredenhagen:1984dc}. Thus,
determining if a given large $N$ algebra can be identified with the algebra of
some causally complete subregion in a higher-dimensional spacetime should be related to determining whether it is of type III$_1$. The diagnostic for a large $N$ algebra where the states $\Y_N$
are thermal for each $N$ was provided in \cite{Furuya:2023fei} and is best
captured by the following:
\begin{theorem}[cf.~Lemma 20 in \cite{Furuya:2023fei}]
    Let $\cY$ be a large $N$ algebra that arises from a thermal state as
    described in the previous section. If the spectral
    function $\r(\w)$ is a measurable function of $\w$, then the algebra is type
   \text{III}$_1$. 
\end{theorem}
Another property that was proved in \cite{Furuya:2023fei} for this case is that
the associated algebras are mixing, i.e., that connected two-point functions
decay to zero in the limit of infinite time. This is connected to the notion of
information loss and more generally with some notion of ergodic behavior
\cite{Gesteau:2023rrx}.
An intuitive explanation for the necessity of type III$_1$ factors with mixing
behavior can be seen as follows. The spectral function $\r(\w)$ is associated
with the Fourier transform of the two-point function. If $\r(\w)$ would not be
continuous, but say a sum of $\d$-functions as in Eq.~\eqref{eq:rho_delta}, then the Fourier transform would
only have support on a discrete set of frequencies and would thus become an
almost-periodic function\footnote{Almost-periodicity means that with increasing time
the function returns arbitrarily close to previous values.}.
It would therefore not be able to decay to zero due
to its almost-periodic behavior. We saw in the case of Eq.~\eqref{eq:rho_delta} that the corresponding depth parameter $\cT$ is finite, so that the full timeband algebra has to be type I since it contains all operators acting on the Hilbert space. As already mentioned, this is the case of vacuum correlators of a CFT on a sphere, whose two-point functions can be computed in the bulk through the extrapolate dictionary \cite{Gubser:1998bc}, where the vacuum two-point functions
at large $N$ are given by vacuum expectation values of free fields in Anti-de
Sitter space (AdS). AdS has the feature that a signal sent out in the bulk will
reach the boundary in finite proper time and be reflected back, traveling back into the bulk. This leads to almost-periodic two-point functions. In particular, because the depth parameter $\cT$ is
finite in this case, the timeband algebra $\cY{}_{[-\cT,\cT]}$ is already the
full algebra $\cY$. In the bulk, this is reflected by the fact that the causal
wedge of a time interval of length $\cT = \pi$ contains a full Cauchy surface
of AdS, thus making the algebra $\cY{}_{[-\pi/2,\pi/2]}$ the full algebra of
operators, which has type I. As we will see below, a similar structure appears
in the $p$-spin model in the spin-liquid state, where we find an infinite set of
peaks in $\r$ that appear equally spaced, thus giving rise to the question
whether in this phase the system exhibits a finite depth parameter or not.
Now we will study the large $N$ spectral functions in spin glasses and ask whether we find
behavior consistent with an emergent radial direction based on the above framework.

\section{Spectral functions in three large $N$ models}\label{sec:numerics}
As the discussion above predisposes, our analysis will be entirely focused on the computation and study of the spectral functions of many-body models. Firstly, we will consider the already well-studied SYK model. In particular, two versions of it, the large $q$ and the $q=4$ SYK. Large $q$ SYK will give us a first indication of what behavior to expect from the spectral function for a model that is considered to have a dual theory.  Later, we will move on to two models that have both a spin liquid and spin glass phase: the spherical $p$-spin and the SU$(M)$ Heisenberg models. Again, we will compute and study the behavior of the spectral function for different parameter regimes of these models, focusing on their spin glass phase, which is the main goal of this work.

\subsection{The SYK model}\label{sec:SYK}
The SYK model \cite{KitaevTalks,Maldacena:2016hyu} is defined at finite $N$ as a quantum mechanical system of $N$ Majorana fermions, i.e., operators $\c_i$ that live on a $N/2$-dimensional Hilbert space that satisfy
\begin{equation}
    \{\c_i,\c_j\} = \d_{ij}. 
\end{equation}
The SYK$_q$ Hamiltonian is given by \begin{equation}
    H = (i)^{q/2} \sum_{1\leq i_1 < \hdots < i_q \leq N} J_{i_1 \hdots i_q} \c_{i_1}\hdots \c_{i_q},
\end{equation}
where $J_{i_1 \hdots i_q}$ are independent Gaussian variables with variance 
\begin{equation}
    \braket{J_{i_1 \hdots i_q}^2} = \frac{J^2 (q-1)!}{N^{q-1}}.
\end{equation}
We consider it as a reference due to its known holographic properties. 
The free Hamiltonian with $J=0$ vanishes, so that 
\begin{equation}
    G(\t) = \frac{1}{2}\text{sgn}(\t).
\end{equation} This is also true for any time-ordered Green's function in the free model, independent of whether time is Euclidean or Lorentzian. The signum function appears due to the anticommutation property of the time-ordering of fermionic fields. 
Following the procedure described in \cite{Trunin:2020vwy} one finds that the disorder-averaged thermal partition function is given by
\begin{equation}
    \begin{aligned}\label{eq:Ieff_SP}
        \braket{Z} &= \int DG d\S e^{-I_{\text{eff}}(G,\S)},\\
        \frac{I_{\text{eff}}(G,\S)}{N} &= - \frac{1}{2} \Tr \log \big( \id
        \partial_\t - \S \big) +\frac{1}{2} \int d\t d\t'\big(\S(\t,\t')G(\t,\t')
        - \frac{J^2}{q}G(\t,\t')^q\big),
    \end{aligned}
\end{equation}
where we have the identity in time $\id(\t,\t') = \d(\t-\t')$.
Varying the action leads to the equations of motion (eom)
\begin{equation}
    \begin{aligned}
        \S(\t,\t') &= J^2 G(\t,\t')^{q-1},\\
        \d(\t'-\t) &= \int d\t \Big(\partial_\t -\S)\Big)(\t'',\t) G(\t,\t').
    \end{aligned}
\end{equation}
Here $G(\t,\t')$ is given by the disorder averaged thermal expectation value
\begin{equation}
    G(\t,\t') = \frac{1}{N} \overline{\sum_i \langle \c_i(\t)\c_i(\t')\rangle_\b}. 
\end{equation}
In deriving this equation, one uses the fact that $G(\tau,\tau')$ is antisymmetric. Taking the Fourier transforms of the above equations, one finds \cite{Maldacena:2016hyu}
\begin{equation}\label{eq:SYK_E_SDE}
    \frac{1}{\hat{G}(k)} = i\w_k - \hat{\S}(k),
\end{equation} 
where $\w_k$ are the Matsubara frequencies
\begin{equation}
    \w_k = \pm\frac{2\pi (k+\frac{1}{2})}{\b}, k = 0,1,2,\hdots
\end{equation}
We can analytically continue Eq.~\eqref{eq:SYK_E_SDE} to real time as explained in appendix \ref{app:analytic_continuation} to obtain the real-time equation
\begin{equation}\label{eq:SYK_L_SDE}
    \frac{1}{G^R(\w)} = -\w - \S^R(\w),
\end{equation}
for the retarded Green's function of Eq.~\eqref{eq:real_time_2pt}.

In the large $q$ limit, one can find an approximate solution to the above system of equations \cite{Maldacena:2016hyu}. We computed the spectral function explicitly for this case in Appendix \ref{app:large_q_syk} and found the asymptotic behavior
\begin{equation}
\r(\w) \sim C \w^{\tfrac{2}{q-1} - 1}e{}^{-\tfrac{\b \w}{2}\big(\tfrac{1}{v} - 1\big)},
\end{equation}
where $0<v<1$ approaches one in the limit $q \rightarrow \infty$. We thus see that SYK is governed by an exponential decay of the spectral function in the large $q$ limit. This was already noted by GL, but not explicitly demonstrated, and raised the question of how to generalize their framework to such superpolynomial decay of the spectral function. It is however expected \cite{Lin2023}, that large $q$ SYK does arise from a dual system. The underlying logic is that the double-scaling limit of SYK, where $\l = 2 \frac{q^2}{N}$ is kept fixed, has a well controlled gravitational description in terms of the chord Hilbert space \cite{Lin:2022rbf}. The large $q$ limit can be achieved from the double scaling limit by taking $\l \rightarrow 0$. Since this limit can be taken on both sides of the duality, the duality should persist. We thus find, that an exponentially decaying spectral function is consistent with an emergent radial direction. As we will see below, such superpolynomial decay appears to be a common feature of the large $N$ limit of systems with quenched disorder away from conformal symmetry. This is a first indication, that the polynomial decay needed to proof Theorem \ref{thm:poly_for_horizon} can not be necessary to have a bulk dual.

\begin{figure}[ht!]
    \centering
    \includegraphics[width=0.85    \linewidth]{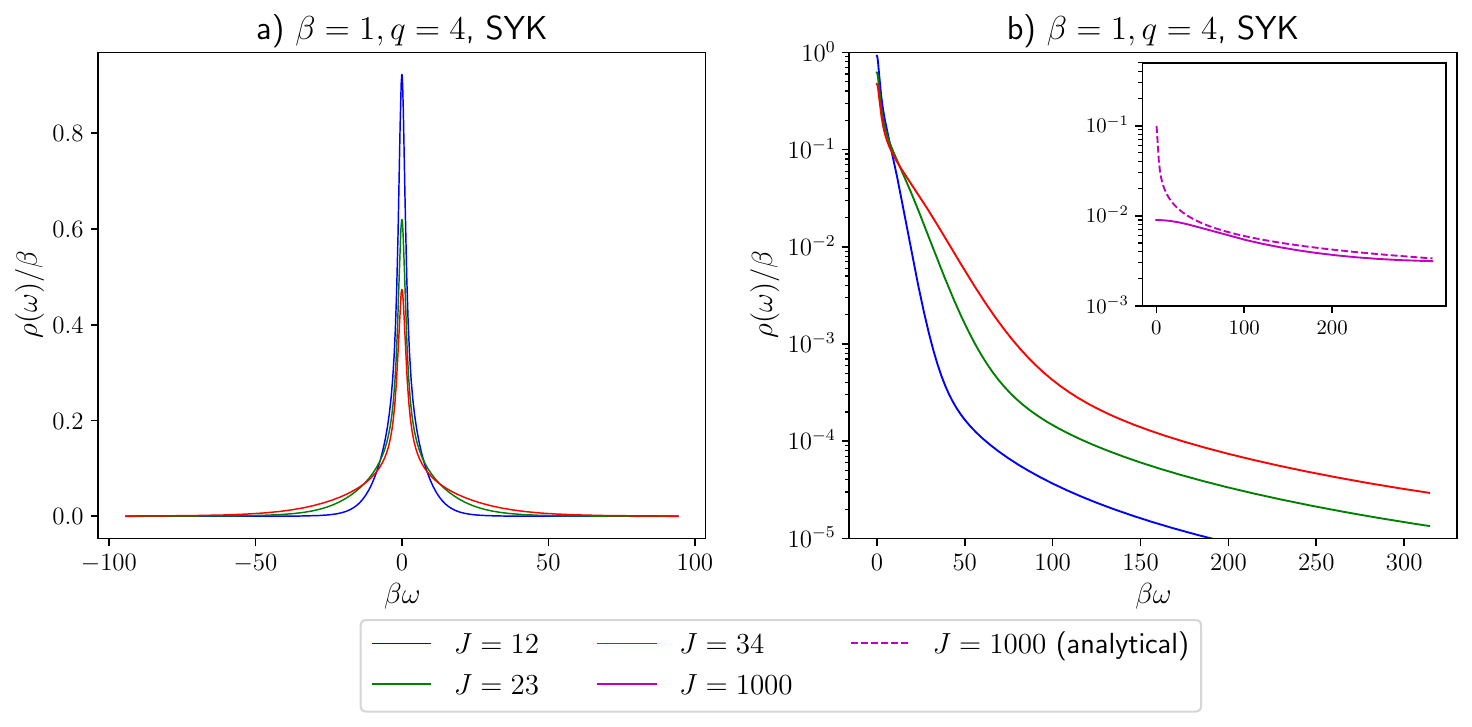}
    \caption{a) Spectral function $\r(\w)$ of the SYK model for varying coupling $J$ at $q=4$, $\beta=1$ with $N=10^5$ points. b) Logarithmic plot of the data in a). The subplot demonstrates how the asymptotics of the numerical data match the polynomial decay of the closed form conformal solution Eq.~\eqref{eq:rho_SYK} for $J=1000$.}
    \label{fig:SYK_rho}
\end{figure}

Then, we numerically solved Eq.~\eqref{eq:SYK_L_SDE} by employing the algorithm of appendix \ref{app:lorentzian_algo} for $q=4$ and increasing values of the coupling $J$. 
We plot the resulting spectral functions in Fig.~\ref{fig:SYK_rho}. 
We see that the asymptotic decay is weaker than exponential for large $\w$. This is expected based on the conformal invariance of the two-point function in the strong coupling limit $J \rightarrow \infty$. We computed the resulting spectral function in Appendix \ref{app:large_J_SYK} and demonstrate that it decays polynomially with large frequencies and thus satisfies the conditions of Theorem \ref{thm:poly_for_horizon} to give rise to an emergent radial direction. This matches the relation to the Schwarzian limit of JT-gravity which is a higher dimensional gravitational model.

\subsection{The spherical $p$-spin model}\label{sec:p_spin}

It is expected that the SYK model does not have a spin glass phase. We refer to \cite{Gur-Ari:2018okm} for an initial analysis and \cite{Anschuetz:2024naj} for a proof up to low temperatures. As we have already addressed above, we are interested in models that exhibit a transition from a spin liquid phase to a spin glass phase, as we want to find connections between the latter and systems of high complexity. The first such model we study is the \textit{spherical $p$-spin model} \cite{Crisanti1992, Cugliandolo2001, Castellani2005,Anous2021}. A significant part of the analysis and spectral function-related results, we present below, can be also found in \cite{Anous2021}. There are some small differences in the numerical procedure we followed (see App.~\ref{app:numerical_algorithm}) and we additionally add a possible physical interpretation to the behavior of the computed spectral functions, based on the main inspiration of this work.

\begin{figure}[h]
\centering
\includegraphics[width=0.3\linewidth]{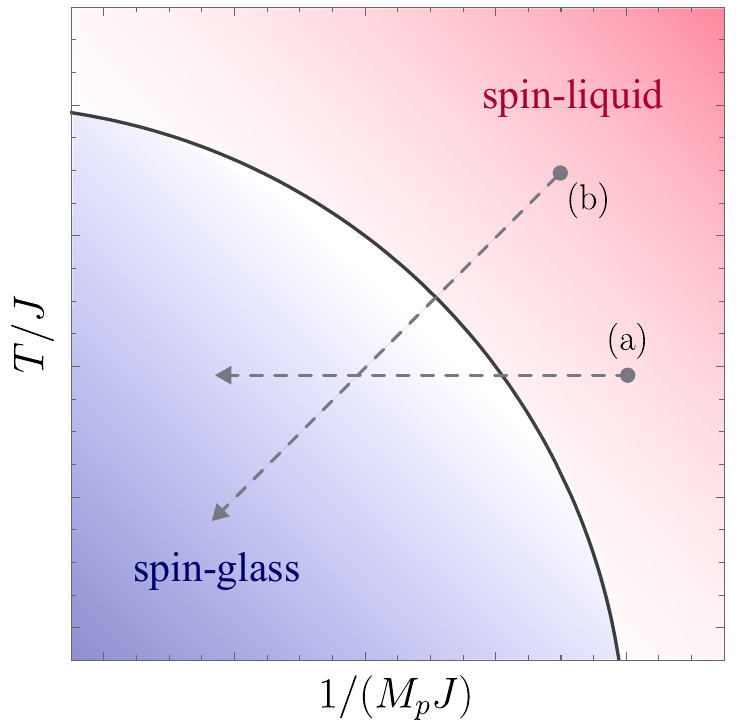}
\caption{A qualitative demonstration of the phase diagram of the spherical $p$-spin model as a function of the global parameters $\{J, T, M_p\}$. For the purposes of our work, the transition between the spin liquid and spin glass phase is determined by the replica symmetry breaking: $u=0$ and $m=1$ corresponds to the spin liquid phase, while for $u>0$ and $m<1$ the state is in the spin glass regime. The dashed arrows show the two adiabatic paths (a) and (b) we followed in our calculation.}
\label{fig:pspin_phasediagram}
\end{figure}

The spherical $p$-spin model is a bosonic version of SYK, where instead of $N$ sites that are each equipped with Majorana fermions one has each equipped with a continuous bosonic degree of freedom $\s_i$ that acts as the position operator on $L^2(\mathbb{R})$. This corresponds to the quantum version of the classical $p$-spin model \cite{Crisanti1992}.
The Hilbert space is given by  
\begin{equation}
    \cH = \big(L^2(\mathbb{R})\big)^{\otimes N}.
\end{equation}
In addition, one imposes the \textit{spherical constraint} 
\begin{equation}\label{eq:spherical_constraint}
    \sum_{i=1}^N \s_i^2 = N,
\end{equation}
i.e., one considers the subspace $\cH_N \subset \cH$ of states $\ket{\Y} \in \cH$ such that 
\begin{equation}
   \sum_{i=1}^N \s_i^2 \ket{\Y}= N \ket{\Y}.
\end{equation}
This is introduced to obtain a stable system, since
otherwise the Hamiltonian defined below can have arbitrarily negative energies \cite{Crisanti1992, Anous2021}, since the $\s_i$ are continuous variables that can take any real value.
The Hamiltonian is given by 
\begin{equation}
    H =\sum_{i_1 < \hdots < i_p} J_{i_1 \hdots i_p} \s_{i_1}\hdots \s_{i_p},
\end{equation}
where again $J_{i_1 \hdots i_p}$ is a Gaussian random variable with variance
\begin{equation}
    \braket{J_{i_1 \hdots i_p}^2} = \frac{J^2 p!}{2N^{p-1}}.
\end{equation}
One gives the bosons a mass $M_p$, so that the partition function for fixed
$J$ is determined by the path integral
\begin{equation}
    \begin{aligned}
    \braket{Z^n} = \int dJ P(J) \int d\s^a_i \prod_{t,a}\d(\sum_i
    \s^a_i\s^a_i(\t) - N ) \exp\Big(-\int_0^\b d\t &
    \big[\frac{M_p}{2} \dot{\s}^a_i(\t) \dot{\s}^a_i(\t) \\
    &+ \sum_a \sum_{i_1 < \hdots
    <i_p}J_{i_1 \hdots i_p} \s_{i_1}^a(\t)\hdots \s_{i_p}^a(\t) ] \Big)
    \end{aligned}
\end{equation}
We consider the $p=3$ spin model. Each point on the phase diagram Fig.~\ref{fig:pspin_phasediagram} of the $p$-spin model is characterized by the three parameters $\{\beta, J, M_p\}$.

The Green's function describing the large $N$ dynamics of the system is then
\begin{equation}
    g(\t,\t') = \overline{\frac{1}{N}\sum_{i = 1}^N \s_i(\t)\s_i(\t')},
\end{equation}
with the overline indicating the ensemble average. 
This only describes the replica-diagonal component. The associated self-energy is given by 
\begin{equation}
    \S_r(\t,\t') = \frac{p}{2}g(\t,\t')^{p-1}.
\end{equation}
An 1-step RSB ansatz, similar to the discussion of Appendix \ref{app:rsb_ansatz}, then leads to
the a set of \textit{equilibrium} equations of motion for the RSB parameters $u$ and $m$ and the Green's function $g(\t,\t')$.
However, a second SG solution has been found, for which the equilibrium equations of motion of parameter $m$ is not satisfied, the \textit{marginally stable SG state}. This was first observed in \cite{Cugliandolo2001} for the spherical $p$-spin model, based on the fact that there exists a zero eigenvalue solution for second order fluctuations in the effective action \cite{Kirkpatrick1987,Cugliandolo1993}. This was argued in \cite{Anous2021} to arise from considering an alternative ensemble, which defines a different state on the system, where replica symmetry is \textit{explicitly} broken. We will consider the marginal spin glass for the rest of the analysis, as it leads to a conformal solution at very low temperature. The equation relating $u$ to the zero mode $\hat{g}(0)$ then takes the form
\begin{equation}\label{eq:marginalSG_u}
    \frac{1}{2}pu^{p-2}(\beta J)^2 - \frac{1}{(p-1)\bigg(\cfrac{\hat{g}(0)}{\beta}-u\bigg)^2} = 0.
\end{equation}
For the marginal SG case, $m$ is then fixed by
\begin{equation}
    m = \frac{p-2}{J\beta u^{p/2}}\sqrt{\frac{2}{p(p-1)}}.
\end{equation}
For simplicity, one defines a new correlation function and self energy
\begin{equation}
\begin{aligned}
    q_r(\t,\t') &:= q(\t,\t') - u,\\
    \S_r(\t,\t') &:= \S_r(\t,\t')-\frac{p}{2}u^{p-1},
\end{aligned}
\end{equation}
for which the Schwinger-Dyson equations simplify to 
\begin{equation}\label{eq:p_spin_E_SDE}
    \frac{1}{\hat{g}_r(k)} -  \frac{1}{\hat{g}_r(0)} = M_p\big(\frac{2\pi k}{\beta}\big)^2 - J^2\Big( \hat{\Sigma}_r(k) - \hat{\Sigma}_r(0) \Big),
\end{equation}
with 
\begin{equation}
    \hat{\S}_r(k) - \hat{\S}_r(0) = \frac{p}{2}\int_0^\beta d\t \Big[\cos(\frac{2 \pi k \t}{\beta}) -1\Big]\big ((g_r(\t)+u)^{p-1}-u^{p-1})\big),
\end{equation}
where the cosine appears since $q$ is symmetric in $\t$ and the subtraction in $u$ arises from the definition of $\S_r$.
The Lorenzian equations of motion for $q(\omega)$ for real frequencies is obtained by analytic continuation as described in App.~\ref{app:analytic_continuation} and gives 
\begin{equation}\label{eq:p_spin_L_SDE}
    \frac{1}{\hat{g}^R_r(\omega)} -  \frac{1}{\hat{g}^R_r(0)} = M_p\omega^2 - J^2\Big( \hat{\Sigma}^R_r(\omega) - \hat{\Sigma}^R_r(0) \Big).
\end{equation}
where the superscript $R$ denotes the retarded functions, see Eq.~\eqref{eq:real_time_2pt}. The difference of the self-energy with its zero mode is given by
\begin{equation}
   \hat{\Sigma}^R_r(\omega) - \hat{\Sigma}^R_r(0) \equiv -\frac{ip}{2}\int_0^\infty dt (e^{-i\omega t} - 1) \big[ (g_r^>(t) + u^{p-1}) - (g_r^>(-t) + u^{p-1})\big]. 
\end{equation}
We will use the following diagnostic to determine, whether the solutions $\{\hat{g}(0), u\}$ are in the SG phase: if $0<m<1$ and $u$ is nonzero then our solutions are indeed in the SG phase, as this condition indicates the breaking of replica symmetry. \\

\begin{figure}[H]
    \centering
    \includegraphics[width=0.85\linewidth]{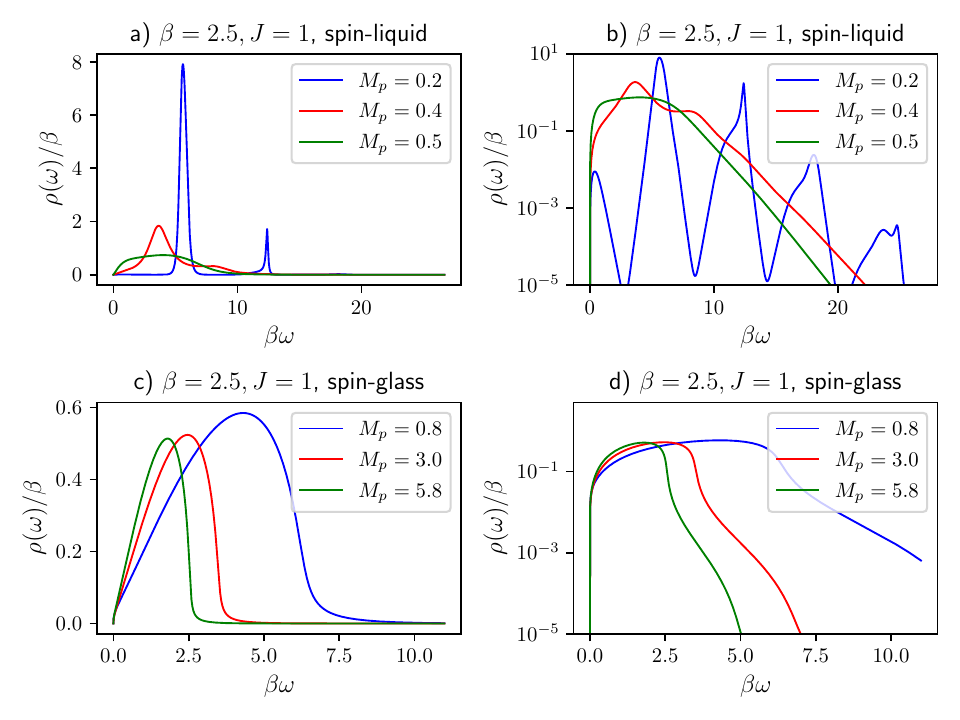}
    \caption{Spectral functions of the spherical $p$-spin model for fixed $\beta=2.5$, $J=1$ and increasing $M_p$. Figures a) and b) show the cases for $M_p=\{0.2, 0.3, 0.4\}$ in the spin liquid phase in linear and log-y scaling respectively. Figures c) and d) show the spin glass cases for $M_p=\{0.8, 3.0, 5.8\}$ in linear and log-y scaling respectively. We considered $N=2\cdot10^5$ points for the calculations above.
    }    \label{fig:Pspin_J1beta2p5}
\end{figure}

\begin{figure}[H]
    \centering
    \includegraphics[width=0.85\linewidth]{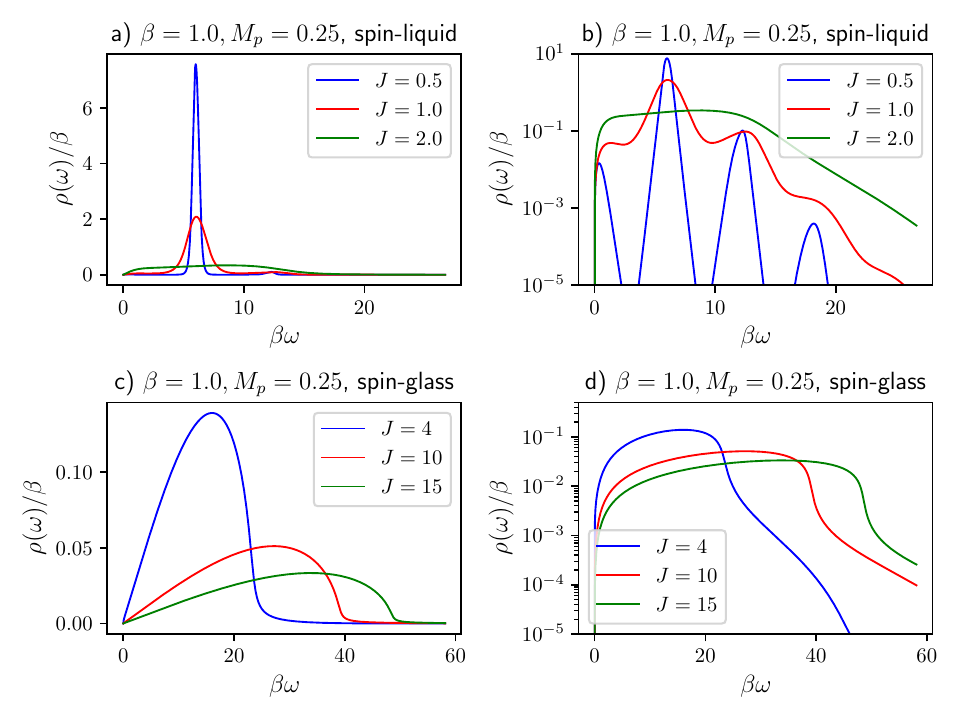}
    \caption{Spectral functions of the spherical p-spin model for $\beta=1$, $M_p=0.25$ and increasing $J$. In the plots a) and b), we show the spin liquid cases for $J=\{0.5, 1, 2\}$ in linear and log-y scaling respectively. The spectral functions in c) and d) represent the spin glass cases for $J=\{4, 10, 15\}$ in linear and log-y scaling respectively. In all calculations presented we used $N=2\cdot 10^5$ points.
    }
    \label{fig:Pspin_b1M0p25}
\end{figure}

So, now we have all the equations needed to apply the numerical algorithm described in Appendix \ref{app:numerical_algorithm}, which will iteratively compute \ref{eq:p_spin_L_SDE} and then the spectral function from $\rho_r(\omega) = 2\,\mathrm{Im}\, g_r^R(\omega) $. The analysis of the spectral functions for the spin liquid and spin glass of the $p$-spin model will be based on two adiabatic processes: 
\begin{itemize}
    \item[(a)] For fixed $\beta=2.5$ and $J=1$, we adiabatically move in the range $M_p\in [0.2, 5.8]$. This corresponds to a \textit{vertical} adiabatic movement on the phase diagram in Fig.~\ref{fig:pspin_phasediagram}, starting from the spin liquid phase and crossing the transition deep into the spin glass phase. In Fig.~\ref{fig:Pspin_J1beta2p5}, we present the results that correspond to the two phases separately.
    \item[(b)] For fixed $\beta=1$ and $M_p=0.25$, we adiabatically move in the $J\in [0.5, 15]$ regime. This process describes an adiabatic movement from the spin liquid to the spin glass phase in a \textit{diagonal} manner in Fig.~\ref{fig:pspin_phasediagram}, moving towards the conformal limit. Again, we present the results corresponding to the two phases separately in Fig.~\ref{fig:Pspin_b1M0p25}.
\end{itemize}

Before we discuss the results, we should note one more time that the spectral functions for similar regimes have already been presented in \cite{Anous2021}. However, our interpretation is focused on the asymptotic behavior of the spectral function, which was largely ignored in previous considerations and that is the relevant property for the emergence of a semiclassical radial direction.

Let us start with the \textit{vertical} adiabatic process, where we have fixed $\beta=2.5$ and $J=1$, in Fig.~\ref{fig:Pspin_J1beta2p5}.

\begin{figure}[H]
    \centering
    \includegraphics[width=0.5\linewidth]{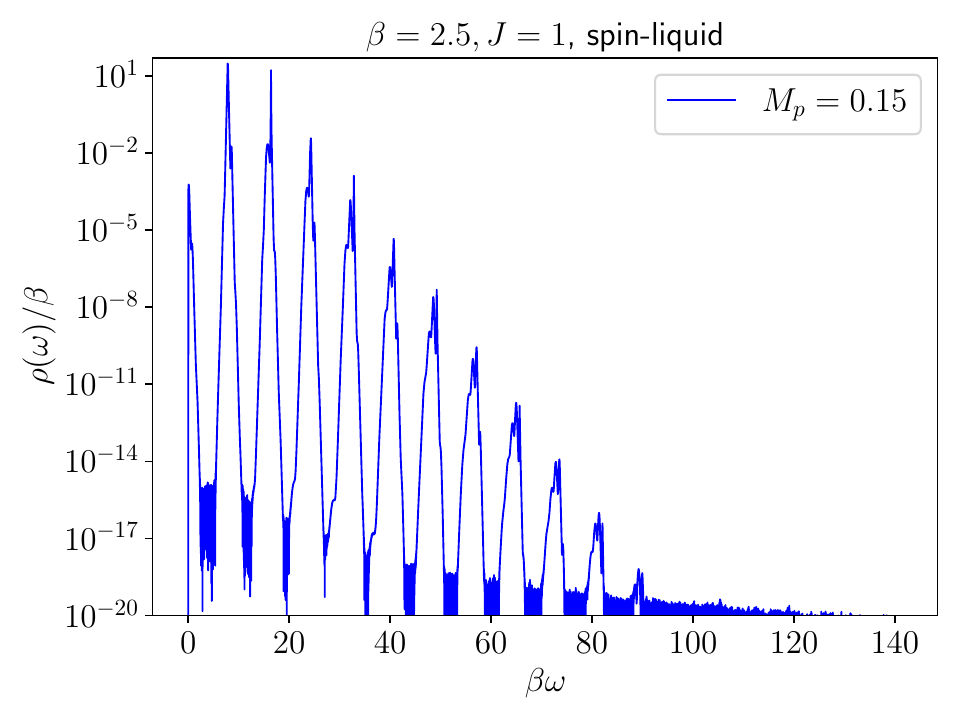}
    \caption{Spectral function for $\beta=2.5, J=1$ and $M_p=0.15$ in a log-y scaling. This parameter regime corresponds to a state deep in the spin-liquid phase of the $p$-spin model. Delta-like peaks are present for all frequencies $\omega$ considered till computational restrictions become significant. In all calculations presented we used $N=5\cdot 10^5$ points.}
    \label{fig:Pspin_deepSL}
\end{figure}

The spectral function of the spin liquid phase is characterized by two distinct behaviors. For the regime between the transition point (close to $M_p=0.6$) and $M\sim 0.4$, the spectral function decays exponentially, as seen in Fig.~\ref{fig:Pspin_J1beta2p5}(b). This indicates that the function shows full support on the frequency space. This is not very surprising, as the spin liquid phase of the spherical $p$-spin model shares many features with the SYK in the large q-limit \cite{Anous2021}, where we observe similar exponential decay, see Fig.~\ref{fig:SYK_rho}. However, the behavior changes drastically when we move deep into the spin liquid phase, for example for $M\leq 0.2$. In this particular regime, the spectral function shows peaks of a small width, that resemble delta-like peaks as we decrease $M_p$. From the log-y plot in Fig.~\ref{fig:Pspin_J1beta2p5}(b), we observe the emergence of them in the whole range of frequencies we considered in our simulation. In Fig.~\ref{fig:Pspin_deepSL} we demonstrate that these peaks are present over the whole range of frequencies up to machine precision, with an exponentially decaying magnitude. This behavior resembles the case of finite depth parameter, where the spectral function is described by an infinite sum of equally space delta functions \eqref{eq:rho_delta}. However, since the peaks have an exponentially decaying coefficient, the Theorem \ref{thm:finite_depth_delta} does not directly apply. It would however be interesting if the spectral function in this case does still allow for a nonzero depth parameter that would lead to a type I algebra. Also note that, in contrast to the case of holographic CFT's such as $\cN = 4$ SYM, where the finite depth parameter appears in the confined phase at low temperatures, this infinite set of peaks appears at \textit{high} temperatures, where the system is in a spin liquid state. Further investigations of this regime are left for future work.

The spectral functions of three instances in the spin glass phase are presented in the lower panel of Fig.~\ref{fig:Pspin_J1beta2p5}. By increasing the parameter $M_p$ we move ``deeper" in the spin glass phase, as it can be seen from the phase diagram in Fig.~\ref{fig:pspin_phasediagram}. The log-y plot (Fig.~\ref{fig:Pspin_J1beta2p5}(d)) reveals an interesting behavior regarding the larger frequencies. As we increase $M_p$, a ``kink" is formed, which we identify to be the almost vertical drop in magnitude of $\r(\w)$. Deep into the spin glass, this drop is followed by a superexponential decay of $\r(\w)$, as seen for example in the tails for $M_p=3$ and $M_p=5.8$. The appearance of the kinks, as we enter the spin glass phase, show signs of compact support regarding the spectral functions, especially deep into the spin glass phase. Following the ideas discussed in subsection \ref{sec:theorems}, in particular Theorem \ref{thm:compact_support}, the spin glass-phase for finite coupling constant $J$ does appear to exhibit a compactly supported spectral function, whose support decreases with increasing $M_p$ and thus is not a good candidate for a many-body physics model that could have a dual theory.  

The above discussion is robust for any $\{M_p, J, \beta\}$-point in the spin glass phase of the spherical $p$-spin model, as long as $J$ and $\beta$ are finite. Thus, the next step is to approach the spin glass phase by increasing the coupling constant $J$, for fixed $\beta$ and $M_p$. This corresponds to approaching the conformal limit of the model. In Fig.~\ref{fig:Pspin_b1M0p25}, we collect all the results regarding this adiabatic movement on the phase diagram.

Similarly to the analysis above, we start from the spin liquid phase for $J=0.5$ increasing $J$. Fig.~\ref{fig:Pspin_b1M0p25}(a, b) show the spectral functions over the real frequencies $\omega$ for three points in the spin liquid phase. As expected, we observe the same behavior as in the previous adiabatic process: there is a regime for which the spectral function decays exponentially (for $J=2$) and a regime where peaks emerge (for $J<0.5$), as we move deeper into the spin liquid phase. 

By increasing $J$ even more, we cross the transition to the spin glass phase (Fig.~\ref{fig:Pspin_b1M0p25}(c, d)). For the cases studied, the appearance of a ``kink" and an almost vertical decay becomes visible. Similarly to the spin glass phase for $J=1$ in Fig.~\ref{fig:Pspin_J1beta2p5}, there is a tail after the first sudden decay, which is characterized by superexponential decay. Therefore, we could claim that for large $J$, the spectral function of the spin glass phase shows signs of compact support. In total, this tendency is robust throughout the entire spin glass phase. This gives a first clue that there is no regime in the spin glass phase of the spherical $p$-spin model for $p=3$ that could give rise to a dual theory, based on the use of the spectral function as a diagnostic in Theorem \ref{thm:compact_support}. It should be noted that in \cite{Anous2021}, an approximate solution for the system approaching the conformal limit was derived, from which the authors extracted an approximate spectral function given by 
\begin{equation}
    \r^\approx_r(\w) = \frac{16 \g^2}{M_p}\Q\Big( \frac{1}{\g}-|\w|\Big) \g \w \sqrt{1-\g^2 \w^2},\hspace{50pt} \g := \sqrt{\frac{M_p \hat{q}_r(0)}{4}}.
\end{equation}
This function has explicit compact support and approaches the exact solution as one increases the coupling. This provides further evidence towards the absence of a nonzero depth parameter. It is interesting to note that the conformal solution is given by \cite{Anous2021} 
\begin{equation} \label{eq:approximate_pspin}
    g(\t) = \frac{8 \pi \g^3}{M_p \b^2 \sin(\frac{\pi \t}{\b})^2}.
\end{equation}
After taking a Fourier transform of Eq.~\eqref{eq:approximate_pspin}, the same way as in large $J$ SYK studied in Appendix  \ref{app:large_J_SYK}, the conformal solution leads to an polynomially decaying spectral function. Thus, it appears that, the conclusion one draws strongly depends on the approximation taken. One feature that is visible from Fig.~\ref{fig:Pspin_b1M0p25} is that the support increases as $J$ increases. It would be interesting to see, if the kink suggested from the approximate solution develops into the polynomial tail of the conformal solution in the limit $J\rightarrow \infty$. Further investigations of this discrepancy are left for future work.

\subsection{The SU($M$) Heisenberg model}\label{sec:SUM}

\begin{figure}[!h]
\centering
\includegraphics[width=0.3\linewidth]{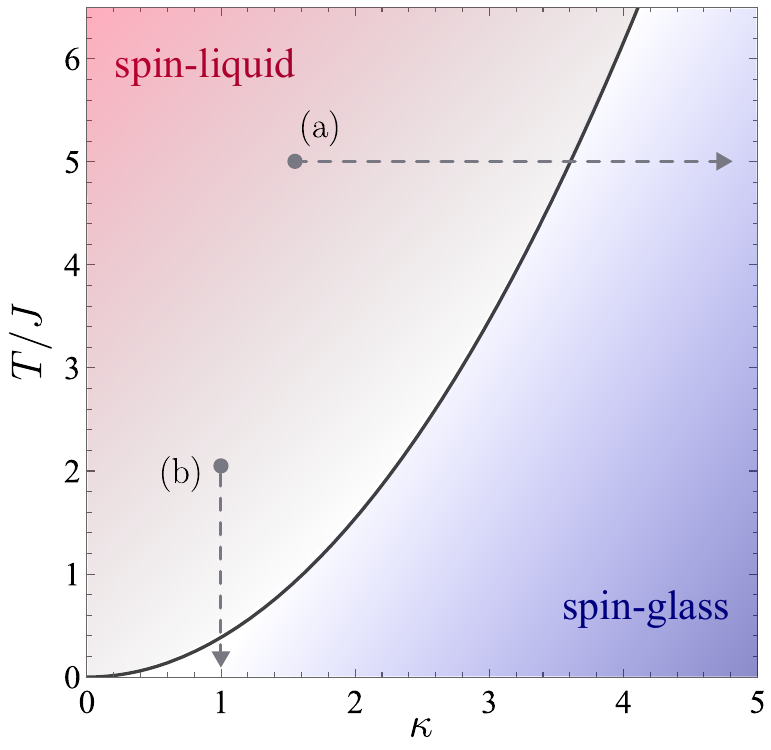}
\caption{Phase diagram of the SU$(M)$ Heisenberg model. The transition line between the two phases is given by $T_c \sim \frac{2}{3\sqrt{3}}J\kappa^2$, as used in \cite{Georges_2001}. However, like in the analysis of the $p$-spin model, we will determine whether a parameter point $\{J, T\}$ lives in the spin glass phase, if the corresponding state breaks the replica symmetry. The two dashed arrows show the adiabatic paths (a) and (b) we followed in our calculations.} 
\label{fig:SY_phases}
\end{figure}

Until this point, we have studied the properties of the spectral functions of the SYK model in the large $q$ limit and the $q=4$ case, as well as of the spherical $p$-spin model. The first one has only a spin liquid phase, the second one has a spin liquid/spin glass phase transition. Based on the observations above, the spin-liquid phases of both models are characterized by spectral function with exponential decaying envelope. On the contrary, the spin glass phase of the $p$-spin model shows a rapid decay for larger frequencies. In this section, we will drift our focus towards the SU($M$) Heisenberg model \cite{Sachdev:1992fk}, also known as the SY model. This is also a model with a spin glass phase, see the phase diagram in Fig.~\ref{fig:SY_phases}. 

The Hamiltonian of the SU$(M)$ Heisenberg model is given by
\begin{equation}
	\mathcal{H}=\frac{1}{\sqrt{NM}}\sum_{i<j=1}^{N}J_{ij}\mathbf{S}_{i}\cdot\mathbf{S}_{j},
\end{equation}
where $\mathbf{S}_{i}$ are SU$(M)$ spins. The model is characterized by all-to-all interactions $J_{ij}$, that are randomly and independently drawn from a Gaussian distribution and satisfy:
\begin{equation}
	\overline{J_{ij}}=0,\quad\overline{J^{2}_{ij}}=J^{2}.
\end{equation}\label{eq:couplings_mean}
We write the SU$(M)$ spin operators as matrices $S_{\beta}^{\alpha}(i) = S_{\alpha}^{\beta}(i)^{\dagger}$ on each site $i$, where $\alpha,\beta=1,\ldots,M$, which are related to the generators of SU($M$).

At this point, we have two representation choices for the SU$(M)$ spins: a bosonic and a fermionic. For the purposes of this work, we will choose the bosonic representation, as we want to access all points of the phase diagram in Fig.~\ref{fig:SY_phases} of the model. In particular, using the fermionic representation would not allow us to access the spin glass phase at all, which is the phase of interest to us.

Thus, we express the spins in terms of bosonic creation/annihilation operators $b^\b(i),b_\b^\dagger(i)$ which leads to
\begin{equation}
    S_{\alpha}^{\beta}(i)=b_{\a}^{\dagger}(i)b^{\b}(i)-\kappa
    \delta_{\a}^{\b}.
\end{equation}
The bosonic operators obey the usual canonical commutation relations: {$[b^{\alpha}(i),b_{\beta}^{\dagger}(j)]=\delta^{\alpha}_{\beta}\delta_{ij}$}. 
For reasons of stability, similar to the case of the $p$-spin model, one introduces a particle number constraint on each site, which leads to
\begin{equation}\label{eq:number_constraint}
	\quad\sum_{\alpha}b_{\alpha}^{\dagger}(i)b^\a (i)=\kappa M.
\end{equation}
The parameter $\k$ serves as a fraction that quantifies the occupancy of bosonic modes. In the limit $M \rightarrow \infty$ it can take every real value $\k >0$.
Under the bosonic representation, we can then express the Hamiltonian as follows:
\begin{equation}\label{eq:SY_bosonicHAM}
	\mathcal{H}=\frac{1}{\sqrt{NM}}\sum_{i<j=1}^{N}J_{ij}\sum_{\alpha,\beta=1}^{M}(b_{\alpha}^{\dagger}(i)b^{\beta}(i)b_{\beta}^{\dagger}(j)b^{\alpha}(j)-\kappa^{2}
    M).
\end{equation}

In Appendix \ref{app:analytical_calculations} we perform the full replica path integral derivation of the effective action, leading to the equilibrium equations satisfied by the model in the limit $M \rightarrow \infty , N \rightarrow \infty $ with $ M \ll N$. The resulting Schwinger-Dyson equations have already been known \cite{Sachdev:1992fk,Georges_2001}, the exact spin glass equations for the replica breaking parameters $u,m$, we have not found in the literature, which are also convention dependent. An analysis of $1/M$ effects in the fermionic system was performed in \cite{Christos:2021wno} and has parallels with our derivation.
The phase diagram was first described in \cite{Georges_2001}. Due to the existence of differing conventions, we rederived them in the fashion used in the $p$-spin model in \cite{Anous2021}. In this section, we will mention them again for completeness, without repeating the derivation. 
The Schwinger-Dyson equations describe the large $N$ limit of the disorder averaged thermal correlation function
\begin{equation}
    g_{r,r'}(\tau,\tau')\equiv\frac{1}{M} \overline{\sum_{\alpha}\langle b_{r}^{\alpha}(\tau)b_{\alpha,r'}^{\dagger}(\tau')\rangle},
\end{equation}
where $r$ are replica indices. In the following, we denote the diagonal component $g_{r,r}(\t,0)$ by $g(\t)$.
The Schwinger-Dyson equation for $g(\t)$ is given by
\begin{equation}\label{eq:SY_E_SDE}
    \frac{1}{\hat{g}_r(k)} = \frac{1}{\hat{g}_r(0)}+\frac{2 \pi i k}{\beta} - \Big(\hat{\Sigma}_r(k)-\hat{\Sigma}_r(0)\Big).
\end{equation}
In addition, the following equilibrium equations must be satisfied by the replica-symmetry breaking parameters $(u,m)$:\begin{equation}\label{eq:SYeom_all}
\begin{aligned}
    \mathrm{EoM}_u: ~~ (m-1)\left(J^{2}\b^{2}u^{3}-\frac{u}{\bigg(\dfrac{\hat{g}(0)}{\beta}-u\bigg)\bigg(\dfrac{\hat{g}(0)}{\beta}+(m-1)u\bigg)}\right)=0, \\
    \mathrm{EoM}_m: ~~ \frac{J^{2}\b^{2}}{4}u^{4}+\frac{1}{m}\frac{u}{\cfrac{\hat{g}(0)}{\b}+(m-1)u}+\frac{1}{m^{2}}\log\left(\frac{\cfrac{\hat{g}(0)}{\b}-u}{\cfrac{\hat{g}(0)}{\b}+(m-1)u}\right)=0.
\end{aligned}
\end{equation}
The equations above will map each point on the phase diagram to a triplet $\{g(0), m, u\}$, following the Euclidean numerical algorithm of Appendix  \ref{app:numerical_algorithm}. Then the Lorentzian algorithm will compute $g_r^R(\omega)$, from:
\begin{equation}\label{eq:SY_L_SDE}
    \frac{1}{\hat{g}^R_r(\omega)} =\frac{1}{\hat{g}^R_r(0)} - \w - \Big(\hat{\Sigma}_r(\omega)-\hat{\Sigma}_r(0)\Big) ,
\end{equation}
via an iterative procedure described in Appendix \ref{app:numerical_algorithm}. Then, the spectral function associated to the bosonic correlator is $\rho_r(\omega) = 2\mathrm{Im} g_r^R(\omega) $. The behavior of $\rho_r(\omega)$ for different regimes of the phase diagram characterizing the SU$(M)$ Heisenberg model in the $M\rightarrow\infty$ limit, we discuss in the following.\\

\begin{figure}[H]
    \centering
    \includegraphics[width=0.85\linewidth]{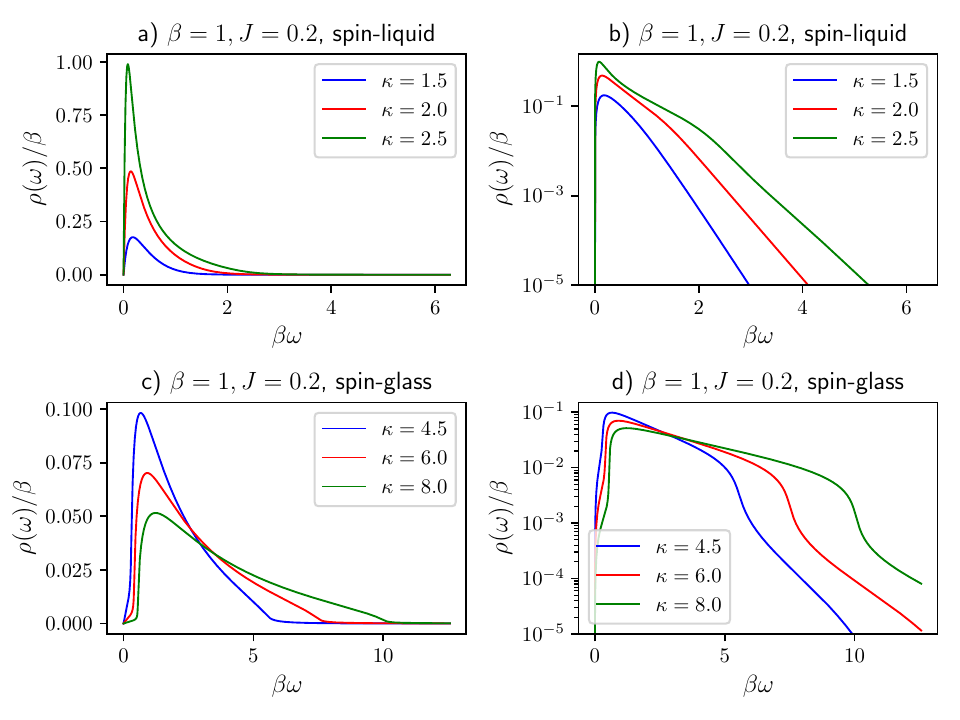}
    \caption{Spectral functions of the SU$(M)$ Heisenberg model for fixed $\beta=1, J=0.2$ and increasing $\kappa$ into the \textit{semiclassical} regime. The cases $\kappa=\{1.5, 2, 2.5\}$ in the spin liquid phase are shown in the plots a) in linear and b) in logarithmic scaling. As we increase $\kappa$, we reach the spin glass phase. The spin glass cases $\kappa= \{4.5, 6, 8\}$ are shown in c) for linear and d) for log-y scaling. In all calculations presented, we used $N=2\cdot 10^5$ points.
    }
    \label{fig:J02SLSG}    
\end{figure}

\begin{figure}[H]
 \includegraphics[width=0.85\linewidth]{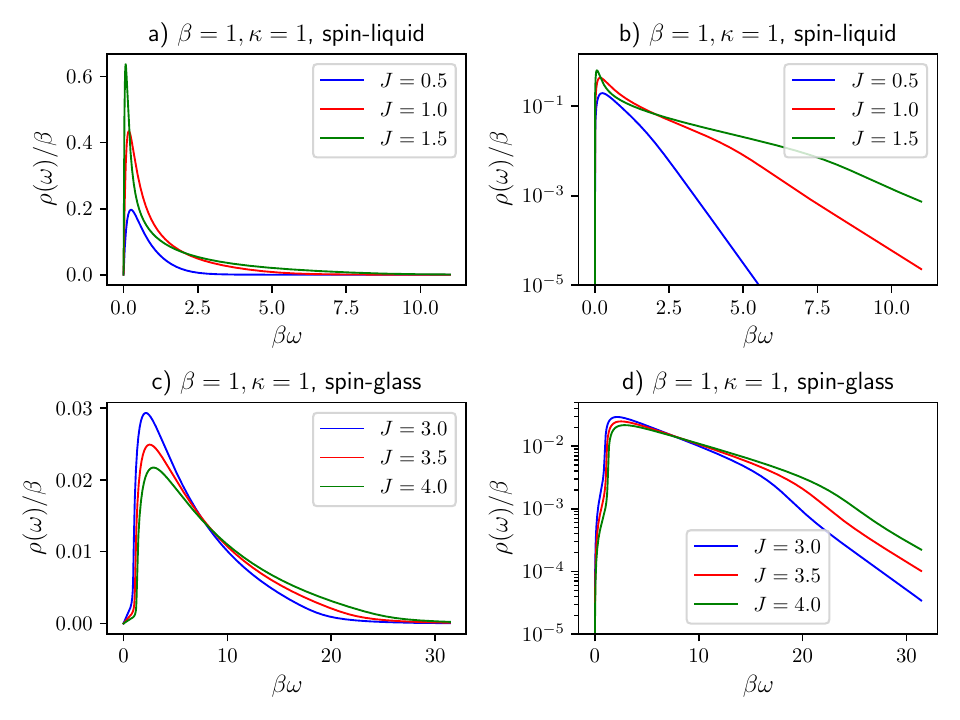}
 \centering
    \caption{Spectral functions of the SU$(M)$ Heisenberg model for fixed $\beta=1, \kappa=1$ and increasing coupling $J$ into the \textit{Quantum spin glass} regime. We show the results in a) in linear and b) in logarithmic scale for the spectral functions corresponding to the spin liquid cases of $J= \{0.5, 1, 1.5\}$. The spin glass cases for $J= \{3, 3.5, 4\}$ are shown in c) in linear and d) in logarithmic scale. In all calculations presented, we used $N=2\cdot 10^5$ points.
    }
    \label{fig:kappa1SLSG}    
\end{figure}

Here, we follow an adiabatic procedure similar to the one for the $p$-spin model above. In particular, we consider the following two adiabatic processes:
\begin{itemize}
    \item[(a)] for fixed $\beta=1$ and $J=0.2$, we increase $\kappa$ adiabatically (moving \textit{horizontally} on the phase diagram \ref{fig:SY_phases}), from $\kappa =1.5$, where the model behaves as a spin liquid, till $\kappa=8$, where the system is deep in the spin glass phase. In Fig.~\ref{fig:J02SLSG}, we present the spectral functions for the spin liquid phase (upper panels) and the spin glass phase (lower panels) separately.
    \item[(b)] for fixed $\beta=1$ and $\kappa=1$, we move adiabatically from the spin liquid phase towards the spin glass phase by increasing the coupling constant $J$. This corresponds to a \textit{vertical} adiabatic process in the phase diagram \ref{fig:SY_phases}. For values between $J\in[0.5,1.5]$ the system lives in the spin liquid phase, Fig.~\ref{fig:kappa1SLSG}(a, b), while for values $J\in[3,4]$, the systems is in the spin glass phase, Fig.~\ref{fig:kappa1SLSG}(c, d). The transition happens around $J \approx 2.3$.
\end{itemize}

Let us start with the first adiabatic process. In Fig.~\ref{fig:J02SLSG}, we show the numerical results for the spectral function as one increases the parameter $\k$. The upper panels correspond to the values of $\kappa$, for which the model behaves as a spin liquid. For all cases studied, the spectral functions decay exponentially (it can be clearly seen in the log-y plot), admitting complete support over the real frequencies. This is the expected  behavior for the spin liquid phase, that resembles the spectral functions of SYK, also a spin liquid model.

Increasing $\kappa$ even more moves us into the spin glass phase. For fixed $J=0.2$, this part of the spin glass phase corresponds to the  \textit{semiclassical regime}  \cite{Georges_2001}. In this regime, for finite $\k$, we observe the emergence of a kink, which can be seen as an almost vertical decay of the spectral function in the logarithmic plot \ref{fig:J02SLSG} d). After this drop, $\r(\w)$ exhibits a superexponential decay that becomes an exponential tail. The existence of a kink is quite similar to the one observed in the spin glass phase of the spherical $p$-spin model, with the difference that the function exhibits a regime of exponential decay that precedes the kink. Due to the order of magnitude drop, that does not analogously appear in the spin liquid phase, we consider it as a sign of compact support at finite $\k$. In addition, we note that as $\k$ increases, the spectral function tends to zero at low frequencies, which we interpret as the emergence of a spectral gap (see Fig. \ref{fig:J02SLSG} c) for $\k = 8$). Furthermore, with increasing $\k$ the support appears to become larger, which is indicated by the position of the kink moving to larger frequencies. It would be interesting to study, whether the limit $\k \rightarrow \infty$ exists, since by extrapolation of the finite $\k$ data, this limit might exhibit noncompact support.

Now, we can discuss the second adiabatic process. In Fig.~\ref{fig:kappa1SLSG} one finds the exact solutions of the spectral function computed in both the spin liquid and marginal spin glass with increasing coupling strength. In the spin liquid phase, in Fig.~\ref{fig:kappa1SLSG}, we see complete support, as well as exponential decay in all parameter regimes. This is similar to SYK, which is a spin liquid as well, and it is the characteristic behavior we have observed in all spin liquid regimes we have studied so far.
 
\begin{figure}[H]
    \centering
    \includegraphics[width=0.8\linewidth]{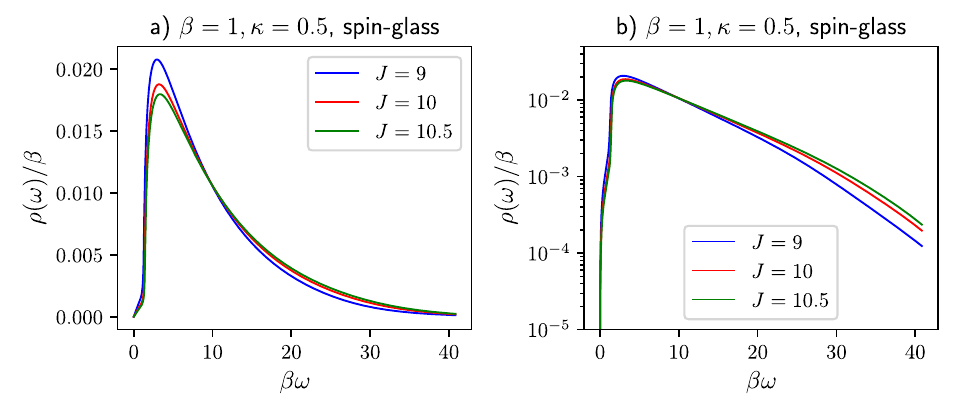}
    \caption{A second example of spectral functions of the SU$(M)$ Heisenberg model for fixed $\beta=1$ and $\kappa=0.5$ in the Quantum spin glass regime. In all calculations presented we used $N=2\cdot 10^5$ points.
    }
    \label{fig:kappa05SLSG}    
\end{figure}

So far, both the spin liquid and spin glass regimes we explored show very similar behavior for both the $p$-spin and SU($M$) Heisenberg models. The spin liquid phase is characterized by exponential decay, while the spin glass phase by compact support over the real frequencies. However, a quite intriguing behavior emerges for fixed $\kappa=1$, as we enter the spin glass phase by increasing the coupling $J$. This region of the spin glass phase is called a \textit{quantum spin glass} \cite{Georges_2001}. In this regime, for all coupling values studied, the spectral function shows a robust exponential decay, thus non-compact over the real frequencies. This behavior is in contrast with the rest of the spin glass regimes we studied above, and instead is similar to the one we observe in the large q limit of the SYK model, although we are deep into a spin glass phase. This is a trend found in the whole spin glass region for $\kappa\in (0,1]$ at finite temperature, see e.g.\ Fig.~\ref{fig:kappa05SLSG}. 

In \cite{Georges_2001} it was argued that for the bosonic model, only for $\k<0.052$ a spin liquid solution exists at zero temperature, which imposes a lower bound up to which the aforementioned behavior is expected.  The similarity with the large q limit of the SYK model may be the first clue that this spin glass regime could be a possible candidate of a spin glass model with a semi-classical spacetime dual.

\section{No detectable emergence for low-energy observables}\label{sec:no_em_low_energy}
In the previous section we demonstrated that, in all cases we studied, thermal states of large $N$ system with quenched disorder have an exponentially decaying tail in their spectral function, if the support is non-compact and there is no emergent conformal symmetry. This is reminiscent of the universal operator growth hypothesis \cite{Parker:2018yvk}, which argues that chaotic systems have associated spectral function that show exponential decay. The arguments leading to this hypothesis are based on the infinite temperature state and also only assert that, chaotic systems have exponentially decaying tails, not that such tails are exclusively found in chaotic systems. The framework of \cite{Gesteau:2024dhj} is only able to make a statement about bulk properties of systems with spectral functions that show at most polynomial decay. Here we want to make a small step forward towards the inclusion of spectral functions with exponential tails. As reviewed in Section \ref{sec:theorems}, the main condition one has to demonstrate for the existence of a relative commutant for the timeband algebra generated by operators in $[-\frac{T}{2},\frac{T}{2}]$ is the existence of a function $g(t)$, which is supported outside of the timeband in question and that is integrable against $\rho(\w)$ in the sense of Eq.~\eqref{eq:L_2_rho}. If $\rho(\w)$ has at most polynomial decay, then every square integrable function is also allowed as a smearing function. If $\rho(\w)$ is exponentially decaying however, we can also smear with functions whose Fourier transforms \textit{grow} with increasing $\w$, as long as the growth does not compensate the decay of $\r$ to hinder the finiteness of the integral in Eq.~\eqref{eq:L_2_rho}. This is a somewhat uncommon feature of such systems, as smearing functions in usual quantum field theories are assumed to be Schwartz function \cite{StreaterWightman2001}, which in particular decay both for large times and large frequencies. Here we want to demonstrate that operators smeared with such functions can not be elements of the relative commutant and thus, if there is a nonzero depth parameter, smearing with these functions will not be able to detect it.
\begin{proposition}
Let the spectral function of a large $N$ system be $\r(\w) = \cO(\exp(-\a \w^\b))$ for any $\a > 0, \b \geq 1$. Then for any timeband $\Delta T := [-T/2,T/2]$, there is no nonzero function $g(t)$, so that the smeared operator $\p(g)$ is in the relative commutant of the algebra in $\D T$, whose Fourier transform $g(\w)$ grows at most polynomially with $\w$.
\end{proposition}
\noindent The proof is basically the same as given by \cite{Gesteau:2024dhj} for Theorem \ref{thm:compact_support} just adapted for the above case of exponential decay instead of compact support.\\

\begin{proof}
Assume that a function $g(t)$ exists, so that the smeared operator $O(g)$ is in the relative commutant for the timeband algebra associated to the interval $\D T$ and its Fourier transform $g(\w) = \cO(\w^k)$ for some constant $k>0$. Following the discussion below Eq. \eqref{eq:commutator}, the relevant function to consider is the Fourier transform of $g(\w) \r(\w)$, which has to vanish inside $\D T$ for $O(g)$ to be in the relative commutant, i.e.
\begin{equation}
    \cF(t) =\int d\w \rho(\w) g(\w) e^{i \w t} = 0,\ \forall t \in \D T.
\end{equation}
Since $\r(\w) = \cO(\exp(-\a \w^\b))$, the product $g(\w)\r(\w)$ with polynomially growing $g(\w)$ is in $L^1(\mathbb{R})$ and in particular we have
\begin{equation}
    g(\w)\r(w) = \cO(\w^k \exp(-\a \w^\b)).
\end{equation}
Because of this decay, the Fourier transform $\cF(t)$ defines a function that is analytic in the interior of the strip $\mathbb{R}+i[-\a,\a]$ for $\beta = 1$ and an entire function for $\beta >1$. Therefore, if $\cF(t)$ vanishes inside $\D T$, by analyticity it has to be zero everywhere, which implies that $g(\w) = 0$ everywhere. Therefore no such nonzero $g(t)$ exists. 
\end{proof}
The previous proposition demonstrates that observables that are able to detect a nontrivial causal structure that emerges from the boundary system need to have superpolynomial growth in frequency, which is a unusual object to consider, as it weighs high energies with increasing strength. It is tempting to define an algebra of operators generated by functions with at most polynomial growth and associate this with observational capabilities of a boundary observer to detect a nontrivial bulk causality. This however appears to be a nontrivial task, since polynomial functions do not form a closed subspace of $L^2(\mathbb{R})$, so that it does not seem to create a von Neumann algebra. Perhaps it might be possible to make such statements for the space $L^2(\r)$ characterised by Eq.~\eqref{eq:L_2_rho} but this goes beyond the scope of this work. Note that we do not associate the presence of an exponential tail in $\r$ with a drastic change in properties of the emergent bulk, since large $q$ SYK has small geometrical fluctuations, but only note that the current framework is not yet capable to quantitatively address this case in accordance with \cite{Gesteau:2024dhj}.

It should be noted that in \cite{Belin:2025nqd}, similar observations towards the incompleteness of the framework proposed by \cite{Gesteau:2024dhj} was made. There, the authors demonstrate that the large $N$ limit of specific correlators in symmetric product orbifolds, \textit{all} converge to the result of the BTZ black hole, despite the impossibility of a local bulk dual at finite $N$, due to the nonchaotic structure of the theories in this regime. Our perspective on this result is that, there is a difference between the large $N$ and finite $N$ system and the framework of \cite{Gesteau:2024dhj} only makes a statement about the former.

\section{Discussion}\label{sec:discussion}
In this work, we numerically studied the large $N$ limits of systems with quenched disorder. Our motivation emerged from two separate ideas. Firstly, recent works \cite{Anninos2011,Kachru:2009xf,Anous2021,Denef:2011ee} tried to bridge high-complexity frameworks, appearing in string theory, with many-body models that show glassy behavior. The arguments are based on shared properties between them. Secondly, Gesteau and Liu \cite{Gesteau:2024dhj} proposed the spectral function as a diagnostic for the emergence of bulk spacetime, based on the algebraic approach to holographic dualities. Here, our main guiding principle was the framework developed in \cite{Gesteau:2024dhj} that proved several theorems based on which one can judge the emergence of a nontrivial bulk causal structure. The main focus was the asymptotic behavior of the spectral functions, which was not studied in detail in the past. Thus, we focused on the numerical computation of large $N$ spectral functions throughout different parameter regimes on the phase diagram of three systems: the SYK, the spherical $p$-spin model and the SU$(M)$ Heisenberg model. The last two of which have a spin glass phase.

We demonstrated that the paradigmatic example, the SYK model at large $q$, exhibits full support in the spectral function. In this regime the spectral function exhibits an exponentially decaying tail as was already noted in \cite{Gesteau:2024dhj}. We numerically compute the spectral function for $q=4$, and demonstrate that at large $J$ it approaches the polynomial decay that is expected from the emergence of conformal symmetry in this limit. Although SYK is expected to be holographic for $q \rightarrow \infty$ the theorem proved by GL does not apply due to the exponential decay. Because of this expected duality, we take the exponential tail as a reference behavior for the study of other models. We consider it as a first indication as to which regimes to explore for a possible holographic dual.

Next, we demonstrated that in the spherical $p$-spin model, the exponential decay is the predominant behavior in the spin liquid phase. However, deeper into the spin liquid phase, the equilibrium state exhibits an infinite set of peaks. This is reminiscent of the situation of holographic CFTs at low temperature, which exhibit spectral functions made from equally spaced $\delta$-functions that lead to finite depth parameter, as asserted by Theorem \ref{thm:finite_depth_delta}. We leave the investigation on whether these peaks become $\d$-functions for future work. It is interesting to note however, that these peaks appear at high-temperature. Since these peaks are accompanied by an exponentially decaying magnitude in contrast to polynomial decay, the theorem \eqref{thm:finite_depth_delta} does not apply. The behavior of the spectral functions in the spin glass phase of the model is more robust: after we cross the phase transition, a ``kink" appears, which indicates compact support. As a result, the whole spin glass phase seems to be unsuitable for a dual theory to exist.

The spin liquid phase of the SU($M$) Heisenberg model gives spectral functions with exponential decay, in agreement with the two other models studied. We argued that our numerics indicate that the semiclassical spin glass phase exhibits a spectral function with compact support due to the existence of a kink, similar to the one seen in the $p$-spin model. However, in the quantum spin glass regime, this changes to exponential decay and non compact support. This is robust throughout this region of the spin glass phase. This is the only spin glass regime, that we studied, in which we find no indication of compact support, thus allowing for the emergence of a bulk.

Motivated by the generic appearance of exponential decay, we proved in Section \ref{sec:no_em_low_energy}, that the smearing functions one usually considers in quantum field theory, namely functions that decay for large frequency, are not able to generate operators that can detect whether a nontrivial bulk emerges. This result also holds for smearing function that polynomially \textit{grow} with energy. This might prove as a path forward to study more generally what boundary properties are associated with bulk emergent structure. 

Our work suggests the following open questions:
\begin{itemize}
    \item In the spin glass phase of the SU($M$) Heisenberg model, for $\kappa \in (0,1]$, the spectral function shows an exponential decay, indicative of non-compact support, as we have already described above. This is the only spin glass region, in the two spin glass models we studied, that does not show evidence of compact support, meaning that one could search for a dual theory there. Although this work gives a first indication of where to search, it does not claim the validity of such a statement. It would be quite interesting to pursue an analysis towards that direction.
    \item Another intriguing regime in the phase diagram of the SU($M$) Heisenberg model is the limit of large $\kappa$. Our numerics suggest that the support becomes larger, as well as the region with exponential decay, as $\kappa$ increases. This raises the question whether there is a $\kappa\rightarrow\infty$ limit for which the decay is exponential and the support is non-compact or full. In this sense, the limit of $\kappa\rightarrow\infty$ could resemble the large $q$ limit of SYK, as they could hypothetically share similar behavior regarding the spectral function. At this point, it is not clear what would be the interpretation of the large number of local bosonic excitations, described by large $\kappa$, in a holographic framework and whether such a limit could provide good analytical control. These questions invite further exploration. 
    \item Understanding the dynamics of spin glasses is an intriguing task. In particular, the connection between chaotic dynamics and glassiness is not yet clear. While the global system exhibits slow relaxation in the spin glass phase, the analysis of OTOCs in the $p$-spin model in \cite{Anous2021} revealed an exponential decay, which is related to fast scrambling and chaotic dynamics. 
    It would be interesting to study the chaotic dynamics in the regimes described in the previous two points, especially if there is any connection with a holographic description. 
    There has been a recent interest in using methods from free probability \cite{Camargo:2025zxr, Vardhan:2025rky} in the study of chaotic systems. An application of these ideas to the above models could be a possible direction of future work.
    \item We predominantly found exponential tails in spectral functions, whenever the support becomes non-compact. As argued above, the framework of \cite{Gesteau:2024dhj} does not yet make quantitative statements in this case. Due to their presence in large $q$ SYK, this should still allow for a semiclassical bulk to emerge. Since this appears to be a generic feature of thermal states, an extension of the framework to this case seems necessary.
\end{itemize}

\section*{Acknowledgments}
We thank Elliott Gesteau, Zhuo-Yu Xian, Eric Anschuetz and Alexander Jahn for fruitful discussions and comments on the draft. We thank Ben Karlsberg for his collaboration during the initial stages of this project. We thank Felix Haehl for communication on their implementation of the numerical procedure. We gratefully acknowledge computing time on the HPC clusters at the physics department of FU Berlin. This work has been supported by the Einstein
Research Unit on Quantum Devices and Berlin Quantum.

\appendix

\appendix

\section{Spectral functions in SYK}
In this section we derive the spectral functions of SYK in the large $q$ and large $J$ limit.

\subsection{Large $q$ SYK}\label{app:large_q_syk}
From \cite{Maldacena:2016hyu} we know that in the large $q$ limit
\begin{equation}
    \S(\t) = J^ 2 2{}^{1-q}\text{sgn}(\t) e{}^{g(\t)} 
\end{equation}
with 
\begin{equation}
    \begin{aligned}       
        e{}^{g(\t)} &= \Big[\frac{\cos\frac{\pi v}{2}}{\cos(\pi v
    (\frac{1}{2}-\frac{|\t|}{\b}))} \Big]^2,\\
                \b \cJ &= \frac{\pi v}{\cos \frac{\pi v}{2}},
    \end{aligned}
\end{equation}
where the second equation fixes $v$ which ranges form zero to one as $\b
\mathcal{J}$ goes from zero to infinity. Thus, for every finite value $\b \cJ$
we have $v<1$.
With the defining equation $\S(\t) = J^2 G(\t){}^{q-1}$ we therefore
approximate the thermal correlator as
\begin{equation}
    G(\t) = \frac{e{}^{g(\t)/(q-1)}}{2}.
\end{equation}
To obtain the spectral function, we take the Fourier transform 
\begin{equation}
    G^>(t) = G(it+\e).
\end{equation}
Therefore
\begin{equation}
    G^>(t) = \frac{1}{2}\Big[\frac{\cos\frac{\pi v}{2}}{\cosh(\pi v
    (\frac{1}{2}-\frac{it+\e}{\b}))} \Big]{}^{\frac{2}{q-1}}.
\end{equation}
We now compute its Fourier transform
\begin{equation}
    \hat{G}^>(\w) = \int dt e{}^{-i\w t} G^>(t).
\end{equation}
We shift the argument by 
\begin{equation}
    T = t C_\b - \e C_\b + i C_2,
\end{equation}
where we defined $C_\b = \frac{\pi v}{\b}, C_2 = \frac{\pi v}{2}$.
Here it is important that $v < 1$ so that the argument in $\cosh$ never reaches
$i \pi \mathbb{Z}$ where $\cosh$ vanishes. Thus the integrand is analytic
inside the strip $\mathbb{R} + i [0, \pi v]$, so that we can shift the
contour to be solely along the real $T$ axis. We then have 
\begin{equation}
    \hat{G}^>(\w) = \frac{1}{2}\frac{\b}{\pi v}\cos(\frac{\pi
    v}{2}){}^{2/(q-1)}e{}^{-\frac{\b \w}{2}} \int_{-\infty}^{\infty} dT
    \frac{e{}^{- i\frac{\w}{C_\b}T}}{(\cosh^2(T))^\a},
\end{equation}
with $\a = \frac{1}{q-1}$ and we let $\e \rightarrow 0$. The integral is given by
\begin{equation}
    \int_{-\infty}^{\infty} dT \frac{e{}^{-i\frac{\w}{C_\b}T}}{(\cosh^2(T))^\a} = 2{}^{2\a-1}
    \frac{\G(\a-\frac{i\w}{2C_\b})\G(\a+\frac{i\w}{2C_\b})}{\G(2\a)}.
\end{equation}
from which we deduce
\begin{equation}
    \r(\w) = (e{}^{\b \w}+1) \frac{\b}{\pi v} \cos(\frac{\p
    v}{\b}){}^{2/(q-1)} 2{}^{\frac{2}{q-1}-2}e{}^{-\frac{\b \w}{2}}
    \frac{\G(\frac{1}{q-1}-\frac{i\w}{2 C_\b})\G(\frac{1}{q-1}+\frac{i\w}{2
    C_\b})}{\G(\frac{2}{q-1})}.
\end{equation}
Now by Stirlings formula
\begin{equation}
    \G(z) \sim \sqrt{2\pi} z{}^{z-\frac{1}{2}} e{}^{-iz}
\end{equation}
In the product $\G(z)\G(\bar{z})$ the oscillatory phase $z^z$ cancels out and we end up with 
\begin{equation}
    |\G(\a+i\w)|^2 \sim 2 \pi |\w|{}^{2\a-1}e{}^{-\pi \w}.
\end{equation}
Therefore we have asymptotically
\begin{equation}
    \r(\w) \sim C |\w|{}^{\frac{2}{q-1}-1}e{}^{-\frac{\b \w}{2}(\frac{1}{v}-1)},
\end{equation}
with some order one constant $C$. We see that for large but finite $q$ we have
exponential decay since $0<v<1$.
\subsection{Large \texorpdfstring{$J$}{J} SYK at finite $q$}\label{app:large_J_SYK}
We want to compute the spectral function for large $J$, i.e. in the conformal
limit as given in \cite{Maldacena:2016hyu}.
The constant $b$ is given by solving
\begin{equation}
    J^2 b^q \pi = (1/2 - \D) \tan \pi \D, \hspace{1cm} \D = \frac{1}{q}.
\end{equation}
The greater Green's function is given by \cite{Maldacena:2016hyu} 
\begin{equation}
    G^>(t) = \braket{\y(t)\y(0)} = b \frac{e{}^{-i \pi
    \D}}{(\frac{\b}{\pi}\sinh(\pi(t-i\e)/\b)){}^{2\D}}.
\end{equation}
Now we want to compute its Fourier transform 
\begin{equation}
    G^>(\w) = \int dt e{}^{-it\w} G^>(t).
\end{equation}
To simplify notation we will drop the overall coefficients and reinsert them in
the end. We thus want to compute the integral 
\begin{equation}
    I(\w) = \int dt \frac{e{}^{-it\w}}{(\sinh(\frac{\pi}{\b}(t-i\e)))^{2\D}}.
\end{equation}
We replace $\sinh(u) = \frac{1}{2} e^u(1-e{}^{-2u})$ and set $u =
\frac{\pi}{\b}(t-i\e)$ so 
that
\begin{equation}
    I(\w) = \frac{ 2^{2\D} \b}{\pi} \int_{-\infty -i\e}^{\infty - i \e} du
    e^{-i\w(\frac{\b}{\pi}u+i\e)} e^{-2\D u} (1-e^{-2u})^{-2\D}.
\end{equation}
Now we want to set $x=e{}^{-2u},\ du = -\frac{dx}{2 x}$ but we need to take
care of the branchcut of $(1-e{}^{-2u})^{-2\D}=e{}^{-2\D \log(1-e{}^{-2u})}$
when crossing $u=0$ to end up with an expression where, if we take $\e
\rightarrow 0$, we do not hit the branchcut. Here we use the principal sheet of
the logarithm with branchcut along the negative real axis. For $\text{Re}(u) >0$, the
function $1-e{}^{-2u}$ will be below the real axis in the interval $(0,1)$,
where the logarithm has no branchcut. For $\text{Re}(u)<0$, $1-e^{-2u}$ is below
the negative real axis, so that as $\e \rightarrow 0$ one hits the branchcut.
For this region, it is thus convenient to use 
\begin{equation}
    \log(1-x) = \log(x-1) - i \pi,
\end{equation}
for which the $r.h.s$ is well defined as $\e \rightarrow 0$ as one then hits
the positive real axis, where the logarithm is well defined.
We then have 
\begin{equation}
    I(\w) = \frac{2^{2\D-1} \b}{\pi} e^{\w \e} \int^{1}_{0}dx x^{\D+i\frac{\w
    \b}{2\pi}-1}(1-x)^{-2\D} +e^{i 2 \pi \D } \int_{1}^{\infty}dx x^{\D+i\frac{\w
    \b}{2\pi}-1}(x-1)^{-2\D}.
\end{equation}
The last integrals can be solved by Mathematica  which returns
\begin{equation}
    \begin{aligned}
    \int_0^1 dx\ x^{\D - 1 + i a} (1 - x)^{-2\D}  
&= \frac{\Gamma(1 - 2\D) \, \Gamma(i a + \D)}{\Gamma(1 + i a - \D)}, 
\quad \text{if } \text{Im}(a) < \text{Re}(\D) < \frac{1}{2},\\
\int_1^{\infty}dx\ x^{\D - 1 + i a} (x - 1)^{-2\D}  
&= \frac{\Gamma(1 - 2\D) \, \Gamma(-i a + \D)}{\Gamma(1 - i a - \D)},\quad
\text{if}\, 2\text{Re}(\D) < 1, \text{Im}(a) < \text{Re}(\D), \text{Im}(a) +
\text{Re}(\D) > 0.
    \end{aligned}
\end{equation}
so that we end with
\begin{equation}
    I(\w) = \frac{2^{2\D-1}\b \G(1-2\D)}{\pi} \Big( \frac{\G(i
    \frac{\w \b}{2\pi}+\D)}{\G(1+i\frac{\w \b}{2\pi}-\D)} + e^{i 2\pi
\D}\frac{\G(\D-i\frac{\w \b}{2\pi})}{\G(1-i\frac{\w \b}{2\pi}-\D)}\Big).
\end{equation}
We can then use Euler's reflection formula 
\begin{equation}
    \G(z) \G(1-z) = \frac{\pi}{\sin(\pi z)}
\end{equation}
to find 
\begin{equation}
    I(\w) = \frac{2^{2\D-1}\b \G(1-2\D)}{\pi^2} \G(\D+i
    \frac{\w \b}{2\pi})\G(\D-i\frac{\w \b}{2\pi}) \Big(\sin\big(\pi(\D-i\frac{\w
        \b}{2\pi})\big) +e^{i2\pi \D} \sin \big(\pi(\D+ i\frac{\w
    \b}{2\pi})\big) \Big).
\end{equation}
We can expand the last term using $\sin(A+B) = \sin(A) \cos(B) + \sin(A)
\cos(B)$ 

\begin{equation}
    \Big(\sin\big(\pi(\D-i\frac{\w
        \b}{2\pi})\big) +e^{i2\pi \D} \sin \big(\pi(\D+ i\frac{\w
    \b}{2\pi})\big) \Big) = e^{\pi \D i} \sin(2\pi\D)e^{-\frac{\b\w }{2}}
\end{equation}
So that in total we have 
\begin{equation}
    I(\w) = \frac{2^{2\D-1}\b \G(1-2\D)}{\pi^2} e^{i\pi \D}\sin(2\pi \D)
    e^{-\frac{\b \w}{2}} \G(\D+i
    \frac{\w \b}{2\pi})\G(\D-i\frac{\w \b}{2\pi}).
\end{equation}
We thus have 
\begin{equation}
    G^>(\w) = \frac{b2^{2\D-1}\pi^{2\D-2}\G(1-2\D)}{\b^{2\D-1}}\sin(2\pi \D)
    e^{-\frac{\b \w}{2}} \G(\D+i
    \frac{\w \b}{2\pi})\G(\D-i\frac{\w \b}{2\pi}).
\end{equation}
Which leads to the spectral function
\begin{equation}\label{eq:rho_SYK}
    \r(\w) = (e^{\b \w}+1)G^>(\w) = \frac{b 2^{2\D}\pi^{2\D-2}\G(1-2\D)}{\b^{2\D-1}}\sin(2\pi \D)
    \cosh(\frac{\b \w}{2}) \G(\D+i
    \frac{\w \b}{2\pi})\G(\D-i\frac{\w \b}{2\pi}).
\end{equation}
For large $\w$ this decays as a polynomial for large $\w$
\begin{equation}
    \r(\w) \sim |\w|^{2\D-1},
\end{equation}
where the exponential growth of $\cosh$ cancels the exponential decay of the $\G$-functions.
\section{Replica analysis of the SU($M$) Heisenberg model}\label{app:analytical_calculations}

In this appendix we give a detailed derivation of the effective action of the bosonic SU(M)-Heisenberg model following the analogous calculation of \cite{Anous2021} for the spherical $p$-spin model. A similar derivation for the fermionic system has been given in \cite{Christos:2021wno} and the resulting Schwinger-Dyson equations have been given in the original paper \cite{Sachdev:1992fk}, which are reproduced here. The derivation follows standard techniques but has not been given in the literature before, especially the exact statement of the equilibrium equations of the spin glass parameter for 1-step RSB $(u,m)$ are necessary for the numerical implementation. Starting with the Hamiltonian of the model and ending up with the computation of the corresponding spectral function. Throughout this appendix, we will use the SU$(M)$ Heisenberg model as a reference. We highly recommend this analysis \cite{Anous2021} for a similar calculation regarding the $p$-spin model. 

\subsection{The effective action}\label{app.SU(M)_effective_action}

As argued in Section \ref{sec:SUM}, after introducing bosonic creation and annihilation operators $b_\a$, the Hamiltonian of the SU($M$) Heisenberg model becomes 
\begin{equation}
	\mathcal{H}=\frac{1}{\sqrt{NM}}\sum_{i<j=1}^{N}J_{ij}(\sum_{\alpha,\beta=1}^{M}b_{\alpha}^{\dagger}(i)b^{\beta}(i)b_{\beta}^{\dagger}(j)b^{\alpha}(j)-\kappa^{2}
    M).
\end{equation}
The partition function at inverse temperature $\beta$ can then be expressed through an path integral as
\begin{equation}
	\begin{split}
	Z[J_{ij}]=\int \prod_{i=1}^{N}\prod_{\alpha=1}^{M}\mathcal{D}&b_{\alpha}(i,\tau)\exp\bigg\{-\int_{0}^{\beta}d\tau\bigg[\sum_{i=1}^{N}\sum_{\alpha=1}^{M}b_{\alpha}^{\dagger}(i,\tau)\partial_{\tau}b^{\alpha}(i,\tau)\\
	&+\frac{1}{\sqrt{NM}}\sum_{i<j=1}^{N}\sum_{\alpha,\beta=1}^{M}J_{ij}S_{\alpha}^{\beta}(i)S_{\beta}^{\alpha}(j)\bigg]\bigg\}.
	\end{split}
\end{equation}
To impose the bosonic number constraint \ref{eq:number_constraint} on each site, we insert it in the form of a delta function expressed as a path integral:
\begin{equation}
	\delta\left(\sum_{\alpha=1}^{M}b_{\alpha}^{\dagger}(i)b^{\alpha}(i)-\kappa M\right)=\int\mathcal{D}\lambda(i,\tau)\exp\left\{-i\int_{0}^{\beta}d\tau\,\lambda(i,\tau)\left[\sum_{\alpha=1}^{M}b_{\alpha}^{\dagger}(i,\tau)b^{\alpha}(i,\tau)-\kappa M\right]\right\}.
\end{equation}
This allows us to rewrite the partition function as:
\begin{equation}
	\begin{aligned}
		Z[J_{ij}]=\int \prod_{i=1}^{N}\prod_{\alpha=1}^{M}\mathcal{D}&b_{\alpha}(i,\tau)\mathcal{D}\lambda(i,\tau)\exp\left\{-\int_{0}^{\beta}d\tau\left[\sum_{i=1}^{N}\sum_{\alpha=1}^{M}b_{\alpha}^{\dagger}(i,\tau)\partial_{\tau}b^{\alpha}(i,\tau)\right.\right.\\
		&\left.\left.+i\lambda(i,\tau)\left(b_{\alpha}^{\dagger}(i,\tau)b^{\alpha}(i,\tau)-\kappa M\right)+\frac{1}{\sqrt{NM}}\sum_{i<j=1}^{N}\sum_{\alpha,\beta=1}^{M}J_{ij}S_{\alpha}^{\beta}(i)S_{\beta}^{\alpha}(j)\right]\right\}.
	\end{aligned}
\end{equation}
From the definition of the partition function we can define the action describing the model as:
\begin{equation}
	\begin{aligned}
		S_{B}&=\int_{0}^{\beta}d\tau\sum_{i=1}^{N}\sum_{\alpha=1}^{M}\left[b_{\alpha}^{\dagger}(i,\tau)\partial_{\tau}b^{\alpha}(i,\tau)+i\lambda(i,\tau)\left(b_{\alpha}^{\dagger}(i,\tau)b^{\alpha}(i,\tau)-\kappa M\right)\right]\\
		S_{J}&=\frac{1}{\sqrt{NM}}\int_{0}^{\beta}d\tau\sum_{i<j=1}^{N}\sum_{\alpha,\beta=1}^{M}J_{ij}S_{\alpha}^{\beta}(i)S_{\beta}^{\alpha}(j),
	\end{aligned}
\end{equation}
so that
\begin{equation}
	Z[J_{ij}]=\int \prod_{i=1}^{N}\prod_{\alpha=1}^{M}\mathcal{D}b_{\alpha}(i,\tau)\mathcal{D}\lambda(i,\tau)\exp\left\{-S_{B}-S_{J}\right\}.
\end{equation}
is satisfied.

\subsection{Introducing replicas}

In order to obtain a disorder independent theory, we need to compute the disorder-averaged partition function. To do so, we introduce $n$ copies of the model, indexed by $r=1,\ldots,n$ and then take the average over the randomly drawn $J_{ij}$, as it is shown in the next steps. Explicitly, we write:
\begin{equation}
	\begin{aligned}
		S_{B}^{r}&=\int_{0}^{\beta}d\tau\sum_{i=1}^{N}\sum_{\alpha=1}^{M}\left[b_{\alpha,r}^{\dagger}(i,\tau)\partial_{\tau}b^{\alpha}_{r}(i,\tau)+i\lambda_{r}(i,\tau)\left(b_{\alpha,r}^{\dagger}(i,\tau)b^{\alpha}_{r}(i,\tau)-\kappa M\right)\right]\\
		S_{J}^{r}&=\frac{1}{\sqrt{NM}}\int_{0}^{\beta}d\tau\sum_{i<j=1}^{N}\sum_{\alpha,\beta=1}^{M}J_{ij}S_{\alpha,r}^{\beta}(i)S_{\beta,r}^{\alpha}(j)\\
		\overline{Z^{n}}&=\int \prod_{i<j=1}^{N}dJ_{ij}\mathbb{P}(J_{ij})\int \prod_{r=1}^{n}\prod_{i=1}^{N}\prod_{\alpha=1}^{M}\mathcal{D}b_{\alpha,r}(i,\tau)\mathcal{D}\lambda_{r}(i,\tau)\exp\left\{-\sum_{r=1}^{n}\left(S_{B}^{r}+S_{J}^{r}\right)\right\}.
	\end{aligned}
\end{equation}
Without loss of generality we assume that the couplings $J_{ij}$ are independently drawn from a Gaussian distribution:
\begin{equation}
	\mathbb{P}(J_{ij})=\frac{1}{J\sqrt{2\pi}}\exp\left(\frac{-J_{ij}^{2}}{2J^{2}}\right),
\end{equation}
and \ref{eq:couplings_mean} conditions are satisfied.
We can perform the integral over the $J_{ij}$'s explicitly. For a single $(i,j)$ pair, we collect all the $J_{ij}$-dependent terms and compute the corresponding integral, $I$, as:
\begin{equation}
	\begin{aligned}
		I&=\frac{1}{J\sqrt{2\pi}}\int dJ_{ij}\exp\left\{\frac{-J_{ij}^{2}}{2J^{2}}+\frac{1}{\sqrt{NM}}\int_{0}^{\beta}d\tau\sum_{\alpha,\beta=1}^{M}\sum_{r=1}^{n}J_{ij}S_{\alpha,r}^{\beta}(i)S_{\beta,r}^{\alpha}(j)\right\}\\
		&=\frac{1}{J\sqrt{2\pi}}\sqrt{\frac{\pi}{1/2J^{2}}}\exp\left\{\frac{J^{2}}{2NM}\int_{0}^{\beta}d\tau\int_{0}^{\beta}d\tau'\,\sum_{r,r'=1}^{n}\sum_{\alpha,\beta=1}^{M}\sum_{\gamma,\delta=1}^{M}\left(S_{\alpha,r}^{\beta}(i,\tau)S_{\beta,r}^{\alpha}(j,\tau)\right)\right.\\
		&\qquad\qquad\qquad\qquad\left.\left(S_{\gamma,r'}^{\delta}(i,\tau')S_{\delta,r'}^{\gamma}(j,\tau')\right)\right\}\\
		&=\exp\left\{\frac{J^{2}}{2NM}\int_{0}^{\beta}d\tau\int_{0}^{\beta}d\tau'\,\sum_{r,r'=1}^{n}\sum_{\alpha,\beta=1}^{M}\sum_{\gamma,\delta=1}^{M}\left[S_{\alpha,r}^{\beta}(i,\tau)S_{\gamma,r'}^{\delta}(i,\tau')\right]\left[S_{\beta,r}^{\alpha}(j,\tau)S_{\delta,r'}^{\gamma}(j,\tau')\right]\right\}.
	\end{aligned}
\end{equation}
where we computed the integral of a Gaussian function, $\int dx\,\exp\left\{-ax^{2}+bx+c\right\}=\sqrt{\frac{\pi}{a}}\exp\left\{\frac{b^{2}}{4a}+c\right\}$, canceled out the prefactors, and rearranged the terms so that the spins that act on the same site are grouped together.

Having computed the integral over the couplings $J_{ij}$, we can write the disorder-averaged partition function as:
\begin{equation}\label{disav}
	\begin{aligned}
		\overline{Z^{n}}&=\int 	\prod_{r=1}^{n}\prod_{i=1}^{N}\prod_{\alpha=1}^{M}\mathcal{D}b_{\alpha,r}(i,\tau)\mathcal{D}\lambda_{r}(i,\tau)\\
		&\exp \Bigg\{-\Bigg( \sum_{r=1}^{n}S_{B}^{r}-\frac{J^{2}}{4NM}\int_{0}^{\beta}d\tau\int_{0}^{\beta}d\tau'\,\sum_{r,r'=1}^{n}\sum_{i,j=1}^{N}\sum_{\alpha,\beta=1}^{M}\sum_{\gamma,\delta=1}^{M}\left[S_{\alpha,r}^{\beta}(i,\tau)S_{\gamma,r'}^{\delta}(i,\tau')\right]\\
		&\left[S_{\beta,r}^{\alpha}(j,\tau)S_{\delta,r'}^{\gamma}(j,\tau')\right]\Bigg) + \cO(N^0)\Bigg\}.
	\end{aligned}
\end{equation}
Note that there is an extra $1/2$ prefactor in the exponential term ($J^2/4NM$) as we changed from summing over $i<j$ to over the full range of both
$i$ and $j$. Here, we do not explicitly write the term
coming from the diagonal contribution $i=j$ in the sum over $(i,j)$, because it is subleading in the large $N$ limit.
We can also write:
\begin{equation}
	\sum_{i,j=1}^{N}S_{\alpha,r}^{\beta}(i,\tau)S_{\gamma,r'}^{\delta}(i,\tau')S_{\beta,r}^{\alpha}(j,\tau)S_{\delta,r'}^{\gamma}(j,\tau')=\left[\sum_{i=1}^{N}S_{\alpha,r}^{\beta}(i,\tau)S_{\gamma,r'}^{\delta}(i,\tau')\right]^{2},
\end{equation}
where the square on the right hand side is the absolute value of the complex numbers satisfying $(S_\a^\b)^\dagger = S^\a_\b$. 

\subsection{Hubbard-Stratonovich transformation}

In the following steps,  we will denote $[Db] = \prod_{\tau, r,\alpha} db^\alpha_{r} db_{\alpha,r}^\dagger(\tau)$. We actually want to perform a complex integral for which one has 
\begin{equation}
    \int [dQ][dQ^*] \exp{(-Q^\dagger M Q +Q^\dagger O + O^\dagger Q)} = \frac{(2\pi)^N}{\det M} \exp{(O^\dagger M^{-1}O)}
\end{equation} 
For that reason we can rewrite \eqref{disav} as 
\begin{equation}
    \overline{Z^n} = \int [Db][D\l] \exp{\big(-\sum_r S^r_B + \frac{J^2 N}{4M} O^\dagger O\big)},
\end{equation}
where 
\begin{equation}
    O^{\beta \delta}_{\alpha \gamma r r'} = \frac{1}{N}\sum_i S^\beta_{\alpha,r}(i,\tau)S^{\delta}_{\gamma,r'}(i,\tau')
\end{equation}
Introducing a Hubbard Stratonovich field $Q$ then gives 
\begin{equation}
    \overline{Z^n} = \int [Db][D\lambda] [DQ]\exp{\Big(-\sum_r S^r_B - \frac{4 M }{J^2 N} Q^\dagger Q + Q^\dagger O + O^\dagger Q\Big)}
\end{equation}
where we used the shorthand notation
\begin{equation}
    Q^\dagger O = \int d\tau d\tau' (Q^\dagger)^{\alpha \delta}_{\beta \gamma,r r'}(\tau,\tau')O^{\alpha \delta}_{\beta \gamma,r r'}(\tau,\tau')
\end{equation}
We rescale $Q$ as $Q \rightarrow \frac{N J^2}{4M}Q$ and obtain:
\begin{equation}
        \overline{Z^n} = \int [Db][D\lambda] [DQ]\exp{\Big(-\sum_r S^r_B - \frac{J^2 N}{4M} Q^\dagger Q + \frac{J^2 N}{4M}Q^\dagger O + \frac{J^2 N}{4M} O^\dagger Q\Big)}
\end{equation}
after we neglected the Jacobian terms from the measure.
Writing everything out one finds
\begin{equation}
	\begin{aligned}
		\overline{Z^{n}}&=\int\prod_{\alpha,\beta,\gamma,\delta}\prod_{rr'}\mathcal{D}Q_{\alpha\gamma,rr'}^{\beta\delta}(\tau,\tau')e^{-S_{HS}}\\
		S_{HS}[Q]&=\frac{J^{2}N}{4M}\int_{0}^{\beta}\int_{0}^{\beta}d\tau d\tau'\sum_{\alpha,\beta,\gamma,\delta}\sum_{r,r'}\left|Q_{\alpha\gamma,rr'}^{\beta\delta}(\tau,\tau')\right|^2-N\log\mathcal{Z}_{f}[Q]\\
		\mathcal{Z}_{f}[Q]&=\int 	\prod_{r=1}^{n}\prod_{\alpha=1}^{M}\mathcal{D}b_{\alpha,r}(\tau)\mathcal{D}\lambda_{r}(i,\tau)\exp\left\{-S_{B}-S_{f}\right\}\\
		S_{B}&=\int_{0}^{\beta}d\tau\sum_{\alpha=1}^{M}\sum_{r}\left[b_{\alpha,r}^{\dagger}(\tau)\partial_{\tau}b_{r}^{\alpha}(\tau)+i\lambda_{r}(\tau)\left(b_{\alpha,r}^{\dagger}(\tau)b_{r}^{\alpha}(\tau)-\kappa M\right)\right]\\
		S_{f}&=-\frac{J^{2}}{2M}\int d\tau d\tau'
        \sum_{\alpha,\beta,\gamma,\delta}\sum_{r,r'}
        Q_{\alpha\gamma,rr'}^{\beta\delta}(\tau,\tau')\left[b_{\alpha,r}^{\dagger}(\tau)b_{r}^{\beta}(\tau)-\kappa\delta_{\alpha}^{\beta}\right]\left[b_{\gamma,r'}^{\dagger}(\tau')b_{r'}^{\delta}(\tau')-\kappa\delta_{\gamma}^{\delta}\right]
	\end{aligned}
\end{equation}

Note that in the large $N$ limit, the actions $S_{B}$ and $S_{f}$ decouple into a sum over the $N$ sites exactly, allowing us to write them as $N\cdot\text{single site action}$. This results to the extra $N$ prefactor of $\log \cZ_f$.

In particular, since the Hubbard-Stratonovich
action $S_{HS}\propto N$, in the large $N$ limit the saddle point approximation
is \textit{exact}, so we just need to find such a saddle point of $S_{HS}/N$.
We assume, following \cite{Christos:2021wno}, that the saddle points in the spin
liquid and spin glass phase do not break the symmetries of the original spin Hamiltonian: in
our case, these are the global $SU(M)$ spin rotation symmetry and the time translation symmetry. We can, therefore, make the ansatz:
\begin{equation}
	Q_{\alpha\gamma,rr'}^{\beta\delta}(\tau,\tau')=\delta_{\alpha}^{\delta}\delta_{\gamma}^{\beta}Q_{rr'}(\tau-\tau'),
\end{equation}\label{eq:Q_symmetries}
where $Q_{rr'}$ is real. Also, since we have reduced the model into a \textit{saddle point problem}, there is no longer a path integral over $Q$ to contend with. Therefore, we can assume that the only components of $Q_{rr'}$, which are truly dependent on $\tau$, are the terms for which $r=r'$.

We can now plug Eq.~\eqref{eq:Q_symmetries} into our expressions for $S_{HS}$ and $S_{f}$. This leads to:
\begin{equation}
	\begin{aligned}
		S_{HS}[Q]&=\frac{J^{2}N}{4M}\int_{0}^{\beta}\int_{0}^{\beta}d\tau d\tau'\sum_{\alpha,\beta,\gamma,\delta}\sum_{r,r'}\left|\delta_{\alpha}^{\delta}\delta_{\gamma}^{\beta}Q_{rr'}(\tau-\tau')\right|^2-N\log\mathcal{Z}_{f}[Q]\\
		&=\frac{J^{2}NM}{4}\int_{0}^{\beta}\int_{0}^{\beta}d\tau d\tau'\sum_{r,r'}\left|Q_{rr'}(\tau-\tau')\right|^{2}-N\log\mathcal{Z}_{f}[Q],\\
		S_{f}&=-\frac{J^{2}}{2M}\int_{0}^{\beta}\int_{0}^{\beta} d\tau d\tau' \sum_{\alpha,\beta,\gamma,\delta}\sum_{r,r'} \delta_{\alpha}^{\delta}\delta_{\gamma}^{\beta}Q_{rr'}(\tau-\tau')\left[b_{\alpha,r}^{\dagger}(\tau)b_{r}^{\beta}(\tau)-\kappa\delta_{\alpha}^{\beta}\right]\left[b_{\gamma,r'}^{\dagger}(\tau')b_{r'}^{\delta}(\tau')-\kappa\delta_{\gamma}^{\delta}\right]\\
		&=-\frac{J^2}{2M}\int_{0}^{\beta}\int_{0}^{\beta}d\tau d\tau'\sum_{\alpha,\beta}\sum_{r,r'}Q_{rr'}(\tau-\tau')\left[b_{\alpha,r}^{\dagger}(\tau)b^{\beta}_{r}(\tau)b_{\beta,r'}^{\dagger}(\tau')b_{r'}^{\alpha}(\tau')-\kappa^{2}M\right]
	\end{aligned}
\end{equation}

\subsection{The $G-\Sigma$ formalism}

The next step in our calculation is to introduce the $G$ and $\Sigma$ variables, similar to the standard SYK formalism. Thus, we set the bosonic propagator:

\begin{equation}\label{eq:bosonic_propagator}
	G_{rr'}(\tau,\tau')\equiv\frac{1}{M}\sum_{\alpha}b_{r}^{\alpha}(\tau)b_{\alpha,r'}^{\dagger}(\tau');
\end{equation}

We insert this in the path integral as a delta-function, represented as an exponential, and then the $\Sigma$ variable appears naturally. Explicitly:
\begin{equation}
		1=\int\prod_{rr'}\mathcal{D}G_{rr'}\mathcal{D}\Sigma_{rr'}\exp\underbrace{\left\{-i\sum_{rr'}\int_{0}^{\beta}\int_{0}^{\beta}d\tau d\tau' \Sigma_{rr'}(\tau,\tau')\left(MG_{r'r}(\tau',\tau)-\sum_{\alpha}b_{r'}^{\alpha}(\tau')b_{\alpha,r}^{\dagger}(\tau)\right)\right\}}_{\equiv -S_{\delta}}
\end{equation}

Doing this allows us to freely replace bosonic bilinears with the $G$ variable. We therefore have:
\begin{equation}
	\begin{aligned}
		\mathcal{Z}_{f}[Q]&=\int 	\prod_{r,r'=1}^{n}\prod_{\alpha=1}^{M}\mathcal{D}b_{\alpha,r}(\tau)\mathcal{D}\lambda_{r}(i,\tau)\mathcal{D}G_{rr'}\mathcal{D}\Sigma_{rr'}\exp\left\{-S_{B}-S_{f}-S_{\delta}\right\}\\
		S_{f}+S_{\delta}&=-\sum_{r,r'}\int_{0}^{\beta}\int_{0}^{\beta}d\tau d\tau'\Bigg\{\frac{J^{2}}{2M}Q_{rr'}(\tau-\tau')\left[M^{2}G_{rr'}(\tau,\tau')G_{r'r}(\tau',\tau)-\kappa^{2}M\right]\\&\hspace{3cm}-i\Sigma_{rr'}(\tau,\tau')\left(MG_{r'r}(\tau',\tau)-\sum_{\alpha}b_{r'}^{\alpha}(\tau')b_{\alpha,r}^{\dagger}(\tau)\right)\Bigg\}\\
		S_{B} &=\int_{0}^{\beta}d\tau\sum_{\alpha=1}^{M}\sum_{r}\left[b_{\alpha,r}^{\dagger}(\tau)\partial_{\tau}b_{r}^{\alpha}(\tau)+i\lambda_{r}(\tau)\left(b_{\alpha,r}^{\dagger}(\tau)b_{r}^{\alpha}(\tau)-\kappa M\right)\right]
	\end{aligned}
\end{equation}
We now group all terms, which still have bosonic variables, together and consider the path integral over them separately:
\begin{equation}
	\begin{aligned}
		\mathcal{Z}_{b}&=\int\prod_{r}\prod_{\alpha}\mathcal{D}b_{\alpha,r}(\tau)\exp\left\{-\sum_{r,r'}\sum_{\alpha}\int_{0}^{\beta}\int_{0}^{\beta}d\tau d\tau'b_{\alpha,r}^{\dagger}(\tau)\partial_{\tau}b_{r}^{\alpha}(\tau)+i\lambda_{r}(\tau)b_{\alpha,r}^{\dagger}(\tau)b_{r}^{\alpha}(\tau)\right.\\&\left.\hspace{3cm}-ib_{\alpha,r}^{\dagger}(\tau)\Sigma_{rr'}(\tau,\tau')b_{r'}^{\alpha}(\tau') \right\}\\
        &=\int\prod_{r}\prod_{\alpha}\mathcal{D}b_{\alpha,r}(\tau)\exp\left\{-\sum_{r,r'}\sum_{\alpha}\int_{0}^{\beta}\int_{0}^{\beta}d\tau d\tau'b_{\alpha,r}^{\dagger}(\tau)\delta_{rr'}\delta(\tau-\tau')\partial_{\tau'}b_{r'}^{\alpha}(\tau')\right.\\&\hspace{3cm}\left.+i\lambda_{r}(\tau)\delta_{rr'}\delta(\tau-\tau')b_{\alpha,r}^{\dagger}(\tau)b_{r'}^{\alpha}(\tau')-ib_{\alpha,r}^{\dagger}(\tau)\Sigma_{rr'}(\tau,\tau')b_{r'}^{\alpha}(\tau') \right\}\\
		&={\det}^{-M}\left[\frac{1}{2\pi}\big(\delta_{rr'}\delta(\tau-\tau')\partial_{\tau'}+i\lambda_{r}(\tau)\d(\t-\t')\delta_{rr'}-i\Sigma_{rr'}(\tau,\tau')\big)\right]\\
        &=\exp(-M\Tr\log \big[ \frac{1}{2\pi}\big(\delta_{rr'}\delta(\tau-\tau')\partial_{\tau'}+i\lambda_{r}(\tau)\d(\t-\t')\delta_{rr'}-i\Sigma_{rr'}(\tau,\tau')\big)\big]),
	\end{aligned}
\end{equation}
where we used
\begin{equation}
    \int_{-\infty}^{\infty}d^{n}xd^{n}x^*\exp\left(-\mathbf{x}^\dagger\cdot\mathbf{A}\cdot\mathbf{x}\right)=\left(\det\frac{\mathbf{A}}{2\pi}\right)^{-1}.
\end{equation}

\noindent Combining the previous expressions, we finally obtain the effective action:

\begin{align}\label{eq:eff_action1}
    \frac{S_{\text{eff}}}{N} = &\frac{J^2M}{4}\int_{0}^{\b}d\t d\t'
    \sum_{r,r'} |Q_{r r'}(\t-\t')|^2 \\
            &+M\Tr\log\frac{1}{2\pi}\big(\delta_{rr'}\delta(\tau-\tau')\partial_{\tau'}+i\lambda_{r}(\tau)\d(\t-\t')\delta_{rr'}-i\Sigma_{rr'}(\tau,\tau')\big)\\
            &+iM\sum_{r,r'}\int_{0}^\b  d \t d\t'\S_{r r'}(\t,\t')G_{r' r}(\t'-\t) -i \k M\sum_{r}\int_{0}^\b d\t \l_r(\t)\\
            &-\frac{J^2M}{2}\int_{0}^{\b} d\t d\t'
            \sum_{r,r'}Q_{r,r'}(\t-\t')G_{r r'}(\t-\t')G_{r'r}(\t'-\t)\\
            &+\frac{J^2 \k^2}{2}\int_{0}^{\b}d \t d \t' \sum_{r,r'}Q_{r
            r'}(\t-\t'),
\end{align}
which leads to the averaged replica partition function via
\begin{equation}
    \overline{Z^n}=\int_{}\cD\l \cD Q \cD G \cD \S  e^{-S_{\text{eff}}}.
\end{equation}

\subsection{The saddle-point approximation}
The next step is to eliminate the matrices $Q_{r r'}(t-t')$ and $\Sigma_{rr'}(\tau,\tau')$ from the replica effective action \ref{eq:eff_action1} above, via the saddle-point approximation, which is exact for $N\rightarrow\infty$, as we also mentioned above. First,  we vary the effective action \ref{eq:eff_action1} over $Q_{r r'}(\t-\t')$ and obtain
\begin{equation}\label{eq:variationQ}
\delta S_\mathrm{eff}\Big|_Q = 0 ~\Rightarrow~ Q_{r r'}(\t-\t') =  G_{r r'}(\t-\t')G_{r'r}(\t'-\t) - \frac{\k^2}{M}
\end{equation}
Next, we vary the effective action \ref{eq:eff_action1} over $\Sigma_{rr'}(\tau,\tau')$. Let us define ${K_{rr'}(\t,\t') = \delta_{rr'}\delta(\tau-\tau')(\partial_{\tau'}+i\lambda_{r}(\tau))}$

Then:
\begin{equation}
\begin{aligned}
   \delta S_\mathrm{eff}\Big|_\Sigma = 0 ~\Rightarrow~ (K-i\S)^{-1}_{r
    r'}(\t,\t') &= G_{r,r'}(\t,\t')\\
    \d_{r,r''}\d(\t-\t'') &= \sum_{r'} \int d\t' (K-i\S)_{r,r'}(\t,\t')
    G_{r',r''}(\t',\t'').
\end{aligned}
\end{equation}

Alternatively, this can be written in the following form
\begin{equation}
    i \sum_{r'}\int d\t' \S_{r r'}(\t,\t') G_{r'r''}(\t',\t'') = -
    \d_{r r''}\d(\t-\t'') + \int d\t' \sum_{r'} \d_{r
    r'}\d(\t-\t')(\partial_{\t'}+i\l_{r}(\t))G_{r',r''}(\t',\t''),
\end{equation}

The variation with respect to $G_{rr'}(\tau-\tau')$ then gives
\begin{equation}
    \begin{aligned}
        \delta S_\mathrm{eff}\Big|_G = 0 ~\Rightarrow~ i\S_{r r'}(\t,\t') &=
        \frac{J^2}{2}G_{r,r'}(\t,\t')(Q_{r,r'}(\t,\t')+Q_{r',r}(\t',\t))\\
        &= J^2G_{r,r'}(\t,\t')Q_{r,r'}(\t,\t')\\
    \end{aligned}
\end{equation}
where we used the symmetry of $Q$ under the simultaneous exchange $r \leftrightarrow r', \t \leftrightarrow \t'$.

\noindent Now, we can write the replicated effective action as a function of the bosonic Green's functions and $\lambda_r(\tau)$ only:

\begin{equation}
\begin{aligned}
\frac{S_{\text{eff}}}{N} = &\frac{J^2M}{4}\int_{0}^{\b}d\t d\t'
    \sum_{r,r'} \big|G_{r r'}(\t-\t')G_{r'r}(\t'-\t) - \frac{\k^2}{M}\big|^2 \\
    &-M\Tr\log \big[G_{r r'}(\t,\t')\big]\\
    &+M\sum_{r,r'}\int_{0}^\b  d \t d\t'\big(
    \delta_{rr'}\delta(\tau-\tau')(\partial_{\tau'}+i\lambda_{r}(\tau))G_{r'
r}(\t'-\t) \big)-M \b n -i \k M\sum_{r}\int_{0}^\b d\t \l_r(\t)\\
    &-\frac{J^2M}{2}\int_{0}^{\b} d\t d\t'
            \sum_{r,r'}\big(G_{r r'}(\t-\t')G_{r'r}(\t'-\t) - \frac{\k^2}{M}\big)G_{r r'}(\t-\t')G_{r'r}(\t'-\t)\\
            &+\frac{J^2 \k^2}{2}\int_{0}^{\b}d \t d \t' \sum_{r,r'}\big(G_{r r'}(\t-\t')G_{r'r}(\t'-\t) - \frac{\k^2}{M}\big).
\end{aligned}
\end{equation}
which reduces to
\begin{equation}\label{eq:eff_action2}
    \begin{aligned}
        \frac{S_{\text{eff}}}{N}=&-M\Tr\log\left(G_{rr'}(\t,\t')\right)\\
        &-\frac{J^{2}M}{4}\int_{0}^{\b}d\t d\t' \sum_{rr'}\left[G_{rr'}(\t,\t')G_{r'r}(\t',\t)-\frac{\k^{2}}{M}\right]^{2}\\
        &+M\int_{0}^{\b}d\t
        d\t'\sum_{rr'}\d_{rr'}\d(\t-\t')\partial_{\t'}G_{r'r}(\t',\t)\\
        &-M\b n+i M\int_{0}^{\b}d\t\sum_{r}(G_{rr}(\t,\t)-\k)\l_{r}(\t)
    \end{aligned}
\end{equation}

\subsection{Schwinger-Dyson equations}
We can derive the Schwinger-Dyson equations by varying the effective action \ref{eq:eff_action2} with respect to the bosonic fields, $G_{rr'}$: 
\begin{equation}
\begin{aligned}
     \delta S_\mathrm{eff}\Big|_G = 0 ~\Rightarrow~  -M(\mathbf{G})^{-1}_{rr'}(\t-\t') & + \big[ M\d_{rr'}\d(\t-\t')(\partial_\t +i\lambda_r(\t)) \big]\\
    &-J^2MG_{rr'}(\t-\t')G^2_{r'r}(\t'-\t) + \frac{J^2\k^2}{2} G_{r'r}(\t'-\t)=0
\end{aligned}
\end{equation}
Multiplying the above equation with the matrix $\textbf{G}$ we find the equations of motion for $G$:

\begin{equation}
\begin{aligned}
     \frac{M}{2}\d_{r' r}\d(\t'-\t) = & + \sum_{r''}\int d\t'' M G_{r' r''}(\t'-\t'')K_{r
    r''}(\t,\t'') \\
    &-\frac{1}{2}\sum_{r''}\int d\t'' J^2M G_{r' r''}(\t'-\t'')G_{r r''}(\t-\t'')G^2_{r''
    r}(\t''-\t)\\
    &+\frac{1}{2} \sum_{r''}\int d\t'' J^2 \k^2 G_{r' r''}(\t'-\t'')G_{r'' r}(\t''-\t),
\end{aligned}
\end{equation}
with
\begin{equation}
K_{r r'}(\t,\t') = \d_{r r'}\d(\t-\t')\partial_{\t'}+i\l_r(\t)\d(\t-\t')\d_{rr'}
\end{equation}

\subsection{The replica symmetry breaking ansatz (RSB-1)}\label{app:rsb_ansatz}

In \cite{Bray_1980}, the authors argued that the matrix elements $Q_{rr'}(\t-\t')$ in \ref{eq:Q_symmetries} are time-dependent only for $r=r'$, meaning that all off-diagonal replica terms are time-independent.  For the purposes of our analysis, we will make the same assumption for the elements of $G_{rr'}(\t-\t')$. In particular we will assume that the SG solutions of the models we study, both the $p$-spin and SU$(M)$ Heisenberg models, satisfy the replica symmetry breaking ansatz at one-step (RSB-1). This means that there is an additional block-diagonal structure that the $G_{rr'}(\t-\t')$ matrices should satisfy, as we show in \eqref{eq:Gmatrix}. The RSB-1 has been already used for the $p$-spin model in \cite{Crisanti1992,Cugliandolo2001,Anous2021}, as well as for the SU$(M)$ Heisenberg model in \cite{Georges_2001}. An example of such an matrix ansatz for block-diagonal matrices of dimension $3\times 3$ is:

\begin{equation}\label{eq:Gmatrix}
\textbf{G}(\t-\t')~=~\left(
\begin{array}{cccccccc}
~g(\tau,\tau')&~u&~u && & &  &\\
~u&~g(\tau,\tau')&~u& & &s & &\cdots \\
~u&~u&~g(\tau,\tau')& & & & & \\
 & & & & ~g(\tau,\tau')&~u&~u &  \\
 & s & & & ~u&~g(\tau,\tau')&~u&  \\
 & & & & ~u&~u&~g(\tau,\tau')& \\
&\vdots & & & &  & & ~\ddots
\end{array}\right)
\end{equation}

\noindent In general, we will assume block diagonal matrices of dimension $m\times m$. We can rewrite the parametrization of $G_{rr'}(\t-\t')$ as follows:
\begin{equation}
    G_{rr'}(\t-\t') = \delta_{r_{1}r'_{1}}[\delta_{r_{0}r'_{0}}g(\t,\t')+u(1-\d_{r_{0}r'_{0}})]+s(1-\d_{r_{1}r'_{1}}),
\end{equation}
where $r=r_{0}r_{1}$, $r_{0}=1,\ldots,m$ keeps track of the element in each $m\times m$ block, and $r_{1}=1,\ldots,n/m$ keeps track of which block we consider.

Next, we insert the ansatz above into the effective action \ref{eq:eff_action2}, assuming  that also $\l_{r}(\t)$ is independent of the replica index:

\begin{equation}\label{eq:eff_action3}
    \begin{aligned}
        \frac{S_{\text{eff}}}{N}=&-M\Bigg\{n\left(1-\frac{1}{m}\right)\log[g(\t,\t')-u]\\
        &+\left(\frac{n}{m}-1\right)\log[g(\t,\t')-u+m(u-s)]+\log[g(\t,\t')-u+m(u-s)+ns]\Bigg\}\\
        &-\frac{J^{2}Mn}{4}\int_{0}^{\b}d\t d\t'\Bigg[g(\t,\t')^{2}g(\t',\t)^{2}+(m-1)u^{4}+(n-m)s^{4}\\
        &-2\frac{\k^{2}}{M}\left(g(\t,\t')g(\t',\t)+(m-1)u^{2}+(n-m)s^{2}\right)+n\left(\frac{\k^{2}}{M}\right)^{2}\Bigg]\\
        &+Mn\int_{0}^{\b}d\t d\t'\d(\t-\t')\partial_{\t}g(\t,\t')\\
        &-M\b n+iMn\int_{0}^{\b}d\t  (g(\t,\t)-\k)\l(\t).
\end{aligned}
\end{equation}

\noindent We can now compute the equation of motion (EoM) for $s$, the parameter that determines the structure of the ansatz outside the block-matrices. Once again, we need to vary \ref{eq:eff_action3} over the corresponding parameter. Then, we have:

\begin{equation}
    \begin{aligned}
    \delta S_\mathrm{eff}\Big|_s = 0 ~\Rightarrow~  &\frac{\beta Mn(n-m) s}{[g(\t,\t')-u+m(u-s)][g(\t,\t')-u+m(u-s)+ns]} \\
    &-J^{2}M\b^{2}n(n-m)s\left(s^{2}-\frac{\k^{2}}{M}\right)=0
    \end{aligned}
\end{equation}

\noindent which is proportional to $s$, so we can set $s=0$. 

The effective action then simplifies to:

\begin{equation}\label{eq:eff_action4}
    \begin{aligned}
        \frac{S_{\text{eff}}}{N}=&-M\Bigg\{n\left(1-\frac{1}{m}\right)\log[g(\t,\t')-u]+\left(\frac{n}{m}\right)\log[g(\t,\t')+u(m-1)]\Bigg\} \\
        &-\frac{J^{2}Mn}{4}\int_{0}^{\b}d\t d\t'\Bigg[g(\t,\t')^{2}g(\t',\t)^{2}+(m-1)u^{4}\\
        &-2\frac{\k^{2}}{M}\left(g(\t,\t')g(\t',\t)+(m-1)u^{2}\right)+n\left(\frac{\k^{2}}{M}\right)^{2}\Bigg]\\
        &+Mn\int_{0}^{\b}d\t d\t'\d(\t-\t')\partial_{\t}g(\t,\t')\\
        &-M\b n+iMn\int_{0}^{\b}d\t  (g(\t,\t)-\k)\l(\t).
\end{aligned}
\end{equation}

\subsection{Equations of motion in momentum space}
As our final goal is to compute the spectral function of our model in frequency space, the rational next step is to write the effective action and compute its EoMs in momentum space. For the rest of the analysis we will use the following convention for the Fourier transforms: 
\begin{equation}\label{eq:fourier_convention}
\begin{aligned}
    F(\t) = \frac{1}{\beta} \sum_k e^{i\frac{2 \pi k}{\beta}\t} \hat{F}(k),\\
    \hat{F}(k) = \int_0^\beta d\t e^{-i\frac{2 \pi k}{\beta}\t}F(\t).
    \end{aligned}
\end{equation}
Under this convention, the time-dependent diagonal elements of $\mathbf{G}(\t-\t')$ in \ref{eq:Gmatrix}, i.e., the $g(\t,\t')$ terms, can be transformed as:

\begin{equation}\label{eq:g_Fourier}
    g(\t,\t') = \frac{1}{\beta} \sum_{k=-\infty}^{+\infty} e^{i\frac{2 \pi k}{\beta}(\t-\t')} \hat{g}(k).
\end{equation}
Plugging \ref{eq:g_Fourier} in the effective action \ref{eq:eff_action4}, we can finally get its form in momentum space, as:
\begin{equation}
\begin{aligned}
    \frac{S_{\text{eff}}}{NMn} = &- \Bigg[\sum_{k\neq 0} \log(\frac{\hat{g}(k)}{\beta}) +\bigg(1-\frac{1}{m}\bigg) \log(\frac{\hat{g}(0)}{\beta}-u) + \frac{1}{m}\log (\frac{\hat{g}(0)}{\beta}+u(m-1))\Bigg]\\
    &-\frac{J^2}{4}\Bigg[\frac{1}{\beta^2}\sum_{k_1,\hdots, k_4} \hat{g}(k_1)\hat{g}(k_2)\hat{g}(k_3)\hat{g}(k_4)\delta(k_1 +k_2 -k_3-k_4) \\
    &+ \beta^2 \bigg((m-1)u^4-2\frac{\kappa^2}{M}(m-1)u^2 + n \frac{\kappa^4}{M^2}\bigg) - 2 \frac{\kappa^2}{M}\sum_k \hat{g}(k)^2 \Bigg]\\ &+\sum_k \frac{2\pi i k}{\beta}\hat{g}(k) - \beta + i\hat{\lambda}(0) \bigg(\frac{1}{\beta} \sum_k \hat{g}(k) - \kappa\bigg).
    \end{aligned}
\end{equation}
Before we compute the EoMs for the different parameters of the model, we need to consider the large $M$ limit. Having written the effective action in the form of $\frac{S_{\text{eff}}}{NMn}$, it is easy to impose the large $M$ limit, by discarding the terms of order $\mathcal{O}(1/M^p)$. Then we can write the effective action at the large $M$ limit as: 

\begin{equation}
\begin{aligned}
    \frac{S_{\text{eff}}}{NMn} = &- \Bigg[\sum_{k\neq 0} \log(\frac{\hat{g}(k)}{\beta}) +\bigg(1-\frac{1}{m}\bigg) \log(\frac{\hat{g}(0)}{\beta}-u) + \frac{1}{m}\log (\frac{\hat{g}(0)}{\beta}+u(m-1))\Bigg]\\
    &-\frac{J^2}{4}\Bigg[\frac{1}{\beta^2}\sum_{k_1,\hdots, k_4} \hat{g}(k_1)\hat{g}(k_2)\hat{g}(k_3)\hat{g}(k_4)\delta(k_1 +k_2 -k_3-k_4) + \beta^2 \Big((m-1)u^4\Big) \Bigg]\\ &+\sum_k \frac{2\pi i k}{\beta}\hat{g}(k) - \beta + i\hat{\lambda}(0) \bigg(\frac{1}{\beta} \sum_k \hat{g}(k) - \kappa\bigg).
    \end{aligned}
\end{equation}
Note that in the computations above we have separated the $k\neq 0$ and $k=0$ terms in the first sum. Now we can calculate the EoMs for $g(k\neq0)$, $g(0)$, $m$ and $u$ by varying the effective action in the momentum space over these variables. Thus, the EoMs for $\hat{g}_k$ for $k\neq0$ are:

\begin{equation}\label{eq:eom_gk_0}
   \delta S_\mathrm{eff}\Big|_{g(k)} = 0 ~\Rightarrow~ \frac{1}{\hat{g}(k)} = i \frac{2\pi k}{\beta}+ i \frac{\hat{\lambda}(0)}{\beta}-\hat{\Sigma}(k),
\end{equation}
where:

\begin{equation}
    \hat{\Sigma}(k) = \frac{J^2}{\b^2}\sum_{k_1,\dots, k_4} \hat{g}(k_1)\hat{g}(k_2)\hat{g}(k_3)\hat{g}(k_4)\delta(k_1 +k_2 -k_3-k_4).
\end{equation}
Here, $\hat{\Sigma}(k)$ is the Fourier transform of the self-energy, which we define in real time as
\begin{equation}
    \Sigma(\t) = J^2 g(\t)^2g(-\t).
\end{equation}
Next, the EoM for $u$ is
\begin{equation}\label{eq:eom_u}
    \delta S_\mathrm{eff}\Big|_u = 0 ~\Rightarrow~ (m-1)\left(J^{2}\b^{2}u^{3}-\frac{u}{\bigg(\dfrac{\hat{g}(0)}{\beta}-u\bigg)\bigg(\dfrac{\hat{g}(0)}{\beta}+(m-1)u\bigg)}\right)=0.
\end{equation}
And the EoM for $m$ is:
\begin{equation}\label{eq:eom_m}
   \delta S_\mathrm{eff}\Big|_m = 0 ~\Rightarrow~ \frac{J^{2}\b^{2}}{4}u^{4}+\frac{1}{m}\frac{u}{\cfrac{\hat{g}(0)}{\b}+(m-1)u}+\frac{1}{m^{2}}\log\left(\frac{\cfrac{\hat{g}(0)}{\b}-u}{\cfrac{\hat{g}(0)}{\b}+(m-1)u}\right)=0.
\end{equation}
Finally, the EoM for $\hat{g}(0)$ is 
\begin{equation}\label{eq:eom_g0}
    \delta S_\mathrm{eff}\Big|_{g(0)} = 0 ~\Rightarrow~ \frac{1}{\beta}\frac{\cfrac{\hat{g}(0)}{\beta}+(m-2)u}{\bigg(\cfrac{\hat{g}(0)}{\beta}-u\bigg)\bigg(\cfrac{\hat{g}(0)}{\beta}+(m-1)u\bigg)} + \hat{\Sigma}(0) - i \frac{\hat{\l}(0)}{\beta}=0.
\end{equation}
Every point on the phase diagram of the SU$(M)$ Heisenberg model satisfies all EoMs \ref{eq:eom_gk}, \ref{eq:eom_u}, \ref{eq:eom_m} and \ref{eq:eom_g0} simultaneously. For the spin liquid phase we expect $u=0$ and $m=1$ for all $J$ and $\kappa$. On the other hand, the spin glass phase is characterized by $0<u\leq\kappa$ and $0<m<1$. The bound for $u$ arises from the constraint Eq.~\eqref{eq:number_constraint} and for integer $n$, $m$ is also an integer that describes the size of the diagonal blocks in the ansatz. In the limit $n \rightarrow 0$, $m$ becomes a continuous number between $0$ and $1$ \cite{Anous2021, Cugliandolo2001}. In order to be able to find the correct solutions $\hat{g}(k)$ numerically for any point on the phase diagram, we need first to massage the EoMs into a more convenient form.

In particular, we will follow some steps shown in \cite{Anous2021} for the $p$-spin model that were motivated by \cite{Cugliandolo2001}. 
We use \eqref{eq:eom_u} to infer
\begin{equation}
    \begin{aligned}
        J^2 \beta^2u^2 &= \frac{1}{\bigg(\cfrac{\hat{g}(0)}{\beta}-u\bigg)\bigg(\cfrac{\hat{g}(0)}{\beta}+(m-1)u\bigg)}\\
        \bigg(\frac{\hat{g}(0)}{\beta}+(m-1)u\bigg) &=  \frac{1}{J^2\beta^2u^2\bigg(\cfrac{\hat{g}(0)}{\beta}-u\ \bigg)}
    \end{aligned}
\end{equation}
This allows us to rewrite \eqref{eq:eom_g0} as
\begin{equation}
    \begin{aligned}
    i\frac{\hat{\l}((0)}{\beta} &= \hat{\Sigma}(0) + J^2\beta u^2\bigg(\frac{\hat{g}(0)}{\beta}+(m-2)u \bigg)\\
       &= \hat{\Sigma}(0) + J^2\beta u^2\bigg(\frac{\hat{g}(0)}{\beta}+(m-1)u - u \bigg)\\
       &= \hat{\Sigma}(0) + J^2\beta u^2\left( \frac{1}{J^2\beta^2u^2\bigg(\cfrac{\hat{g}(0)}{\beta}-u\ \bigg)} - u \right)\\
       &= \hat{\Sigma}(0) + \left( \frac{1}{\beta\bigg(\cfrac{\hat{g}(0)}{\beta}-u\ \bigg)} - J^2\beta u^3 \right).
    \end{aligned}
\end{equation}
Now we redefine both the self-energy and $g(0)$ variables, such that: 
\begin{equation}
\begin{aligned}
    \hat{\Sigma}_r(0) &= \hat{\Sigma}(0) - J^2 \beta u^3,\\
    \hat{g}_r(0) &= \hat{g}(0) - \beta u.
\end{aligned}
\end{equation}
This leads to 
\begin{equation}
    i\frac{\hat{\l}(0)}{\beta} = \hat{\Sigma}_r(0) + \frac{1}{\hat{g}_r(0)}.
\end{equation}
Finally, the equations of motion for $k \neq 0$ become
\begin{equation}\label{eq:eom_gk}
    \frac{1}{\hat{g}_r(k)} = \frac{1}{\hat{g}_r(0)}+\frac{2 \pi i k}{\beta} - \Big(\hat{\Sigma}_r(k)-\hat{\Sigma}_r(0)\Big).
\end{equation}

As we will see in the section discussing the numerical algorithm below, it is the equation above that we will iterate adiabatically for every point on the phase diagram, till we converge to the correct $\hat{g}_r(k)$. 

\subsection{Lorentzian two point function}

Given that the final goal is to calculate the spectral function that describes the model under specific $T, J, \kappa$ parameter values, we need to write the EoM for the Euclidean $g(k)$ in frequencies $\omega$, which are conjugate to the real time $t$. 

First, for a Euclidean time-ordered correlator $g(\t)$, we define the real-time correlators from analytic continuation as:

\begin{equation}
\begin{aligned}
        g^>(t) &= g(\t = it + \e)\\
        g^<(t) &= g(\t = -it + \e).
\end{aligned}
\end{equation}

\noindent The retarded propagator is defined as:

\begin{equation}
g^R(t) = -i \Q(t) \Big[g^>(t)-g^<(t)\Big].
\end{equation}

If one considers the Fourier coefficients of the Euclidean correlator as analytic functions of frequency, one has
\begin{equation}
    \hat{g}(\w) = \int_{-\infty}^{\infty}dt e^{-i\w t}g(t)
\end{equation}

Additionally, the retarded correlator, $g^R(\omega)$, and the Euclidean correlators, $g_E$, satisfy:
\begin{equation}\label{eq:retarded_euclidean}
    \hat{g}^R(\w) = - g_E(-iw-0^+)
\end{equation}
The relation above is quite useful, as it will let us compute $g^R(\omega)$, from the values $\{u, m, g(k=0)\}$ that satisfy the triplet of Euclidean equations of motion \ref{eq:eom_u}, \ref{eq:eom_m} and \ref{eq:eom_g0}, derived above. In particular, the analytic continuation of \eqref{eq:eom_gk} is the following: 

\begin{equation}
    \frac{1}{\hat{g}^R(\w)} = -\w -  i\frac{\l^R(0)}{\beta} - \hat{\Sigma}^R(\w)
\end{equation}
As explained in \ref{app:analytic_continuation}, the real-time self-energy is:
\begin{equation}
    \Sigma^R(t) = -i J^2 \Theta(t) [g^>(t)^2g^<(-t)-g^<(t)^2g^>(-t)]
\end{equation}
The Fourier transform of it will give $\Sigma^R(\omega)$.
Based on relation \eqref{eq:retarded_euclidean}, we can read $\Sigma^R(\omega=0)$ from the Euclidean EoM for $g(0)$ \eqref{eq:eom_g0}, as:
\begin{equation}
    \frac{1}{\beta}\frac{-\cfrac{\hat{g}(0)}{\beta}+(m-2)u}{\bigg(-\cfrac{\hat{g}(0)}{\beta}-u\bigg)\bigg(-\cfrac{\hat{g}(0)}{\beta}+(m-1)u\bigg)} - \hat{\Sigma}^R
    (0) + i \frac{\hat{\l}(0)}{\beta}=0
\end{equation}

Finally, we can express $g^R_r(\omega) \equiv g^R_r(\omega) - u$ as the analytic continuation of \eqref{eq:eom_gk} as follows:

\begin{equation}
    \frac{1}{\hat{g}_r(\omega)} =\frac{1}{\hat{g}_r(0)} - \w - \Big(\hat{\Sigma}_r(\omega)-\hat{\Sigma}_r(0)\Big) 
\end{equation}

where we know already the values of $g^R_r(0)$ and $\Sigma^R(0)$ from the corresponding Euclidean parameters.

\section{Details on the spectral function}\label{app:spectral_function}
In this appendix we derive standard facts about the relation between imaginary time Green's function and their analytic continuation to real time for both Fermions and Bosons as needed for our simulations.
We use the following conventions for Fourier transforms in real and imaginary times
\begin{equation}\label{eq:fourier_transform_convention}
    \begin{aligned}  
        \hat{F}(\w) &= \int dt e^{-i\w t} F(t),\ 
        F(t) = \int \frac{d\w}{2\pi} e^{i\w t}\hat{F}(\w),\\
        \hat{F}(k) &= \int_0^\b d\t e^{-i \w_k \t } F(\t),\        F(\t) = \frac{1}{\b} \sum_k e^{i\w_k \t} \hat{F}(k).
    \end{aligned}
\end{equation}
Define the bosonic/fermionic spectral function as 
\begin{equation}
    \r^{\pm}(\w) = \frac{2\pi}{Z} \sum_m |b_{m+\w,m}|^2 e^{-\b e_m}(1 \pm e^{-\b \w}
    ),
\end{equation}
where $ b_{m n} = \braket{m|b|n}$ are the respective matrix elements in the
energy eigenbasis and the $-$(+) sign is the bosonic (fermionic) definition. 
The following representation is useful for later purposes

\begin{equation}\label{eq:rho_delta_def}
\r^\pm(\w) = \frac{2\pi}{Z} \sum_{m,n} |b_{nm}|^2 (e^{-\b e_m}\pm e^{-\b e_n})
\d(e_n-e_m-\w).
\end{equation}
One can directly see that for Fermions 
\begin{equation}
    \r \geq 0
\end{equation}
and for Bosons
\begin{equation}
    \w \rho(\w) \geq 0.
\end{equation}
The spectral function can be related to the Fourier transforms of the following functions 
\begin{equation}\label{eq:real_time_2pt}
    g^>(t) = \braket{ b(t) b^\dagger(0)},\\
    g^<(t) = \braket{ b^\dagger(0) b(t)},\\
    g^R(t) = -i \Q(t) \braket{[b(t),b^\dagger(0)]_\pm}.
\end{equation}
We find the following relations between their Fourier components 
\begin{equation}\label{eq:lorentzian_correlators}
    \begin{aligned}
        g^<(\w) &= e^{\b \w} q^>(\w),\\
        g^>(\w) &= n^\pm_B(\w) \r^\pm(\w),\\
        n_B^\pm(\w) &= \frac{1}{e^{\b \w}\pm 1},
    \end{aligned}
\end{equation}
where the $\pm$-sign in the last line defines the
Fermi-Dirac/Bose-Einstein distribution.

Next, we want to discuss the relation between these real-time correlators and the euclidean two-point function.
The time-ordered thermal Green's function is given by 
\begin{equation}
    g(\t) = \Q(\t) \braket{ b(\t) b^\dagger} \pm \Q(-\t) \braket{b^\dagger
    b(\t)}.
\end{equation}
For Hermitian $b$ this defines a (anti-) symmetric function around $\t = 0$ for (fermions) bosons.
We can take its Fourier transform and find 
\begin{equation}\label{eq:gE_w_rho}
    \hat{g}(k) = \mp \int \frac{dw'}{2\pi} \frac{\rho^\pm(\w')}{\w'-i\w_k} = \mp \G(\w_k),
\end{equation}
where we defined the analytic function
\begin{equation}
    \G(z) = \int \frac{d\w}{2\pi}\frac{\rho(\w')}{\w'-iz}.
\end{equation}
Explicitly, the integral is given by 
\begin{equation}
    \begin{aligned}
        \int \frac{d\w'}{2\pi} \frac{\r(\w')}{\w'-i\w_k } &= \sum_{m n}
        \frac{1}{Z}|b_{mn}|^2 (e^{-\b e_m} \pm e^{-\b e_n})
        \int d\w'\frac{\d(e_n-e_m-\w')}{-i\w_k + \w'} \\
        &= \sum_{m n} |b_{m n}|^2 \frac{e^{-\b e_m} \pm e^{-\b e_n}}{e_n-e_m-i\w_k},
    \end{aligned}
\end{equation}
which can directly be compared with the Fourier transform of $g(\t)$.
Summing up the Fourier coefficients \eqref{eq:gE_w_rho} we obtain the alternative representation
\begin{equation}\label{eq:euclidean_}
    g(\t) = \int \frac{d\w}{2 \pi} \frac{e^{\w \t}}{e^{\b\w}-1}\rho(\w),
\end{equation}
for $0<t<\b$ and periodic extension outside this interval, i.e.
\begin{equation}\label{eq:g_tau_negative}
    g(-\t) = g(\b-\t) = \int \frac{d\w}{2\pi}\frac{e^{-\w \t}}{1-e^{-\b\w}}\r(\w).
\end{equation}
This should be contrasted with the expression found in \cite{Anous2021}, which coincides with Eq.~\eqref{eq:g_tau_negative} but is used for positive $\t$. This is only true for correlators, where the underlying operator is Hermitian. Our expressions are correct in the general case.
Now we compare this to the Fourier transform for the
retarded Green's function 
\begin{equation}\label{eq:retarded_fct}
    g^{\pm,R}(t) = -i \Q(t) \braket{[b(t),b^\dagger(t)]_\pm}.
\end{equation}
for which we find
\begin{equation}\label{eq:analytic continuation_retarded}
    g^{\pm,R}(\w) = \pm \int \frac{d\w}{2\pi}\frac{\rho^\pm(\w')}{\w'-i(-i\w -\e)} = \pm \G(-i\w-\e).
\end{equation}
This in turn implies 
\begin{equation}\label{eq:euclidean_lorentzian_zero_mode}
    \hat{g}(k = 0) = -g^R(\w = 0),
\end{equation}
so that the zero modes between the Euclidean and Lorentzian two-point functions are related.
Using the Sokhotski–Plemelj  formula
\begin{equation}
    \int d\w \frac{f(\w)}{\w+i\e} = -i \pi f(0) + \cP (\int d\w
    \frac{f(\w)}{\w}),
\end{equation}
with $\cP$ the principal value we find
\begin{equation}\label{eq:g_retarded_full}
    g^{\pm,R}(\w) = \frac{\mp}{2\pi}\Big(i \pi \r^{\pm}(\w) + \cP\Big[\int d\w' \frac{\r^{\pm}(\w-\w')}{\w'}\Big] \Big).
\end{equation}

\section{Analytic continuation of equations of motion}\label{app:analytic_continuation}
Here we will describe how to derive the real-time version of the
Schwinger-Dyson equations from the imaginary time equations.
The standard method to obtain real-time Schwinger-Dyson equations is to
introduce a Schwinger-Keldysh contour \cite{fetter2003quantum} along a complex
valued time-parameter. However, the derivation of the effective action from
which the large $N$ Schwinger-Dyson equations are derived will proceed by the
same method as in the euclidean case and one will end up with a function of the
schematic form
\begin{equation}\label{eq:SK_action}
    S{}_{SK} = \int dt dt' F\big(G(t,t'),G(t',t)\big) + \S(t,t')G(t,t') +
    \hdots,
\end{equation}
where the parameter $t,t'$ now goes along the Schwinger-Keldysh contour and $F$
is a function that depends on the specific model of interest. In the contour of
relevance for the thermal two-point functions, the contour will have three
segments 
\begin{equation}
    \t \in [0,+i\infty) \cup (+i\infty,0] \cup (0,\b),
\end{equation}
of which we call the first two the \textit{plus}-and \textit{minus}-branch
respectively. Here we have put the three branches to start at $t= 0$. To include negative time arguments, one should instead let the branches include relevant negative times as well. The definition of the real-time two-point function in terms of contour ordered Green's function is inclusive of these cases.
The correlation functions appearing in the path-integral with effective action
\eqref{eq:SK_action} represent \textit{contour-ordered} correlation function,
i.e., every product of operators in the effective action at different times
should be identified with
\begin{equation}
    \tilde{G}(t,t') = \braket{\cT_C(O(t)O^\dagger(t'))}.
\end{equation}
For example, if $t$ is on the plus-branch and $t'$ is on the minus-branch, this corresponds
to 
\begin{equation}
    \tilde{G}(t_+,t'_-) = \pm \braket{O^\dagger(t')O(t)},
\end{equation}
and analogous for other combinations, the operators are ordered from right to
left, where leftmost operators are \textit{later} on the contour and the prefactor is determined from whether $O$ is bosonic/fermionic. This gives a direct way to define the functions $g^>,g^<$ from the previous appendix via 
\begin{equation}
\begin{aligned}
    g^>(t) &:= \pm \tilde{G}(t_-,0_+),\\
    g^<(t) &:= \pm \tilde{G}(t_+,0_-).
    \end{aligned}
\end{equation}
As we saw in \ref{eq:analytic continuation_retarded}, the retarded Green's function is related to the euclidean Green's function by an analytic continuation of the form $\w \rightarrow - i \w$. We can apply the same procedure to the Schwinger-Dyson equations to obtain the real time equations.
which in turn allows us to find an expression for the self energy $\S(\t)$
\begin{equation}
    \S(\t) = \int \frac{d\w}{2 \pi} \frac{e^{\w \t}\rho^\S(\w)}{e^{\b\w}-1},
\end{equation}
with
\begin{equation}\label{eq:rho_sigma}
    \r^\S(\w) = (e^{\b\w}-1)\int \frac{d^3\w}{(2\pi)^3} \frac{\r(\w_1)\r(\w_2)\r(\w_3)}{(e^{\b\w_1}-1)(e^{\b\w_2}-1)(e^{-\b\w_3}-1)} \d(w-(\w_1+\w_2-\w_3))
\end{equation}
We then find that $\hat{\S}(k)$ is given by \eqref{eq:gE_w_rho} with $\r$ replaced by $\r^\S$.
This defines a function that one can analytically continue to real frequencies. Then the euclidean equation
\begin{equation}
    \frac{1}{\hat{g}(k)} = \text{Poly}(\w_k) + J^2 \hat{\S}(k),
\end{equation}
is replaced by
\begin{equation}
    \frac{1}{\G(z)} = \text{Poly}(z) + J^2 \G^\S(z),
\end{equation}
which is evaluated at real time frequencies to obtain the Lorentzian SDE's. Here $\text{Poly}(z)$ indicates the polynomial that appears in the Schwinger-Dyson equation of the model in question.
To obtain the real-time equations one just replaces every occurrence of $\hat{g}(k)$ with $-\hat{g}^R(\w)$, $\hat{\Sigma}(k)$ with $-\hat{\Sigma}^R(\w)$ and $\w_k$ with $-i\w$, where 
$g^R(t)$ is defined in \eqref{eq:real_time_2pt}. We see that one needs to have access to the self-energy in real time as well, i.e., to objects like 
\begin{equation}
    \S^>(t) = \S(t_-,0_+),
\end{equation}
etc to compute 
\begin{equation}
    \S^R(t) = -i\Q(t) (\S^>(t) \pm \S^<(t) ),
\end{equation}
where the sign stays fermionic/bosonic dependent on the fundamental field.
We obtain $\S^{>}$ and $\S^<$ from $\G(t,t')$ by using the euclidean partition function.
For example if, as in the SU(M)-Heisenberg model, one has in imaginary time
\begin{equation}
    \S(\t,\t') = G(\t,\t')^2G(t',t),
\end{equation}
then in real-time
\begin{equation}
    \S^>(t) = g(t_-,0_+)^2g(0_+,t_-) = g(t_-,0_+)^2g(-t_+,0_-)= g^>(t)^2g^<(-t),
\end{equation}
where in the last equality we used time-translation invariance for the $g^<$ function.
Similarly we have
\begin{equation}
    \S^<(t) = g^<(t)^2 g^>(-t)
\end{equation}
Using the expression for $g^>(t),g^<(t)$ in terms of $\rho(\w)$, one can check explicitly that the Fourier transform of $\S^R(t)$ is the analytic continuation of \eqref{eq:gE_w_rho} with $\rho$ replaced by $\rho^\S$ of Eq.~\eqref{eq:rho_sigma}.
\section{Numerical method}\label{app:numerical_algorithm}

In this section, we describe the numerical methods, we used to compute the
spectral function for any parameter value of the model under study. This is a slightly altered version of the algorithm described in appendix G of
\cite{Anous2021} and adapted from
\cite{Maldacena:2016hyu}.
The main reason we need to rely on numerics is that an analytic solution of the
Schwinger-Dyson equations (SDE) is only attainable in specific limits, e.g. strong
coupling, and we are interested in their spectral functions away from these
limits. In short, we first solve the euclidean SDEs to obtain values for the
spin glass parameters $u,m$ so that they can be used as inputs for the
Lorentzian algorithm that then computes the spectral function $\r(\w)$.
Furthermore the euclidean algorithm gives us access to the euclidean two-point
function $g(\t)$, whose Fourier transform can also be computed from the
spectral function via the analytic function $\G(z)$ evaluated at Matsubara
frequencies in the complex plane as described in Appendix \ref{app:spectral_function}. Therefore, given the spectral function
obtained from the Lorentzian algorithm, we can check for consistency with the
solution found by the euclidean algorithm. In particular we found that
achieving convergence in imaginary time is much easier than in real time, so
that it provides a good check on the reliability of the real time solution. 
\subsection{The general procedure}

In this section, we provide a sketch of the procedure we follow to obtain the corresponding spectral function from numerical data. A review of all functions necessary is given in Appendix \ref{app:spectral_function}. The systems we study in this article are defined by
    Hamiltonians with random all-to-all interactions of the general form 
    \begin{equation}
        H = f(q) \sum_{i_1 \hdots i_q} J_{i_1 \hdots i_q} O^{i_1}\hdots O^{i_q}, 
    \end{equation}
    where $f(q)$ is a function needed to make the Hamiltonian Hermitian,
    depending on the structure of the operator $O$, if it is bosonic or
    fermionic for example. Usually, there are $N$ different $O_i$ operators, where $N$ is the number of sites. This is not accurate for the SY model, where there is additionally a parameter $M$ that describes the number of degrees of freedom per site, in which case we take the limit $1\ll M \ll N$. Due to factorization in this limit, we will still refer to this as the large $N$ limit. To determine the spectral function, we need to determine the Fourier transform of correlators such as
    \begin{equation}\label{eq:g_greater}
        G^{>}(t) = \braket{O(t) O^\dagger(0)}.
    \end{equation}
    We will now describe a general procedure that allows one to obtain correlators similar to \eqref{eq:g_greater}. All of the system we study satisfy large $N$ factorization. The steps are as follows:\\
    
    1) First, one determines the Schwinger-Dyson equations (SDEs) for euclidean two-point functions
    \begin{equation}
        G(\t) = \braket{ \cT_\t O(\t) O^\dagger(0)},
    \end{equation}
    where $\cT_\t$ is the thermal time-ordering. These fully describe the system as $N \rightarrow \infty$.
    These are obtained as configurations that minimize the average free energy
    \begin{equation}
        \bar{F} = - \braket{\log Z},\ Z = \Tr e^{-\b H},
    \end{equation}
    where the bracket $\braket{}$ denotes the disorder average over the
    couplings $J$. After the disorder average, the resulting effective action reduces to a single-site effective action with an overall prefactor $N$. This extensivity in $N$ justifies a saddle point approximation. The Saddle-point approximation results in an equation for $G(\t)$ that is of the form
    \begin{equation}
        G(\w_k) = \text{Poly}(\w_k) - J^2 \S(\w_k),
    \end{equation}
    where $\text{Poly}(\w)$ is a low-degree complex polynomial in the frequency. The full derivation of these equations for the SY model is given in Appendix \ref{app.SU(M)_effective_action}. Analogous derivations can be found in \cite{Bray_1980, Anous2021, Trunin:2020vwy, Christos:2021wno}. We solve these equations numerically as described in Appendix \ref{app:numerical_algorithm} to have a reference against which we can check our Lorentzian solutions as well as to obtain the replica symmetry breaking parameters $u,m$ in the spin glass phase.\\
    
    2) After obtaining the Euclidean solutions, we need to analytically continue them to obtain the Lorentzian versions. We demonstrate in appendix \ref{app:spectral_function}, that $G(\w_k)$ admits
    an analytic continuation from imaginary Matsubara to real frequencies, see
    e.g. \cite{fetter2003quantum} for a textbook treatment. This allows us to
    determine the Schwinger-Dyson equations for the retarded Green's function
    \begin{equation}
        G^R(t) = -i \Q(t) \braket{[O(t),O^\dagger(0)]_\pm}, 
    \end{equation}
    which we solve numerically as described in \ref{app:numerical_algorithm} and extract the spectral function using 
    \begin{equation}
        \r(\w) = \mp 2 \text{Im} G^R(\w).
    \end{equation}
    Here $\pm$ stands for Fermions and Bosons respectively.
\subsection{Finding initial guesses and moving adiabatically}
In the algorithm below we want to iteratively determine a solution to the SDEs,
for which one needs an initial guess. The procedure we followed is that in the
euclidean case, we start out in a parameter regime in which the system is in
the spin liquid phase, where no replica-symmetry breaking occurs and
convergence is usually achieved easier. There, for the first set of external
parameters we start with an ansatz for the Fourier coefficients
$\hat{g}{}^{(0)}(k)$ that is
given by the free solution, the solution of the SDE at vanishing coupling. Then
we iterate as described below and obtain a solution
$\hat{g}{}^{(\text{n})}(k)$ for
the current set of parameters, where $n$ labels the number of iterations. This solution is then used as an ansatz for the next
set of parameter values. We found that for the euclidean equations, convergence
is not strongly dependent on the initial guess, also if one starts in the
spin glass phase. However, this is not true for the real-time equations. We
therefore, as suggested in a footnote in \cite{Anous2021}, start for a set
of parameter values where convergence is easily achieved for imaginary time and
then slowly change the parameters until reaching the final set of parameter
values. This procedure breaks down if crossing a discontinuous
phase-transition, so care has to be taken when choosing the path along which
one moves in parameter-space.

\subsection{The Euclidean algorithm}\label{app:euclidean_algo}
The goal of the algorithm is to generate a numerical solution to the large $N$
Schwinger-Dyson equation in terms of Fourier coefficients $\hat{g}(k)$ at bosonic - or
fermionic Matsubara frequencies. We start with an initial guess as
described in the previous section, which determines $\hat{g}{}^{(0)}(k)$, similarly one starts with
initial guesses for $u$ and $m$. 

\begin{itemize}
    
\item[1)] Compute $g{}^{(n)}(\t)$ by Fourier transformation and from it the self energy
    $\S{}^{(n)}(\t)$ according to the model under study.

\item[(2)] Compute the Fourier coefficients $\hat{\S}{}^{(n)}(k)$
\item[(3)] Update the Fourier coefficients $\hat{g}{}^{(n+1)}(k)$ for $k \neq 0$ by
    setting
    \begin{equation}
        \hat{g}_r^{(n+1)}(k) = (1-x) \hat{g}_r^{(n)}(k) +
        \frac{x}{\text{SDE}(\w_k,\hat{g}_r^{(n)}(0),\hat{g}_r^{(n)}(k),\hat{\S}{}^{(n)}(k))}
    \end{equation}
    where the numerator is the r.h.s. of the SDE in the form
    \begin{equation}
        \hat{g}(k) = \frac{1}{\text{SDE}}
    \end{equation}
    as usually presented. Here $x$ is a parameter that we initially set to $x = 0.5$ and iteratively adapt as described in $(5)$. The relevant equations are Eq.~\eqref{eq:SYK_E_SDE} for the SYK model, Eq.~\eqref{eq:p_spin_E_SDE} for the $p$-spin model and Eq.~\eqref{eq:SY_E_SDE} for the SU($M$) Heisenberg model.
\item[(4)] If the model is bosonic, determine the zero mode $\hat{g}^{(n+1)}(0)$ from
    the normalization condition as explained in \ref{app:normalization_euclidean}.
\item[(5)] Compute the norm 
\begin{equation}
    \D{}^{(n+1)} :=\sum_k (\hat{g}{}^{(n+1)}(k)-\hat{g}{}^{(n)}(k))^2
\end{equation}
if $\D{}^{(n+1)} > \D{}^{(n)}$ divide $x$ by $2$.
\item[(6)] Solve the spin glass equations for $u,m$ with $\hat{g}{}^{(n+1)}(0)$. In this step, in the $p$-spin model usually several solutions for $u$ apply. For the case we study ($p=3$) this equation has three solutions, one with negative $m$ that is excluded by $m>0$. In \cite{Cugliandolo2001} it is argued that, given the other two solutions, one should use the larger one (the right branch) that corresponds to a physical value. In the SU(M) Heisenberg model in the equilibrium state, the solution is unique.
\item[(7)] With the updated value of $u$, determine the new zero mode $\hat{g}_r(0)$ from the normalization condition as explained in \ref{app:normalization_euclidean}.
\item[(8)] Repeat from (1).
\end{itemize}

\subsubsection{Euclidean zero mode update}\label{app:normalization_euclidean}
As described above, for the bosonic systems we study, the SDE's explicitly depend on the zero mode $\hat{g}(0)$ that is not fixed by the equilibrium equations. The approach advocated in \cite{Anous2021} is to determine the zero mode by enforcing that the constraints defining the models are satisfied. For the $p$-spin model, the spherical constraint Eq.~\eqref{eq:spherical_constraint} enforces the following constraint on the Fourier modes
\begin{equation}
    \sum_k \hat{g}_r(k) = \beta(1-u),
\end{equation}
so that, after obtaining a new set of Fourier modes $\hat{g}(k)^{(n+1)}$ for $k \neq 0$, we subtract their sum from the r.h.s. with the value of $u$ used in the iteration step (3), to determine a new zero mode $\hat{g}(k)$. For the SU(M) Heisenberg model, the fundamental operator is not Hermitian, so that one has to take into account the commutation relations of the creation operator $f_\a$. This non-hermiticity is encoded in disagreeing right- and left-limits of the time ordered correlator
\begin{equation}
    \begin{aligned}
        g(0^+) &= \frac{1}{M}\sum_\a b^\a b_\a^{\dagger} = \frac{1}{M}(\k M + M) = \k +1  \\
        g(0^-) &= \k,
    \end{aligned}
\end{equation}
which is based on the form of the bosonic propagator \eqref{eq:bosonic_propagator}. Numerically, we only have access to a finite number of Fourier coefficients. Additionally, we can not really compute either the right or left-limit but only represent the sum  $ \sum_k \hat{g}(k)$.
Such a sum of Fourier coefficients is known to converge to the average of the right and left limits, so that for the determination of the zero-mode $\hat{g}(0)$, 
we solve 
\begin{equation}
    \sum_k \hat{g}_r(k) = \beta (\k + \frac{1}{2}-u)
\end{equation} 
\subsection{The Lorentzian algorithm}\label{app:lorentzian_algo}
The Lorentzian algorithm aims to compute the spectral function $\r(\w)$ by
iterating the Lorentzian Schwinger-Dyson equations for the retarded two-point
function by an analogous procedure as the Euclidean algorithm. Here it is
assumed that, if one is solving in the spin glass phase that the euclidean
algorithm ran previously to find the values of the spin glass parameters $u,m$.
We start by assuming that an initial guess $\r{}^{(0)}(\w)$ has been prepared
which we henceforth denote by $\r{}^{(n)}(\w)$ since this generalizes to any
finite step of the iteration. In the spin liquid regime convergence is generically fast, so that one can start with a well behaving initial guess, e.g. a differentiable function that satisfies the constraints of the model, e.g. antisymmetric for bosonic Hermitian fields.
\begin{itemize}    
\item[1)] Compute the Lorentzian functions $g{}^{>,(n)}(t),g{}^{<,(n)}(t)$ as a
    Fourier
    transform from \eqref{eq:lorentzian_correlators}. Use them to compute the
    self-energy $\S(t)$ for the model under study.
\item[(2)] Compute the Fourier coefficients $\hat{\S}{}^{R,(n)}(\w)$
\item[(3)] Compute the retarded Fourier coefficients $\hat{g}{}^{R,(n)}(\w)$ from
    \eqref{eq:g_retarded_full}.
\item[(4)] Update the Fourier coefficients $\hat{g}{}^{R,(n+1)}(\w)$ for $\w \neq 0$ by
    setting
    \begin{equation}
        \hat{g}{}^{R,(n+1)}(\w) = (1-x) \hat{g}{}^{R,(n)}(\w) +
        \frac{x}{\text{SDE}(\w,\hat{g}{}^{>,(n)}(0),\hat{g}{}^{<,(n)}(k),\hat{\S}{}^{R,(n)}(k))}
    \end{equation}
    where the numerator is the r.h.s. of the Lorentzian SDE in the form
    \begin{equation}
        \hat{g}^R(k) = \frac{1}{\text{SDE}}
    \end{equation}
    as usually presented. The relevant equations are Eq.~\eqref{eq:SYK_L_SDE} for the SYK model, Eq.~\eqref{eq:p_spin_L_SDE} for the $p$-spin model and Eq.~\eqref{eq:SY_L_SDE} for the SU($M$) Heisenberg model.
\item[(5)] Extract the updated spectral function from the imaginary part of
    $g{}^{R,(n+1)}(\w)$ using \eqref{eq:g_retarded_full}.
\item[(6)] If the model is bosonic, determine the zero mode $g{}^{(n+1)}_0$ from
    the normalization condition implied by the model as described in \ref{app:lorentzian_zero_mode} with Eq.~\eqref{eq:zero_mode_update_pspin} for the spherical $p$-spin model and Eq.~\eqref{eq:zero_mode_update_SY} for the SU($M$) Heisenberg model. 
\item[(7)] Compute the norm
\begin{equation}
    \D{}^{(n+1)} :=\sum_k (\hat{g}{}^{(n+1)}(k)-\hat{g}{}^{(n)}(k))^2
\end{equation}
if $\D{}^{(n+1)} > \D{}^{(n)}$ divide $x$ by $2$. 
\item[(8)] Repeat from (1).
\end{itemize}
\subsubsection{Lorentzian zero mode update}\label{app:lorentzian_zero_mode}
Similar to the Euclidean algorithm, one needs the zero mode $\hat{g}^R(0)$ to update the remaining Fourier coefficients. We found that the following procedure leads to good results whose zero modes computed using \eqref{eq:g_retarded_full} with the converged spectral function does indeed agree with the zero mode obtained from the Euclidean algorithm, while maintaining correct normalization of the spectral function.
The correct spectral function is normalized such that in the $p$-spin model
\begin{equation}
    \int \frac{d\w}{2\pi} \frac{\rho_r(\w)}{e^{\b\w}-1} = 1-u.
\end{equation}
The left hand side is equal to $\sum_k \hat{g}_r(k)$ by \eqref{eq:analytic continuation_retarded}, so that we have the constraint
\begin{equation}
    1-u = \int_{-\infty}^{\infty} \frac{d\w}{2\pi } \frac{\r_r(\w)}{e^{\b
    \w}-1} = \frac{1}{\b} \Big( \hat{g}_r(0) + \sum_{k \neq 0} \hat{g}_r(k) \Big).
\end{equation}
To obtain an expression for the zero mode, we want an integral expression for
\begin{equation}
 I = \sum_{k \neq 0} \hat{g}_r(k)
\end{equation}
to subtract it from $\b(1-u)$ to obtain a value for the zero mode. We want to determine $I$ purely from the spectral function, to use it in the Lorentzian algorithm.
We can rewrite it as \cite{fetter2003quantum}
\begin{equation}
    I = \b \int_{\cC}\frac{dz}{2\pi i} \frac{ q_r(z)}{e^{\b z}-1},
\end{equation}
where the contour $\cC$ goes anticlockwise around the positive imaginary an
negative imaginary axis and 
the numerator $e^{\b z}-1$ has poles at $z =\frac{ 2\pi k}{\b}$ for integers $k$.
Deforming the contour to $\cC'$ which goes clockwise around the real axis, and plugging in the analytic continuation of $q_r(k)$ we have
\begin{equation}
I = \b \int_\mathbb{R} \frac{d\w}{2\pi} \r_r(\w) \int_{\cC'} \frac{dz}{2\pi i}\frac{1}{(e^{\b z}-1)(z+\w-i\e)}
\end{equation}
Evaluating the residues at $z = 0, z = -\w+i\e$ we obtain 
\begin{equation}
    \begin{aligned}
        I &= \b \int \frac{d\w}{2\pi} \r_r(\w) (-1) \big(\frac{1}{\b(\w-i\e)} +
    \frac{1}{e^{-\b \w}-1}\big)\\
          &= - \int \frac{d\w}{2\pi}\frac{ \r_r(\w)}{\w -i \e} + \b \int
          \frac{d\w}{2\pi} \frac{\r_r(\w)}{1-e^{-\b \w}} 
    \end{aligned}
\end{equation}
Now using the Sokhotski–Plemelj formula on the first term gives, with $\r_r(0) = 0$,
\begin{equation}
    I = - \frac{1}{2\pi} \cP\big[ \int d\w \frac{\r_r(\w)}{\w} \big]+ \b \int
          \frac{d\w}{2\pi} \frac{\r_r(\w)}{1-e^{-\b \w}}
\end{equation}
The normalization condition thus gives 
\begin{equation}
    \b(1-u) - I = \hat{g}_r(0) 
\end{equation}
so that 
\begin{equation}
    \hat{g}_r(0) = \b(1-u) + \frac{1}{2\pi} \cP\big[ \int d\w \frac{\r_r(\w)}{\w}
    \big] - \b \int
          \frac{d\w}{2\pi} \frac{\r_r(\w)}{1-e^{-\b \w}}.
\end{equation}
In the case of correctly normalized $\r$ the first and third term cancel, so that one has
\begin{equation}
    \hat{g}_r(0) = \frac{1}{2\pi} \cP\big[ \int d\w \frac{\r_r(\w)}{\w}
    \big].
\end{equation}
For the $p$-spin model, we thus update the zero mode using 
\begin{equation}\label{eq:zero_mode_update_pspin}
    g_r^{R,(n+1)}(0) = g_r^{R,(n)}(0) - \b(1-u) + \b \int
          \frac{d\w}{2\pi} \frac{\r_r(\w)}{e^{\b \w}-1},
\end{equation}
where we used \eqref{eq:euclidean_lorentzian_zero_mode} to relate the Euclidean zero mode to the Lorentzian zero mode.
For the SU($M$) Heisenberg model, the analogous logic leads to 
\begin{equation} \label{eq:zero_mode_update_SY}
    g_r^{R,(n+1)}(0) = g_r^{R,(n)}(0) - \b(\k-u) + \b \int
          \frac{d\w}{2\pi} \frac{\r_r(\w)}{1-e^{-\b \w}}.
\end{equation}
Here the denominator in the last term is flipped relative to \eqref{eq:zero_mode_update_pspin}, because $\r(\w)$ is not antisymmetric and the expression above comes from the \textit{left-limit} $g_r(0^-)$ in Euclidean time.

\bibliographystyle{jhep}
\bibliography{references}

@article{Hartnoll:2009sz,
    author = "Hartnoll, Sean A.",
    editor = "Uranga, A. M.",
    title = "{Lectures on holographic methods for condensed matter physics}",
    eprint = "0903.3246",
    archivePrefix = "arXiv",
    primaryClass = "hep-th",
    doi = "10.1088/0264-9381/26/22/224002",
    journal = "Class. Quant. Grav.",
    volume = "26",
    pages = "224002",
    year = "2009"
}

@article{Herzog:2009xv,
    author = "Herzog, Christopher P.",
    title = "{Lectures on Holographic Superfluidity and Superconductivity}",
    eprint = "0904.1975",
    archivePrefix = "arXiv",
    primaryClass = "hep-th",
    reportNumber = "PUPT-2297",
    doi = "10.1088/1751-8113/42/34/343001",
    journal = "J. Phys. A",
    volume = "42",
    pages = "343001",
    year = "2009"
}

@article{Sachdev:2010uj,
    author = "Sachdev, Subir",
    editor = "van Beijeren, Henk and M{\'e}zard, Marc",
    title = "{Strange metals and the AdS/CFT correspondence}",
    eprint = "1010.0682",
    archivePrefix = "arXiv",
    primaryClass = "cond-mat.str-el",
    doi = "10.1088/1742-5468/2010/11/P11022",
    journal = "J. Stat. Mech.",
    volume = "1011",
    pages = "P11022",
    year = "2010"
}

@article{Faulkner:2010tq,
    author = "Faulkner, Thomas and Polchinski, Joseph",
    title = "{Semi-Holographic Fermi Liquids}",
    eprint = "1001.5049",
    archivePrefix = "arXiv",
    primaryClass = "hep-th",
    reportNumber = "NSF-KITP-10-010",
    doi = "10.1007/JHEP06(2011)012",
    journal = "JHEP",
    volume = "06",
    pages = "012",
    year = "2011"
}

@article{Faulkner:2009wj,
    author = "Faulkner, Thomas and Liu, Hong and McGreevy, John and Vegh, David",
    title = "{Emergent quantum criticality, Fermi surfaces, and AdS(2)}",
    eprint = "0907.2694",
    archivePrefix = "arXiv",
    primaryClass = "hep-th",
    reportNumber = "MIT-CTP-4050",
    doi = "10.1103/PhysRevD.83.125002",
    journal = "Phys. Rev. D",
    volume = "83",
    pages = "125002",
    year = "2011"
}

@article{Sachdev:2011wg,
    author = "Sachdev, Subir",
    title = "{What can gauge-gravity duality teach us about condensed matter physics?}",
    eprint = "1108.1197",
    archivePrefix = "arXiv",
    primaryClass = "cond-mat.str-el",
    doi = "10.1146/annurev-conmatphys-020911-125141",
    journal = "Ann. Rev. Condensed Matter Phys.",
    volume = "3",
    pages = "9--33",
    year = "2012"
}

@article{Ryu:2006bv,
    author = "Ryu, Shinsei and Takayanagi, Tadashi",
    title = "{Holographic derivation of entanglement entropy from AdS/CFT}",
    eprint = "hep-th/0603001",
    archivePrefix = "arXiv",
    reportNumber = "NSF-KITP-06-11, NSF-KITP-06-11",
    doi = "10.1103/PhysRevLett.96.181602",
    journal = "Phys. Rev. Lett.",
    volume = "96",
    pages = "181602",
    year = "2006"
}

@article{Sekino:2008he,
    author = "Sekino, Yasuhiro and Susskind, Leonard",
    title = "{Fast Scramblers}",
    eprint = "0808.2096",
    archivePrefix = "arXiv",
    primaryClass = "hep-th",
    reportNumber = "SU-ITP-08-18, OIQP-08-08, SU-ITP-08/18, OIQP-08-08",
    doi = "10.1088/1126-6708/2008/10/065",
    journal = "JHEP",
    volume = "10",
    pages = "065",
    year = "2008"
}

@article{Shenker:2013pqa,
    author = "Shenker, Stephen H. and Stanford, Douglas",
    title = "{Black holes and the butterfly effect}",
    eprint = "1306.0622",
    archivePrefix = "arXiv",
    primaryClass = "hep-th",
    reportNumber = "SU-ITP-13-08",
    doi = "10.1007/JHEP03(2014)067",
    journal = "JHEP",
    volume = "03",
    pages = "067",
    year = "2014"
}

@article{Maldacena:2013xja,
    author = "Maldacena, Juan and Susskind, Leonard",
    title = "{Cool horizons for entangled black holes}",
    eprint = "1306.0533",
    archivePrefix = "arXiv",
    primaryClass = "hep-th",
    doi = "10.1002/prop.201300020",
    journal = "Fortsch. Phys.",
    volume = "61",
    pages = "781--811",
    year = "2013"
}

@article{Maldacena:2015waa,
    author = "Maldacena, Juan and Shenker, Stephen H. and Stanford, Douglas",
    title = "{A bound on chaos}",
    eprint = "1503.01409",
    archivePrefix = "arXiv",
    primaryClass = "hep-th",
    doi = "10.1007/JHEP08(2016)106",
    journal = "JHEP",
    volume = "08",
    pages = "106",
    year = "2016"
}

@article{Susskind:2014rva,
    author = "Susskind, Leonard",
    title = "{Computational Complexity and Black Hole Horizons}",
    eprint = "1403.5695",
    archivePrefix = "arXiv",
    primaryClass = "hep-th",
    doi = "10.1002/prop.201500092",
    journal = "Fortsch. Phys.",
    volume = "64",
    pages = "24--43",
    year = "2016",
    note = "[Addendum: Fortsch.Phys. 64, 44--48 (2016)]"
}

@article{Pastawski:2015qua,
    author = "Pastawski, Fernando and Yoshida, Beni and Harlow, Daniel and Preskill, John",
    title = "{Holographic quantum error-correcting codes: Toy models for the bulk/boundary correspondence}",
    eprint = "1503.06237",
    archivePrefix = "arXiv",
    primaryClass = "hep-th",
    doi = "10.1007/JHEP06(2015)149",
    journal = "JHEP",
    volume = "06",
    pages = "149",
    year = "2015"
}

@inbook{Hartnoll:2011fn,
    author = "Hartnoll, Sean A.",
    editor = "Horowitz, Gary T.",
    title = "{Horizons, holography and condensed matter}",
    booktitle = "{Black holes in higher dimensions}",
    eprint = "1106.4324",
    archivePrefix = "arXiv",
    primaryClass = "hep-th",
    pages = "387--419",
    year = "2012",
    publisher ="Cambridge University Press "
}

@article{Witten:2021jzq,
    author = "Witten, Edward",
    title = "{Why does quantum field theory in curved spacetime make sense? And what happens to the algebra of observables in the thermodynamic limit?}",
    eprint = "2112.11614",
    archivePrefix = "arXiv",
    primaryClass = "hep-th",
    doi = "10.1007/978-3-031-17523-7\_11",
    year = "2022"
}

@article{Witten:2018zxz,
    author = "Witten, Edward",
    title = "{APS Medal for Exceptional Achievement in Research: Invited article on entanglement properties of quantum field theory}",
    eprint = "1803.04993",
    archivePrefix = "arXiv",
    primaryClass = "hep-th",
    doi = "10.1103/RevModPhys.90.045003",
    journal = "Rev. Mod. Phys.",
    volume = "90",
    number = "4",
    pages = "045003",
    year = "2018"
}

@article{Witten:2021unn,
    author = "Witten, Edward",
    title = "{Gravity and the crossed product}",
    eprint = "2112.12828",
    archivePrefix = "arXiv",
    primaryClass = "hep-th",
    doi = "10.1007/JHEP10(2022)008",
    journal = "JHEP",
    volume = "10",
    pages = "008",
    year = "2022"
}

@article{Duetsch:2002hc,
    author = "Duetsch, Michael and Rehren, Karl-Henning",
    title = "{Generalized free fields and the AdS - CFT correspondence}",
    eprint = "math-ph/0209035",
    archivePrefix = "arXiv",
    doi = "10.1007/s00023-003-0141-9",
    journal = "Annales Henri Poincare",
    volume = "4",
    pages = "613--635",
    year = "2003"
}

@book{Friedli_Velenik_2017, place={Cambridge}, title={Statistical Mechanics of Lattice Systems: A Concrete Mathematical Introduction}, publisher={Cambridge University Press}, author={Friedli, Sacha and Velenik, Yvan}, year={2017}}

@article{Maldacena:1997re,
    author = "Maldacena, Juan Martin",
    title = "{The Large $N$ limit of superconformal field theories and supergravity}",
    eprint = "hep-th/9711200",
    archivePrefix = "arXiv",
    reportNumber = "HUTP-97-A097, HUTP-98-A097",
    doi = "10.4310/ATMP.1998.v2.n2.a1",
    journal = "Adv. Theor. Math. Phys.",
    volume = "2",
    pages = "231--252",
    year = "1998"
}

@article{Witten:1998qj,
    author = "Witten, Edward",
    title = "{Anti de Sitter space and holography}",
    eprint = "hep-th/9802150",
    archivePrefix = "arXiv",
    reportNumber = "IASSNS-HEP-98-15",
    doi = "10.4310/ATMP.1998.v2.n2.a2",
    journal = "Adv. Theor. Math. Phys.",
    volume = "2",
    pages = "253--291",
    year = "1998"
}

@article{Susskind:1994vu,
    author = "Susskind, Leonard",
    title = "{The World as a hologram}",
    eprint = "hep-th/9409089",
    archivePrefix = "arXiv",
    reportNumber = "SU-ITP-94-33",
    doi = "10.1063/1.531249",
    journal = "J. Math. Phys.",
    volume = "36",
    pages = "6377--6396",
    year = "1995"
}

@article{tHooft:1993dmi,
    author = "'t Hooft, Gerard",
    title = "{Dimensional reduction in quantum gravity}",
    eprint = "gr-qc/9310026",
    archivePrefix = "arXiv",
    reportNumber = "THU-93-26",
    journal = "Conf. Proc. C",
    volume = "930308",
    pages = "284--296",
    year = "1993"
}

@article{Anninos2011,
    author = "Anninos, Dionysios and Anous, Tarek and Barandes, Jacob and Denef, Frederik and Gaasbeek, Bram",
    title = "{Hot Halos and Galactic Glasses}",
    eprint = "1108.5821",
    archivePrefix = "arXiv",
    primaryClass = "hep-th",
    doi = "10.1007/JHEP01(2012)003",
    journal = "JHEP",
    volume = "01",
    pages = "003",
    year = "2012"
}

@article{Edwards1975,
    author = "Edwards, S. F. and Anderson, P. W.",
    title = "{Theory of spin glasses}",
    doi = "10.1088/0305-4608/5/5/017",
    journal = "J. Phys. F",
    volume = "5",
    number = "5",
    pages = "965",
    year = "1975"
}

@article{Anninos2015,
    author = "Anninos, Dionysios and Anous, Tarek and Denef, Frederik and Peeters, Lucas",
    title = "{Holographic Vitrification}",
    eprint = "1309.0146",
    archivePrefix = "arXiv",
    primaryClass = "hep-th",
    doi = "10.1007/JHEP04(2015)027",
    journal = "JHEP",
    volume = "04",
    pages = "027",
    year = "2015"
}

@article{Anous2021,
    author = "Anous, Tarek and Haehl, Felix M.",
    title = "{The quantum p-spin glass model: a user manual for holographers}",
    eprint = "2106.03838",
    archivePrefix = "arXiv",
    primaryClass = "hep-th",
    doi = "10.1088/1742-5468/ac2cb9",
    journal = "J. Stat. Mech.",
    volume = "2111",
    pages = "113101",
    year = "2021"
}

@article{Maldacena1999,
    author = "Maldacena, Juan Martin and Michelson, Jeremy and Strominger, Andrew",
    title = "{Anti-de Sitter fragmentation}",
    eprint = "hep-th/9812073",
    archivePrefix = "arXiv",
    reportNumber = "HUTP-98-A088, UCSBTH-98-8",
    doi = "10.1088/1126-6708/1999/02/011",
    journal = "JHEP",
    volume = "02",
    pages = "011",
    year = "1999"
}

@article{Brown:2017jil,
    author = "Brown, Adam R. and Susskind, Leonard",
    title = "{Second law of quantum complexity}",
    eprint = "1701.01107",
    archivePrefix = "arXiv",
    primaryClass = "hep-th",
    doi = "10.1103/PhysRevD.97.086015",
    journal = "Phys. Rev. D",
    volume = "97",
    number = "8",
    pages = "086015",
    year = "2018"
}

@article{Cotler:2017jue,
    author = "Cotler, Jordan and Hunter-Jones, Nicholas and Liu, Junyu and Yoshida, Beni",
    title = "{Chaos, Complexity, and Random Matrices}",
    eprint = "1706.05400",
    archivePrefix = "arXiv",
    primaryClass = "hep-th",
    doi = "10.1007/JHEP11(2017)048",
    journal = "JHEP",
    volume = "11",
    pages = "048",
    year = "2017"
}

@article{Song:2017pfw,
    author = "Song, Xue-Yang and Jian, Chao-Ming and Balents, Leon",
    title = "{Strongly Correlated Metal Built from Sachdev-Ye-Kitaev Models}",
    eprint = "1705.00117",
    archivePrefix = "arXiv",
    primaryClass = "cond-mat.str-el",
    doi = "10.1103/PhysRevLett.119.216601",
    journal = "Phys. Rev. Lett.",
    volume = "119",
    number = "21",
    pages = "216601",
    year = "2017"
}

@article{Saad:2018bqo,
    author = "Saad, Phil and Shenker, Stephen H. and Stanford, Douglas",
    title = "{A semiclassical ramp in SYK and in gravity}",
    eprint = "1806.06840",
    archivePrefix = "arXiv",
    primaryClass = "hep-th",
    month = "6",
    year = "2018"
}

@article{Kachru:2009xf,
    author = "Kachru, Shamit and Karch, Andreas and Yaida, Sho",
    title = "{Holographic Lattices, Dimers, and Glasses}",
    eprint = "0909.2639",
    archivePrefix = "arXiv",
    primaryClass = "hep-th",
    reportNumber = "NSF-KITP-09-173, SU-ITP-09-42",
    doi = "10.1103/PhysRevD.81.026007",
    journal = "Phys. Rev. D",
    volume = "81",
    pages = "026007",
    year = "2010"
}

@inproceedings{Denef:2011ee,
    author = "Denef, Frederik",
    title = "{TASI lectures on complex structures}",
    booktitle = "{Theoretical Advanced Study Institute in Elementary Particle Physics}: {String theory and its Applications: From meV to the Planck Scale}",
    eprint = "1104.0254",
    archivePrefix = "arXiv",
    primaryClass = "hep-th",
    doi = "10.1142/9789814350525_0007",
    pages = "407--512",
    month = "4",
    year = "2011"
}

@article{Parcollet:1999itf,
    author = "Parcollet, Olivier and Georges, Antoine",
    title = "{Non-Fermi-liquid regime of a doped Mott insulator}",
    doi = "10.1103/PhysRevB.59.5341",
    journal = "Phys. Rev. B",
    volume = "59",
    number = "8",
    pages = "5341",
    year = "1999"
}

@article{Mulder1981,
  title = {Susceptibility of the $\mathrm{Cu}\mathrm{Mn}$ spin-glass: Frequency and field dependences},
  author = {Mulder, C. A. M. and van Duyneveldt, A. J. and Mydosh, J. A.},
  journal = {Phys. Rev. B},
  volume = {23},
  issue = {3},
  pages = {1384--1396},
  numpages = {0},
  year = {1981},
  month = {Feb},
  publisher = {American Physical Society},
  doi = {10.1103/PhysRevB.23.1384},
  url = {https://link.aps.org/doi/10.1103/PhysRevB.23.1384}
}

@article{Jafferis:2015del,
    author = "Jafferis, Daniel L. and Lewkowycz, Aitor and Maldacena, Juan and Suh, S. Josephine",
    title = "{Relative entropy equals bulk relative entropy}",
    eprint = "1512.06431",
    archivePrefix = "arXiv",
    primaryClass = "hep-th",
    reportNumber = "NSF-KITP-15-162, NSF-KITP-15-162",
    doi = "10.1007/JHEP06(2016)004",
    journal = "JHEP",
    volume = "06",
    pages = "004",
    year = "2016"
}

@article{Furuya:2023fei,
    author = "Furuya, Keiichiro and Lashkari, Nima and Moosa, Mudassir and Ouseph, Shoy",
    title = "{Information loss, mixing and emergent type III$_{1}$ factors}",
    eprint = "2305.16028",
    archivePrefix = "arXiv",
    primaryClass = "hep-th",
    doi = "10.1007/JHEP08(2023)111",
    journal = "JHEP",
    volume = "08",
    pages = "111",
    year = "2023"
}

@article{Gesteau:2023rrx,
    author = "Gesteau, Elliott",
    title = "{Emergent spacetime and the ergodic hierarchy}",
    eprint = "2310.13733", 
    archivePrefix = "arXiv",
    primaryClass = "hep-th",
    doi = "10.1103/PhysRevD.110.106005",
    journal = "Phys. Rev. D",
    volume = "110",
    number = "10",
    pages = "106005",
    year = "2024"
}

@article{Fredenhagen:1984dc,
    author = "Fredenhagen, Klaus",
    title = "{On the Modular Structure of Local Algebras of Observables}",
    reportNumber = "CPT-84/P-1604",
    doi = "10.1007/BF01206179",
    journal = "Commun. Math. Phys.",
    volume = "97",
    pages = "79",
    year = "1985"
}

@article{Perlmutter:2016pkf,
    author = "Perlmutter, Eric",
    title = "{Bounding the Space of Holographic CFTs with Chaos}",
    eprint = "1602.08272",
    archivePrefix = "arXiv",
    primaryClass = "hep-th",
    doi = "10.1007/JHEP10(2016)069",
    journal = "JHEP",
    volume = "10",
    pages = "069",
    year = "2016"
}

@article{El-Showk:2011yvt,
    author = "El-Showk, Sheer and Papadodimas, Kyriakos",
    title = "{Emergent Spacetime and Holographic CFTs}",
    eprint = "1101.4163",
    archivePrefix = "arXiv",
    primaryClass = "hep-th",
    doi = "10.1007/JHEP10(2012)106",
    journal = "JHEP",
    volume = "10",
    pages = "106",
    year = "2012"
}

@article{Chandrasekaran:2022eqq,
    author = "Chandrasekaran, Venkatesa and Penington, Geoff and Witten, Edward",
    title = "{Large N algebras and generalized entropy}",
    eprint = "2209.10454",
    archivePrefix = "arXiv",
    primaryClass = "hep-th",
    doi = "10.1007/JHEP04(2023)009",
    journal = "JHEP",
    volume = "04",
    pages = "009",
    year = "2023"
}

@article{Camargo:2025zxr,
    author = "Camargo, Hugo A. and Fu, Yichao and Jahnke, Viktor and Pal, Kuntal and Kim, Keun-Young",
    title = "{Quantum Signatures of Chaos from Free Probability}",
    eprint = "2503.20338",
    archivePrefix = "arXiv",
    primaryClass = "hep-th",
    month = "3",
    year = "2025"
}

@article{Chemissany:2025vye,
    author        = {Chemissany, Wissam and Gesteau, Elliott and Jahn, Alexander and Murphy, Daniel C. and Shaposhnik, Leo},
    title         = {On Infinite Tensor Networks, Complementary Recovery and Type II Factors},
    journal       = {Journal of Physics A: Mathematical and Theoretical},
    year          = {2025},
    eprint        = {2504.00096},
    archivePrefix = {arXiv},
    primaryClass  = {hep-th},
    month         = {3},
    url           = {http://iopscience.iop.org/article/10.1088/1751-8121/ae0edd},
    doi           = {10.1088/1751-8121/ae0edd}
}

@article{Heemskerk:2009pn,
    author = "Heemskerk, Idse and Penedones, Joao and Polchinski, Joseph and Sully, James",
    title = "{Holography from Conformal Field Theory}",
    eprint = "0907.0151",
    archivePrefix = "arXiv",
    primaryClass = "hep-th",
    reportNumber = "NSF-KITP-09-110",
    doi = "10.1088/1126-6708/2009/10/079",
    journal = "JHEP",
    volume = "10",
    pages = "079",
    year = "2009"
}

@book{Peskin:1995ev,
    author = "Peskin, Michael E. and Schroeder, Daniel V.",
    title = "{An Introduction to quantum field theory}",
    doi = "10.1201/9780429503559",
    isbn = "978-0-201-50397-5, 978-0-429-50355-9, 978-0-429-49417-8",
    publisher = "Addison-Wesley",
    address = "Reading, USA",
    year = "1995"
}

@article{Kudler-Flam:2023hkl,
    author = "Kudler-Flam, Jonah and Leutheusser, Samuel and Rahman, Adel A. and Satishchandran, Gautam and Speranza, Antony J.",
    title = "{Covariant regulator for entanglement entropy: Proofs of the Bekenstein bound and the quantum null energy condition}",
    eprint = "2312.07646",
    archivePrefix = "arXiv",
    primaryClass = "hep-th",
    doi = "10.1103/PhysRevD.111.105001",
    journal = "Phys. Rev. D",
    volume = "111",
    number = "10",
    pages = "105001",
    year = "2025"
}

@article{Sewell:1982zz,
    author = "Sewell, Geoffrey L.",
    title = "{Quantum fields on manifolds: PCT and gravitationally induced thermal states}",
    doi = "10.1016/0003-4916(82)90285-8",
    journal = "Annals Phys.",
    volume = "141",
    pages = "201--224",
    year = "1982"
}

@article{Bisognano:1975ih,
    author = "Bisognano, J. J and Wichmann, E. H.",
    title = "{On the Duality Condition for a Hermitian Scalar Field}",
    doi = "10.1063/1.522605",
    journal = "J. Math. Phys.",
    volume = "16",
    pages = "985--1007",
    year = "1975"
}

@article{Wiesbrock:1992mg,
    author = "Wiesbrock, H. W.",
    title = "{Half sided modular inclusions of von Neumann algebras}",
    reportNumber = "SFB-288-17",
    doi = "10.1007/BF02098019",
    journal = "Commun. Math. Phys.",
    volume = "157",
    pages = "83--92",
    year = "1993",
    note = "[Erratum: Commun.Math.Phys. 184, 683--685 (1997)]"
}

@book{Haag:1996hvx,
    author = "Haag, Rudolf",
    title = "{Local Quantum Physics}",
    doi = "10.1007/978-3-642-61458-3",
    isbn = "978-3-540-61049-6, 978-3-642-61458-3",
    publisher = "Springer",
    address = "Berlin",
    series = "Theoretical and Mathematical Physics",
    year = "1996"
}

@article{Dong:2016eik,
    author = "Dong, Xi and Harlow, Daniel and Wall, Aron C.",
    title = "{Reconstruction of Bulk Operators within the Entanglement Wedge in Gauge-Gravity Duality}",
    eprint = "1601.05416",
    archivePrefix = "arXiv",
    primaryClass = "hep-th",
    reportNumber = "NSF-KITP-16-005",
    doi = "10.1103/PhysRevLett.117.021601",
    journal = "Phys. Rev. Lett.",
    volume = "117",
    number = "2",
    pages = "021601",
    year = "2016"
}

@book{fetter2003quantum,
  author    = {Fetter, Alexander L. and Walecka, John Dirk},
  title     = {Quantum Theory of Many-Particle Systems},
  series    = {Dover Books on Physics},
  publisher = {Dover Publications},
  address   = {New York},
  year      = {2003},
  month     = jun,
  day       = {20},
  edition   = {Reprint edition},
  isbn      = {978-0486428277}
}

@book{sunder_invitation1987,
  author    = {Sunder, V. S.},
  title     = {An Invitation to {von} {Neumann} Algebras},
  series    = {Universitext},
  edition   = {1},
  publisher = {Springer–Verlag},
  address   = {New York, NY, USA},
  year      = {1987},
  isbn      = {978-0-387-96356-3},
  doi       = {10.1007/978-1-4613-8669-8},
  url       = {https://doi.org/10.1007/978-1-4613-8669-8},
  pages     = {172},
}

@article{Rehren:1999jn,
    author = "Rehren, Karl-Henning",
    title = "{Algebraic holography}",
    eprint = "hep-th/9905179",
    archivePrefix = "arXiv",
    doi = "10.1007/PL00001009",
    journal = "Annales Henri Poincare",
    volume = "1",
    pages = "607--623",
    year = "2000"
}

@article{Faulkner:2022ada,
    author = "Faulkner, Thomas and Li, Min",
    title = "{Asymptotically isometric codes for holography}",
    eprint = "2211.12439",
    archivePrefix = "arXiv",
    primaryClass = "hep-th",
    month = "11",
    year = "2022",
note=""
}

@article{Kitaev:2017awl,
    author = "Kitaev, Alexei and Suh, S. Josephine",
    title = "{The soft mode in the Sachdev-Ye-Kitaev model and its gravity dual}",
    eprint = "1711.08467",
    archivePrefix = "arXiv",
    primaryClass = "hep-th",
    doi = "10.1007/JHEP05(2018)183",
    journal = "JHEP",
    volume = "05",
    pages = "183",
    year = "2018"
}

@article{Gesteau:2024rpt,
  author  = {Gesteau, Elliott and Liu, Hong},
  title   = {Toward stringy horizons},
  archive = {hep-th}, 
  eprint  = {2408.12642}, year    = {2024},
  month   = aug,
  note    = {MIT-CTP/5751}
}

@article{Leutheusser:2021frk,
    author = "Leutheusser, Samuel Aaron Wehlau and Liu, Hong",
    title = "{Emergent Times in Holographic Duality}",
    eprint = "2112.12156",
    archivePrefix = "arXiv",
    primaryClass = "hep-th",
    reportNumber = "MIT-CTP/5382",
    doi = "10.1103/PhysRevD.108.086020",
    journal = "Phys. Rev. D",
    volume = "108",
    number = "8",
    pages = "086020",
    year = "2023"
}

@article{Leutheusser:2021qhd,
    author = "Leutheusser, Samuel and Liu, Hong",
    title = "{Causal connectability between quantum systems and the black hole interior in holographic duality}",
    eprint = "2110.05497",
    archivePrefix = "arXiv",
    primaryClass = "hep-th",
    reportNumber = "MIT-CTP/5335",
    doi = "10.1103/PhysRevD.108.086019",
    journal = "Phys. Rev. D",
    volume = "108",
    number = "8",
    pages = "086019",
    year = "2023"
}

@article{Leutheusser:2022bgi,
    author = "Leutheusser, Sam and Liu, Hong",
    title = "{Subregion-subalgebra duality: Emergence of space and time in holography}",
    eprint = "2212.13266",
    archivePrefix = "arXiv",
    primaryClass = "hep-th",
    doi = "10.1103/PhysRevD.111.066021",
    journal = "Phys. Rev. D",
    volume = "111",
    number = "6",
    pages = "066021",
    year = "2025"
}

@article{tHooft:1973alw,
    author = "'t Hooft, Gerard",
    editor = "Taylor, J. C.",
    title = "{A Planar Diagram Theory for Strong Interactions}",
    reportNumber = "CERN-TH-1786",
    doi = "10.1016/0550-3213(74)90154-0",
    journal = "Nucl. Phys. B",
    volume = "72",
    pages = "461",
    year = "1974"
}

@article{Kajuri:2020vxf,
    author = "Kajuri, Nirmalya",
    title = "{Lectures on Bulk Reconstruction}",
    eprint = "2003.00587",
    archivePrefix = "arXiv",
    primaryClass = "hep-th",
    doi = "10.21468/SciPostPhysLectNotes.22",
    journal = "SciPost Phys. Lect. Notes",
    volume = "22",
    pages = "1",
    year = "2021"
}

@article{Gubser:1998bc,
    author = "Gubser, S. S. and Klebanov, Igor R. and Polyakov, Alexander M.",
    title = "{Gauge theory correlators from noncritical string theory}",
    eprint = "hep-th/9802109",
    archivePrefix = "arXiv",
    reportNumber = "PUPT-1767",
    doi = "10.1016/S0370-2693(98)00377-3",
    journal = "Phys. Lett. B",
    volume = "428",
    pages = "105--114",
    year = "1998"
}

@article{Hamilton:2005ju,
    author = "Hamilton, Alex and Kabat, Daniel N. and Lifschytz, Gilad and Lowe, David A.",
    title = "{Local bulk operators in AdS/CFT: A Boundary view of horizons and locality}",
    eprint = "hep-th/0506118",
    archivePrefix = "arXiv",
    reportNumber = "BROWN-HET-1448, CU-TP-1130",
    doi = "10.1103/PhysRevD.73.086003",
    journal = "Phys. Rev. D",
    volume = "73",
    pages = "086003",
    year = "2006"
}

@article{Hamilton:2006az,
    author = "Hamilton, Alex and Kabat, Daniel N. and Lifschytz, Gilad and Lowe, David A.",
    title = "{Holographic representation of local bulk operators}",
    eprint = "hep-th/0606141",
    archivePrefix = "arXiv",
    reportNumber = "CU-TP-1149",
    doi = "10.1103/PhysRevD.74.066009",
    journal = "Phys. Rev. D",
    volume = "74",
    pages = "066009",
    year = "2006"
}

@article{Festuccia:2006sa,
    author = "Festuccia, Guido and Liu, Hong",
    title = "{The Arrow of time, black holes, and quantum mixing of large N Yang-Mills theories}",
    eprint = "hep-th/0611098",
    archivePrefix = "arXiv",
    reportNumber = "MIT-CTP-3783",
    doi = "10.1088/1126-6708/2007/12/027",
    journal = "JHEP",
    volume = "12",
    pages = "027",
    year = "2007"
}

@article{Festuccia:2005pi,
    author = "Festuccia, Guido and Liu, Hong",
    title = "{Excursions beyond the horizon: Black hole singularities in Yang-Mills theories. I.}",
    eprint = "hep-th/0506202",
    archivePrefix = "arXiv",
    reportNumber = "MIT-CTP-3641",
    doi = "10.1088/1126-6708/2006/04/044",
    journal = "JHEP",
    volume = "04",
    pages = "044",
    year = "2006"
}

@article{Gesteau:2024dhj,
    author = "Gesteau, Elliott and Santilli, Leonardo",
    title = "{Explicit large $N$ von Neumann algebras from matrix models}",
    eprint = "2402.10262",
    archivePrefix = "arXiv",
    primaryClass = "hep-th",
    month = "2",
    year = "2024"
}

@article{Christos:2021wno,
    author = "Christos, Maine and Haehl, Felix M. and Sachdev, Subir",
    title = "{Spin liquid to spin glass crossover in the random quantum Heisenberg magnet}",
    eprint = "2110.00007",
    archivePrefix = "arXiv",
    primaryClass = "cond-mat.str-el",
    doi = "10.1103/PhysRevB.105.085120",
    journal = "Phys. Rev. B",
    volume = "105",
    number = "8",
    pages = "085120",
    year = "2022"
}

@article{Sachdev:1992fk,
    author = "Sachdev, Subir and Ye, Jinwu",
    title = "{Gapless spin fluid ground state in a random, quantum Heisenberg magnet}",
    eprint = "cond-mat/9212030",
    archivePrefix = "arXiv",
    reportNumber = "PRINT-93-0077",
    doi = "10.1103/PhysRevLett.70.3339",
    journal = "Phys. Rev. Lett.",
    volume = "70",
    pages = "3339",
    year = "1993"
}

@article{Georges_2001,
   title={Quantum fluctuations of a nearly critical Heisenberg spin glass},
   volume={63},
   ISSN={1095-3795},
   url={http://dx.doi.org/10.1103/PhysRevB.63.134406},
   DOI={10.1103/physrevb.63.134406},
   number={13},
   journal={Physical Review B},
   publisher={American Physical Society (APS)},
   author={Georges, A. and Parcollet, O. and Sachdev, S.},
   year={2001},
   month=mar }

@article{Maldacena:2016hyu,
    author = "Maldacena, Juan and Stanford, Douglas",
    title = "{Remarks on the Sachdev-Ye-Kitaev model}",
    eprint = "1604.07818",
    archivePrefix = "arXiv",
    primaryClass = "hep-th",
    doi = "10.1103/PhysRevD.94.106002",
    journal = "Phys. Rev. D", 
    volume = "94",
    number = "10",
    pages = "106002",
    year = "2016"
}

@article{Sherrington:75,
  title = {Solvable Model of a Spin-Glass},
  author = {Sherrington, David and Kirkpatrick, Scott},
  journal = {Phys. Rev. Lett.},
  volume = {35},
  issue = {26},
  pages = {1792--1796},
  numpages = {0},
  year = {1975},
  month = {Dec},
  publisher = {American Physical Society},
  doi = {10.1103/PhysRevLett.35.1792},
  url = {https://link.aps.org/doi/10.1103/PhysRevLett.35.1792}
}

@article{Parisi:79,
  title = {Infinite Number of Order Parameters for Spin-Glasses},
  author = {Parisi, G.},
  journal = {Phys. Rev. Lett.},
  volume = {43},
  issue = {23},
  pages = {1754--1756},
  numpages = {0},
  year = {1979},
  month = {Dec},
  publisher = {American Physical Society},
  doi = {10.1103/PhysRevLett.43.1754},
  url = {https://link.aps.org/doi/10.1103/PhysRevLett.43.1754}
}

@article{Parisi:83,
  title = {Order Parameter for Spin-Glasses},
  author = {Parisi, Giorgio},
  journal = {Phys. Rev. Lett.},
  volume = {50},
  issue = {24},
  pages = {1946--1948},
  numpages = {0},
  year = {1983},
  month = {Jun},
  publisher = {American Physical Society},
  doi = {10.1103/PhysRevLett.50.1946},
  url = {https://link.aps.org/doi/10.1103/PhysRevLett.50.1946}
}

@book{mezard1986spin,
  title     = {Spin Glass Theory and Beyond: An Introduction to the Replica Method and Its Applications},
  author    = {M. M\'ezard and G. Parisi and M. A. Virasoro},
  series    = {Lecture Notes in Physics},
  publisher = {World Scientific},
  year      = {1986},
  month     = nov,
  pages     = {476},
  doi       = {10.1142/0271}
}

@article{Trunin:2020vwy,
    author = "Trunin, Dmitrii A.",
    title = "{Pedagogical introduction to the Sachdev{\textendash}Ye{\textendash}Kitaev model and two-dimensional dilaton gravity}",
    eprint = "2002.12187",
    archivePrefix = "arXiv",
    primaryClass = "hep-th",
    doi = "10.3367/UFNe.2020.06.038805",
    journal = "Usp. Fiz. Nauk",
    volume = "191",
    number = "3",
    pages = "225--261",
    year = "2021"
}

@article{Bray_1980,
doi = {10.1088/0022-3719/13/24/005},
url = {https://dx.doi.org/10.1088/0022-3719/13/24/005},
year = {1980},
month = {aug},
publisher = {},
volume = {13},
number = {24},
pages = {L655},
author = {A J Bray and M A Moore},
title = {Replica theory of quantum spin glasses},
journal = {Journal of Physics C: Solid State Physics}
}

@misc{KitaevTalks,
  author       = {Kitaev, Alexei},
  title        = {A Simple Model of Quantum Holography},
  howpublished = {Talks at the Kavli Institute for Theoretical Physics, University of California, Santa Barbara},
  note         = {April 7, 2015 and May 27, 2015. 
                  Available at \href{http://online.kitp.ucsb.edu/online/entangled15/kitaev/}{KITP Talk 1} 
                  and \href{http://online.kitp.ucsb.edu/online/entangled15/kitaev2/}{KITP Talk 2}.},
  year         = {2015}
}

@article{Castellani2005,
    author = "Castellani, Tommaso and Cavagna, Andrea",
    title = "{Spin-glass theory for pedestrians}",
    doi = "10.1088/1742-5468/2005/05/P05012",
    journal = "J. Phys. A",
    volume = "2005",
    number = "05",
    pages = "P05012",
    year = "2005"
}

@article{Kirkpatrick1987,
  title = {p-spin-interaction spin-glass models: Connections with the structural glass problem},
  author = {Kirkpatrick, T. R. and Thirumalai, D.},
  journal = {Phys. Rev. B},
  volume = {36},
  issue = {10},
  pages = {5388--5397},
  numpages = {0},
  year = {1987},
  month = {Oct},
  publisher = {American Physical Society},
  doi = {10.1103/PhysRevB.36.5388},
  url = {https://link.aps.org/doi/10.1103/PhysRevB.36.5388}
}

@article{Cugliandolo1993,
  title = {Analytical solution of the off-equilibrium dynamics of a long-range spin-glass model},
  author = {Cugliandolo, L. F. and Kurchan, J.},
  journal = {Phys. Rev. Lett.},
  volume = {71},
  issue = {1},
  pages = {173--176},
  numpages = {0},
  year = {1993},
  month = {Jul},
  publisher = {American Physical Society},
  doi = {10.1103/PhysRevLett.71.173},
  url = {https://link.aps.org/doi/10.1103/PhysRevLett.71.173}
}

@article{Gur-Ari:2018okm,
    author = "Gur-Ari, Guy and Mahajan, Raghu and Vaezi, Abolhassan",
    title = "{Does the SYK model have a spin glass phase?}",
    eprint = "1806.10145",
    archivePrefix = "arXiv",
    primaryClass = "hep-th",
    doi = "10.1007/JHEP11(2018)070",
    journal = "JHEP",
    volume = "11",
    pages = "070",
    year = "2018"
}

@article{Anschuetz:2024naj,
    author = "Anschuetz, Eric R. and Chen, Chi-Fang and Kiani, Bobak T. and King, Robbie",
    title = "{Strongly Interacting Fermions Are Nontrivial yet Nonglassy}",
    eprint = "2408.15699",
    archivePrefix = "arXiv",
    primaryClass = "quant-ph",
    doi = "10.1103/cbqf-d24r",
    journal = "Phys. Rev. Lett.",
    volume = "135",
    number = "3",
    pages = "030602",
    year = "2025"
}

@article{Lin2023,
    author = "Lin, Henry W. and Stanford, Douglas",
    title = "{A symmetry algebra in double-scaled SYK}",
    eprint = "2307.15725",
    archivePrefix = "arXiv",
    primaryClass = "hep-th",
    doi = "10.21468/SciPostPhys.15.6.234",
    journal = "SciPost Phys.",
    volume = "15",
    number = "6",
    pages = "234",
    year = "2023"
}

@article{Lin:2022rbf,
    author = "Lin, Henry W.",
    title = "{The bulk Hilbert space of double scaled SYK}",
    eprint = "2208.07032",
    archivePrefix = "arXiv",
    primaryClass = "hep-th",
    doi = "10.1007/JHEP11(2022)060",
    journal = "JHEP",
    volume = "11",
    pages = "060",
    year = "2022"
}

@article{Crisanti1992,
    author = "Crisanti, A. and Sommers, H. -J.",
    title = "{The spherical p-spin interaction spin glass model: the statics}",
    doi = "10.1007/BF01309287",
    journal = "Z. Phys. B",
    volume = "87",
    number = "3",
    pages = "341--354",
    year = "1992"
}

@article{Cugliandolo2001,
  title = {Imaginary-time replica formalism study of a quantum spherical p-spin-glass model},
  author = {Cugliandolo, Leticia F. and Grempel, D. R. and da Silva Santos, Constantino A.},
  journal = {Phys. Rev. B},
  volume = {64},
  issue = {1},
  pages = {014403},
  numpages = {26},
  year = {2001},
  month = {Jun},
  publisher = {American Physical Society},
  doi = {10.1103/PhysRevB.64.014403},
  url = {https://link.aps.org/doi/10.1103/PhysRevB.64.014403}
}

@article{Zlokapa:2025bbm,
    author = "Zlokapa, Alexander and Kiani, Bobak T. and Anschuetz, Eric R.",
    title = "{Average-case quantum complexity from glassiness}",
    eprint = "2510.08497",
    archivePrefix = "arXiv",
    primaryClass = "quant-ph",
    month = "10",
    year = "2025"
}

@article{Jensen:2024dnl,
    author = "Jensen, Kristan and Raju, Suvrat and Speranza, Antony J.",
    title = "{Holographic observers for time-band algebras}",
    eprint = "2412.21185",
    archivePrefix = "arXiv",
    primaryClass = "hep-th",
    doi = "10.1007/JHEP06(2025)242",
    journal = "JHEP",
    volume = "06",
    pages = "242",
    year = "2025"
}

@article{Lashkari:2024lkt,
    author = "Lashkari, Nima and Leung, Kwing Lam and Moosa, Mudassir and Ouseph, Shoy",
    title = "{Modular Intersections, Time Interval Algebras and Emergent AdS$_2$}",
    eprint = "2412.19882",
    archivePrefix = "arXiv",
    primaryClass = "hep-th",
    month = "12",
    year = "2024"
}

@article{Vardhan:2025rky,
    author = "Vardhan, Shreya and Wang, Jinzhao",
    title = "{Free mutual information and higher-point OTOCs}",
    eprint = "2509.13406",
    archivePrefix = "arXiv",
    primaryClass = "quant-ph",
    month = "9",
    year = "2025"
}

@article{Parker:2018yvk,
    author = "Parker, Daniel E. and Cao, Xiangyu and Avdoshkin, Alexander and Scaffidi, Thomas and Altman, Ehud",
    title = "{A Universal Operator Growth Hypothesis}",
    eprint = "1812.08657",
    archivePrefix = "arXiv",
    primaryClass = "cond-mat.stat-mech",
    doi = "10.1103/PhysRevX.9.041017",
    journal = "Phys. Rev. X",
    volume = "9",
    number = "4",
    pages = "041017",
    year = "2019"
}

@book{StreaterWightman2001,
  author    = {Raymond F. Streater and Arthur S. Wightman},
  title     = {PCT, Spin and Statistics, and All That},
  series    = {Princeton Landmarks in Mathematics and Physics},
  publisher = {Princeton University Press},
  address   = {Princeton, NJ},
  year      = {2001},
  edition   = {Revised},
  isbn      = {978-0-691-07062-9},
  pages     = {224},
  doi       = {10.1515/9781400884230}
}

@article{Belin:2025nqd,
    author = "Belin, Alexandre and Bintanja, Suzanne and Castro, Alejandra and Knop, Waltraut",
    title = "{Symmetric product orbifold universality and the mirage of an emergent spacetime}",
    eprint = "2502.01734",
    archivePrefix = "arXiv",
    primaryClass = "hep-th",
    reportNumber = "YITP-SB-2025-01",
    doi = "10.1007/JHEP05(2025)190",
    journal = "JHEP",
    volume = "05",
    pages = "190",
    year = "2025"
}
\end{document}